\documentclass[%
 aip,
 amsmath,amssymb,
 reprint,%
]{revtex4-1}

\usepackage{graphicx}
\usepackage{dcolumn}
\usepackage{xcolor}
\usepackage[utf8]{inputenc}
\usepackage[T1]{fontenc}
\usepackage{mathptmx}
\usepackage{etoolbox}

\makeatletter
\def\@email#1#2{%
 \endgroup
 \patchcmd{\titleblock@produce}
  {\frontmatter@RRAPformat}
  {\frontmatter@RRAPformat{\produce@RRAP{*#1\href{mailto:#2}{#2}}}\frontmatter@RRAPformat}
  {}{}
}%
\makeatother
\begin{document}

\preprint{AIP/123-QED}

\title[Quantum-enhanced super-sensitivity of MZI using squeezed Kerr state]{Quantum-enhanced super-sensitivity of Mach-Zehnder interferometer using squeezed Kerr state}

\author{Dhiraj Yadav}\homepage{These authors contributed equally to this work.}
\author{Gaurav Shukla}\homepage{These authors contributed equally to this work.}
\author{Priyanka Sharma}
 \author{Devendra Kumar Mishra$^{*,}$}
\email{kndmishra@gmail.com}

\affiliation{Department of Physics, Institute of Science, Banaras Hindu University, Varanasi-221005, India}
\date{\today}

\begin{abstract}
We study the phase super-sensitivity of a Mach-Zehnder interferometer (MZI) with the squeezed Kerr state (SKS) and coherent states as the inputs. We discuss the lower bound in phase sensitivity by considering the quantum Fisher information (QFI) and corresponding quantum Cramér–Rao bound (QCRB). With the help of single intensity detection (SID), intensity difference detection (IDD) and homodyne detection (HD) schemes, we find that our scheme gives better sensitivity in both the lossless as well as in lossy conditions as compared to the combination of well-known results of inputs as coherent plus vacuum, coherent plus squeezed vacuum and double coherent state as the inputs. Because of the possibility of the generation of SKS with the present available quantum optical techniques, we expect that SKS may be an alternative nonclassical resource for the improvement in the phase super-sensitivity of the MZI under realistic scenarios. 
\end{abstract}

\maketitle

\section{\label{sec:level1}Introduction}

In the science and technology of metrology, the central task is to perform accurate measurement of certain parameters \cite{1976Helstrom}. By exploiting the peculiar properties of quantum mechanics \cite{RevModPhys.90.035005,2015PrOpt..60..345D, Lawrie2019, 1976Helstrom} quantum metrology deals with the precision measurements of such a parameter  \cite{doi:10.1080/00107510802091298}.  For precision measurement of certain parameters, which are not measurable directly via conventional techniques, phase estimation \cite{RevModPhys.90.035005,2015PrOpt..60..345D, Lawrie2019, 1976Helstrom} via optical interferometers plays an important role. In order to perform the phase measurement scheme, usually, SU(2) or SU(1,1) based interferometers \cite{PhysRevA.33.4033} are used. SU(2)-type interferometers, like Michelson interferometer (MI) and MZI, are based on the passive type beam splitters while, SU(1,1)-type interferometers are based on the active elements, e.g., optical parametric amplifiers (OPAs), in place of the beam splitters \cite{PhysRevA.33.4033, doi:10.1063/5.0004873, Hudelist2014, PhysRevLett.117.013001}.

Theoretically as well as experimentally, it is found that the performance of the interferometer maximally depends on the input light sources \cite{doi:10.1080/00107510802091298}. If we compare the performance of the interferometer, which depends maximally on the input light, then the ascending order of the performance would be thermal lights, coherent lights and maximally by the nonclassical lights \cite{doi:10.1080/00107510802091298, Shukla:22}.

Non-classical lights are the class of lights which are only understood by the quantum mechanical theories \cite{ Mishra_2021}, e.g., single photon state \cite{PhysRevD.9.853}, squeezed states \cite{Walls1983}, twins Fock states \cite{Friedrichs1954MathematicalAO}, Schr\"{o}dinger’s cat states \cite{PhysRevLett.57.13, PhysRevA.45.6811, RevModPhys.85.1103}, N00N states \cite{PhysRevLett.85.2733, RevModPhys.81.865, PhysRevA.78.063828, PhysRevA.67.053801, PhysRevA.90.045804}, etc. The well-known combination of coherent and squeezed vacuum as the input states \cite{PhysRevD.23.1693, PhysRevLett.99.223602, Barnett2003} became a famous choice, for their good performance in the low as well as high-power range \cite{Gard2017, PhysRevA.100.063821, PhysRevLett.111.173601} of interferometry and, also, due to its very recent application in the gravitational wave detection \cite{Acernese_2014, Oelker:14, Mehmet_2018, PhysRevLett.121.173601, PhysRevLett.123.231107}. Since the seminal work by Caves \cite{PhysRevD.23.1693}, four decades ago, squeezing-assisted optical interferometry \cite{Breitenbach1997, Andersen_2016} become a centrepiece of theoretical \cite{PhysRevLett.111.173601, Shukla:21, PhysRevA.98.043856, PhysRevA.96.052118} and experimental \cite{Lawrie2019} quantum metrology. An MZI injected by an intense light in coherent state (which is an eigenstate of the annihilation operator, denoted as $|\beta\rangle$) at one input port and squeezed-vacuum state at the other input port can attain the phase sensitivity $\Delta\phi=e^{-r}/\sqrt{\Bar{n}}
$, where $r (\geq 0)$ is the squeezing parameter, $\Bar{n}$ is the total average number of photons inside the interferometer and $\phi$ is the relative phase shift between the two arms of the interferometer \cite{PhysRevD.23.1693}. This scheme can beat the shot-noise limit (SNL), $\Delta\phi_{SNL}=1/\sqrt{\Bar{n}}$, by an amount depending on the squeezing parameter $r$. To date, squeeze factors \footnote{In experiment, decibel [dB] is a common unit of squeezing. The degree of squeezing in dB is calculated according to $10\ \text{log}_{10}e^{-2r}$, where $r$ is the squeezing parameter.} of more than 10 dB have been observed in several experiments \cite{SCHNABEL20171, PhysRevLett.117.110801, Schonbeck:18}. Interestingly, for the similar intensity, MZI can achieve Heisenberg limit (HL), $\Delta\phi_{HL}=1/{\Bar{n}}$, for the coherent and squeezed-vacuum states at the inputs \cite{PhysRevLett.100.073601}. 

In order to achieve the HL, we must have squeezing in such amount that $e^{-r}\approx1/\sqrt{\Bar{n}}$. Experimently it is very tough for higher values of photons \cite{SCHNABEL20171, PhysRevLett.117.110801, Schonbeck:18}. In 2000, Dowling group proposed a new type of state known as NOON state \cite{PhysRevLett.85.2733} by which one can easily attain the HL. But for higher photon numbers, generation of NOON state is very much challenging \cite{PhysRevLett.121.160502, PhysRevA.103.013315}. So, this has led to open a new area of research having a significant amount of work in the optimisation and generation of the nonclassical light \cite{Andersen_2016}. In order to generate the nonclassical light, special types of nonlinear materials \cite{Mishra2020} and techniques \cite{Yadav2021} are being used. For example, parametric processes in second-order $\chi^{(2)}$ media generate squeezing and entanglement \cite{Dirmeier:20}. On the other hand, the Kerr effect occurring in third-order nonlinear $\chi^{(3)}$ media is being used to perceive quantum nondemolition measurements \cite{PhysRevA.46.1499, RevModPhys.68.755}, to generate quantum superpositions \cite{PhysRevLett.57.13, doi:10.1080/09500348714550721} as well as squeezing \cite{PhysRevA.53.1096} and entanglement \cite{PhysRevA.73.012306, PhysRevLett.86.4267}. The most common methods of generating nonclassical light are parametric downconversion (PDC) \cite{Dirmeier:20}, four-wave mixing \cite{PhysRevA.78.043816, https://doi.org/10.48550/arxiv.2201.10935} and the Kerr effect \cite{Hosaka:15}. Unlike the former two approaches, squeezing via the Kerr interaction \cite{PhysRevLett.66.153, PhysRevLett.81.2446} is inherently phase-matched, which allows for flexibility in the wavelength of the probe light. These features meant the utilisation of the Kerr effect is a robust and flexible approach. The Kerr interaction requires high optical powers to reach sufficient nonlinearity and this is commonly achieved by using ultrashort pulses \cite{Bergman:91, Bergman:94}. However, this requires careful control of the pulses, since dispersion can act to spread out the pulse and, therefore reduce the nonlinearity. Control of pulse spreading may be achieved by generating optical solitons, where the nonlinearity and dispersion are perfectly balanced \cite{PhysRevLett.66.153, Yu2001SolitonSA}. For Kerr squeezing, the possibility of using materials such as optical fibre lends significant flexibility and it does not require a cavity \cite{PhysRevLett.66.153} to enhance the strength of interaction as well as simplify the experimental requirements.  For high nonlinearities and good transparency, chalcogenide and tellurite like glasses \cite{Anashkina:20, ANASHKINA2021104843, math10193477} are good candidates for optical fibre fabrications.

However, there is a problem in identifying the Kerr squeezing in direct detection experiments on the resulting fields. The nonlinear index of refraction of the medium modifies the phases of the number states in the initial coherent state but photon statistics of the field remain Poissonian \cite{PhysRevA.49.2033}. This produces a squeezing ellipse in phase space that is oriented in an oblique direction, neither in the direction of the phase nor the direction of the amplitude quadrature \cite{sizmann1999v}. Consequently, it has been difficult to identify Kerr squeezing. A method to observe Kerr squeezing in fibre was proposed by Kitagawa and Yamamoto \cite{PhysRevA.34.3974}. They used an asymmetric Sagnac loop to displace the squeezing ellipse in phase space so that the short axis lined up with the amplitude quadrature \cite{PhysRevLett.81.2446, Krylov:98}. Another method proposed by Gerry \& Grobe \cite{PhysRevA.49.2033}. They applied the Kerr state to a two-photon parametric process such as degenerate parametric down-conversion or degenerate four-wave mixing. The resulting radiation is termed as `squeezed Kerr states (SKS)', since the unitary evolution operator in the interaction picture for these processes is of the form of squeeze operator \cite{PhysRevA.49.2033}. It is shown that the effect of the Kerr medium on photon statistics becomes readily apparent by the application of the squeeze operator to a Kerr state. Statistics of SKS are sub-Poissonian as well as super-Poissonian and its quadrature squeezing is improved. Further, higher-order nonclassical properties of SKS have been studied by Mishra \cite{MISHRA20103284}. 

So, based on the nonclassical properties of the SKS \cite{PhysRevA.49.2033, MISHRA20103284}, we are motivated to study the improvement in phase sensitivity by using the SKS at the input of MZI. We will also compare our results with the previous studies and results. To the best of our knowledge, there is no such study has occurred.

Since, the total phase shift in MZI arms is $\phi=\phi_{us} + \phi_{es}$, where $\phi_{us}$ is phase change due to unknown source and $\phi_{es}$ is phase change due to controllable experimental setup. But, $\phi_{us} << \phi_{es}$, so we can write $\phi\approx\phi_{es}$ and for simplicity, throughout the paper, we can ignore the suffix of $\phi_{es}$ and denote it as $\phi$. Therefore, for precise observation, one must adjust the phase difference between the arms of the MZI nearest to $\phi$. We perform the SID, IDD and HD schemes \cite{Gard2017, PhysRevA.98.043856} and we calculate the optimal solutions of phase sensitivity for all these detection schemes. In this paper, we are focused on the single parameter case, i.e., the phase change only in one arm of the interferometer. So, with the help of the single parameter quantum Fisher information (QFI) and its associated quantum Cramér–Rao bound (QCRB)\cite{PhysRevLett.72.3439, BRAUNSTEIN1996135, PhysRevLett.111.173601}, we analyse the better performance in our MZI setup.

The paper is organized as follows. In section \ref{section 2}, we discuss the basics of interferometry by using different types of detection schemes and briefly discuss the SKS. Section \ref{section 3} describes the phase sensitivity of MZI with SKS and coherent states as the inputs of MZI under the lossless condition. Section \ref{section 4} describes the phase sensitivity of MZI with SKS and coherent states as the inputs of MZI under lossy conditions. In Section \ref{section 5}, we conclude our results.

\section{Basics of phase sensing and parameter estimation with MZI and generation of SKS}\label{section 2}
In this section, we will discuss the basics of phase estimation and parameter estimation with MZI and the generation of SKS. Here, we will see the operator transformations under the MZI action and the method of standard error propagation formula to calculate the phase sensitivity of the interferometer for different detection schemes. We will discuss the lower bound in phase sensitivity by considering the single parameter quantum Fisher information (QFI) and corresponding quantum Cramér–Rao bound (QCRB). Further, we will review the interaction of coherent light with Kerr media followed by the squeezer.

\subsection{Interferometery with MZI and detection operators}\label{subA}
\begin{figure}
\includegraphics[width=7cm, height=6cm]{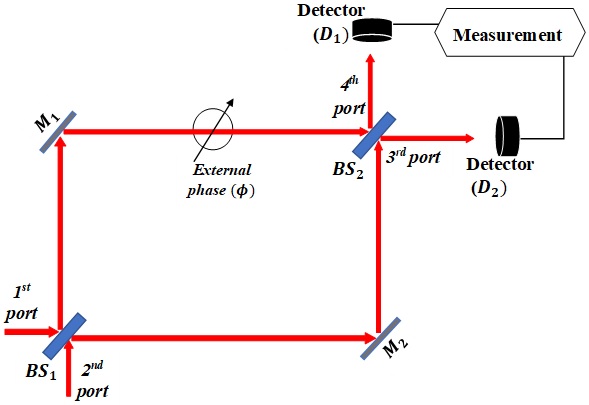}
\caption{\label{fig:1} The schematic block diagram of MZI having two input and two output ports including two 50:50 beam splitters ($BS_1$ and $BS_2$), two mirrors ($M_1$ and $M_2$) and two detectors ($D_1$ and $D_2$).}
\end{figure}

A standard MZI setup, as shown in Fig. \ref{fig:1}, consists of two input ports ($1^{st}$ port and $2^{nd}$ port) and two output ports ($3^{rd}$ port and $4^{th}$ port) associated with two 50:50 beam splitters, two mirrors and two detectors. 
Interferometry is a three-step process: probe state preparation, state evolution and measurement. In probe preparation, two input states are mixed via the first beam splitter $BS_1$ and the input/output mode transformations follow the SU(2) transformation \cite{gerry_knight_2004, loudon2000quantum}. So, input/output  relations via the $BS_1$ is written as \cite{loudon2000quantum}
\begin{equation}
    \begin{pmatrix}
    \hat{a}_{1_{out}}\\
    \hat{a}_{2_{out}}
\end{pmatrix}=\begin{pmatrix}
    i\mathcal{R}_1&\mathcal{T}_1\\
    \mathcal{T}_1&i\mathcal{R}_1
\end{pmatrix}
\begin{pmatrix}
    \hat{a}_{1_{in}}\\
    \hat{a}_{2_{in}}
\end{pmatrix}.
\label{eq:3}
\end{equation}
Here $i=\sqrt{-1}$, $\mathcal{R}_1$ ($\mathcal{T}_1$) represents the reflection (transmission) coefficient and $\hat{a}_{1_{in}, 2_{in}}$ ($\hat{a}_{1_{out}, 2_{out}}$) are the input (output) annihilation operators of the $BS_1$. Probe state, during the propagation inside the interferometer, experiences the phase change and the resulting phase change between the probes is, say $\phi$. So we consider the phase, $\phi$, in any one of the arms as a single parameter case and allow the probes to recombine via the second beam splitter $BS_2$. Transformation of annihilation operators, again, follows the relation, Eq. \eqref{eq:3}. Therefore, working with the two 50:50 beam splitters, i.e., $\mathcal{R}_{1,2}=\mathcal{T}_{1,2}=1/\sqrt{2}$, and considering phase change during reflection (transmission) equal to $\pi/2\ (0)$, the relation between output and input annihilation operators are written as,
\begin{eqnarray}
\hat{a}_{3}=-e^{\frac{i\phi}{2}}\left(-\sin\left(\frac{\phi}{2}\right)\hat{a}_{1}+\cos\left(\frac{\phi}{2}\right)\hat{a}_{2}\right)\label{2q},\\
\hat{a}_{4}=-e^{\frac{i\phi}{2}}\left(\cos\left(\frac{\phi}{2}\right)\hat{a}_{1}+\sin\left(\frac{\phi}{2}\right)\hat{a}_{2}\right).
\label{1q}
\end{eqnarray}
Here $\phi$ is the phase difference between the two arms and $\hat{a}_{1,2}$ ($\hat{a}_{3,4}$) are the input (output) annihilation operators of the MZI. As a final step, data collected by the detectors at the output of MZI are analysed by using statistical protocols and formulae. From the standard error propagation formula, the phase sensitivity of the interferometer reads
\begin{equation}
    \Delta\phi=\frac{\Delta\hat{L}(\phi)}{\left|\frac{\partial\langle\hat{L}(\phi)\rangle}{\partial\phi}\right|}.
    \label{a1}
\end{equation}
Here $\hat{L}(\phi)$ is an observable containing information about the phase change and $\Delta\hat{L}(\phi)$ is the standard deviation of $\hat{L}(\phi)$ defined by,
\begin{equation}
    \Delta\hat{L}(\phi)=\sqrt{\langle\hat{L}^2(\phi)\rangle - \langle\hat{L}(\phi)\rangle^2}.
\end{equation}
Here, $\langle...\rangle$ is the expectation value of the operator with respect to the state $|\psi\rangle_{in}=|\psi_1\rangle\otimes|\psi_2\rangle$, where $|\psi_1\rangle$ and $|\psi_2\rangle$ represent the input states at ports $1^{st}$ and $2^{nd}$, respectively. 

We are considering three detection schemes: SID, IDD, and HD. In quantum mechanics, there must be an operator associated with the observable. Similarly, each detection scheme is also associated with an operator. For example, the operator associated with the SID scheme is
\begin{equation}
    \hat{L}_{sid}(\phi) \equiv \hat{N}_3,\label{6l}
\end{equation}
for IDD scheme
\begin{equation}
    \hat{L}_{idd}(\phi) \equiv \hat{N}_3- \hat{N}_4,\label{7l}
\end{equation}
and HD scheme perform at $3^{rd}$ port having quadrature operator
\begin{equation}
    \hat{L}_{hd}(\phi)=\frac{1}{\sqrt{2}}(\hat{a}_{3}+\hat{a}^\dagger_{3}),\label{8l}
\end{equation}
Here, $\hat{N}_3(=\hat{a}^\dagger_3\hat{a}_3)$ and $\hat{N}_4(=\hat{a}^\dagger_4\hat{a}_4)$ are the photon number operators for the $3^{rd}$ and $4^{th}$ ports of the MZI, respectively and $\hat{a}_{3}$ ($\hat{a}^\dagger_{3}$) is the annihilation (creation) operator associated with the $3^{rd}$ port.

\subsection{Quantum parameter estimation}
In an estimation procedure, our task is to estimate the value of a parameter, the total phase change in the arms of the MZI in the present case, from the data collected by $n$ measurements, say $\{v_1, v_2,...,v_n\}$. For measurement, we consider an operator $\hat{O}$. The estimated value of the
parameter will be characterized by the statistical error $\delta\phi$, whose lower bound is the Cramér-Rao bound (CRB) \cite{cramer1999mathematical,doi:10.1142/S0219749909004839},
\begin{equation}
    \delta\phi^2\geq\frac{1}{nF(\phi)}.
\end{equation}
Here $n$ stands for the number of measurements and $F(\phi)$ denotes the classical Fisher information (CFI), defined by
\begin{equation}
    F(\phi)=\left\langle\left(\frac{\partial\ln{p(v|\phi)}}{\partial\phi}\right)^2\right\rangle.
    \label{218}
\end{equation}
Here, $p(v|\phi)$ is the probability that the outcome of a measurement is $v$ when the value of the parameter is $\phi$ and $\langle...\rangle$ is the expectation value over the probability distribution $p(v|\phi)$. If we consider the quantum system, then $p(v|\phi)=\text{Tr}(\hat{\rho}_\phi\hat{\Pi}_v)$, where
$\hat{\rho}_\phi$ is the density operator and $\hat{\Pi}_v$ is the positive operator-valued measure (POVM) operator for the outcome $v$. By introducing the symmetric logarithmic derivative (SLD), $\hat{L}_\phi$, defined by $2\partial_\phi\hat{\rho}_\phi=\hat{L}_\phi\hat{\rho}_\phi+\hat{\rho}_\phi \hat{L}_\phi$, Eq. \eqref{218} can be written as,
\begin{equation}
    F(\phi)=\left \langle\frac{\text{Re}[\text{Tr}(\hat{\rho}_\phi\hat{\Pi}_v \hat{L}_\phi)]^2}{\text{Tr}(\hat{\rho}_\phi\hat{\Pi}_v)}\right\rangle.
\end{equation}
By maximizing $F(\phi)$ over all possible quantum measurements on the quantum system, we obtain the QFI as \cite{doi:10.1142/S0219749909004839}:
\begin{equation}
    {F}_Q=\text{Tr}(\hat{\rho}_\phi \hat{L}_\phi^2),
\end{equation}
and, thus, QCRB is \cite{PhysRevLett.111.173601}
\begin{equation}
\Delta\phi_{QCRB} \geq \frac{1}{\sqrt{{F_Q}}}\label{q13}.
\end{equation}
This gives us the ultimate precision achievable on the estimation of $\phi$ independent of a quantum measurement. Density operator, $\hat{\rho}_\phi$, of a mixed state can be written in terms of the complete basis, $\{|i\rangle$\}, as $\hat{\rho}_\phi=\sum_ip_i|i\rangle\langle i|$ with $p_i\geq0$ and $\sum_ip_i=1$. Therefore, QFI can be written as \cite{PhysRevLett.72.3439, 2015PrOpt..60..345D}
\begin{equation}
    F_Q=\sum\limits_{i,i'}\frac{2}{p_i+p_{i'}}|\langle i|\partial_\phi\hat{\rho}_\phi|i'\rangle|^2.\label{314}
\end{equation}
The density operator for a pure state $|\psi\rangle$ is $\hat{\rho}=|\psi\rangle\langle\psi|$. For this case, the Eq. \eqref{314} becomes \cite{PhysRevLett.72.3439, PhysRevA.102.013704}
\begin{equation}
    {F}_Q=4\left(\langle \partial_\phi \psi | \partial_\phi \psi \rangle - |\langle \partial_\phi \psi | \psi \rangle|^2\right)
    \label{eq11}.
\end{equation}
Therefore, in order to find the single parameter QFI we use the Eq. \eqref{eq11}. Where $|\psi\rangle$ is the state just before the second beam splitter (Fig. \ref{fig:1}) and $\partial_{\phi}=\partial/\partial\phi$. So, by using the transformations given in section \ref{subA}, after some straightforward calculations, Eq. \eqref{eq11} can be written as \cite{PhysRevLett.111.173601, PhysRevA.102.013704}
\begin{equation}
    \begin{split}
        F_Q=2(g_1+g_2)+g_5+g_4-(g_1-g_2)^2-g_6+g^2_{10}\\
        +2i(g_{11}+g_{12}+g_{10}\times(g_1+g_2-1)),\label{q16}
    \end{split}
\end{equation}
Here, $g_i$ with $i={1,2,...,12}$ are given in Eq.~\eqref{14} of Appendix \ref{appendix A} and complete expressions for the expectation values and their relations are given in the Appendix \ref{appendix}.

\subsection{Interaction of coherent light with Kerr medium followed by the squeezer: squeezed-Kerr state (SKS)}
The Kerr effect, also known as a quadratic electro-optic effect, is a change in the refractive index of a material medium in response to an applied electromagnetic field. In the Kerr medium, the electromagnetic field interacts with the material medium having third-order nonlinearity where the refractive index is intensity dependent \cite{Agarwal_2012, gerry_knight_2004}. Hamiltonian, $\hat{H}$, of this quantum mechanical system can be written as \cite{Agarwal_2012, gerry_knight_2004} 
\begin{equation}
    \hat{H}=\hslash{\omega}\hat{a}^\dagger\hat{a}+\hslash{\chi^{(3)}}\hat{a}^{\dagger2}\hat{a}^{2}, 
\end{equation}
where $\hslash$ is the Dirac constant, $\omega$ is the frequency, $\hat{a}~(\hat{a}^\dagger)$ is annihilation (creation) operator of the oscillator and $\chi^{(3)}$ is the third-order susceptibility of the Kerr medium. The operator associated with the Kerr medium is
\begin{equation}
    \hat{U}_K(\gamma)=\exp[-i\gamma\hat{n}(\hat{n}-1)],
\end{equation}
 where $\hat{n}$ $(=\hat{a}^{\dagger}\hat{a})$ is the photon number operator and,
 \begin{equation}
     \gamma=\chi^{(3)}{L}/v,\label{K19}
 \end{equation}
 with $L$ is the length of the Kerr medium and $v$ is the velocity of the electromagnetic field into the Kerr medium. As we can see from Eq. \eqref{K19}, $\gamma$ tells about the interaction time of the electromagnetic field with $\chi^{(3)}$ material medium, so, we can call $\gamma$ as Kerr interaction coefficient. Factor $\gamma$ plays an important role in our discussion to understand the effect of the Kerr medium on the phase sensitivity of the MZI.
 
 The squeezing operator for a single-mode electromagnetic field is  \cite{gerry_knight_2004}
 \begin{equation}
     \hat{S}(\zeta)\equiv\exp[\frac{1}{2}(\zeta\hat{a}^{\dagger2}-\zeta^{*}\hat{a}^{2})],
 \end{equation}
 where, $\zeta=re^{i\theta}$, with $r$ as the squeezing parameter and $\theta$ gives the phase information of the squeezing.
 
 In order to find the SKS, simply, inject the coherent state into the  Kerr medium followed by squeezing. So, injecting the light beam in coherent state $|\beta\rangle$ through the material having $\chi^{(3)}$ non-linearity results Kerr state which can be written as
 \begin{equation}
     |\psi_{K}\rangle=\hat{U}_{K}(\gamma)|\beta\rangle.
 \end{equation}
 Now, the application of the  squeezing operator on the Kerr state gives us the SKS, i.e., 
 \begin{equation}
     |\psi_{SK}\rangle=\hat{S}(\zeta)\hat{U}_{K}(\gamma)|\beta\rangle.
 \end{equation}
 
\begin{figure}
\includegraphics[width=8.5cm, height=8cm]{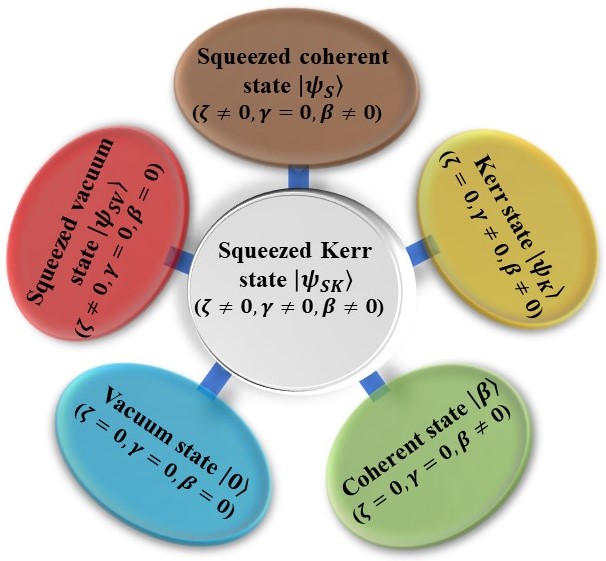}
\caption{\label{fig:1s} Diagram shows the special cases of SKS.}
\end{figure}

 Let, $\hat{a}$ be the field operator for the coherent state, and then the field operator associated with SKS can be written as
 \begin{equation}
     \hat{a}(\zeta,\gamma)=\hat{U}^{\dagger}_{K}(\gamma)\hat{S}^{\dagger}(\zeta)\hat{a}\hat{S}(\zeta)\hat{U}_{K}(\gamma).
     \label{6a}
 \end{equation}
 Since, from the Baker–Campbell–Hausdorff formula \cite{messiah1999quantum}, we can write the squeezing field operator
 \begin{eqnarray}
 \hat{b}(\zeta)\equiv\hat{S}(\zeta)\hat{b}\hat{S}(\zeta)
 =\hat{b}\cosh{r}+e^{\iota\theta}\hat{b}^{\dagger}\sinh{r}\label{27l},
 \end{eqnarray}    
and Kerr state field operator as
\begin{eqnarray}
\hat{c}(\gamma)\equiv\hat{U_{K}}^{\dagger}(\gamma)\hat{c}\hat{U}_{K}(\gamma)=e^{-2\iota\gamma\hat{c}^{\dagger}\hat{c}}\hat{c}\label{28l}.
\end{eqnarray}
Therefore, the field operator for the SKS can be written as
\begin{eqnarray}
\hat{a}(\zeta,\gamma)=e^{-2i\gamma\hat{a}^{\dagger}\hat{a}}\hat{a}\cosh{r}+e^{i\theta}\hat{a}^{\dagger}e^{2i\gamma\hat{a}^{\dagger}\hat{a}}\sinh{r}.\label{eq10}
\end{eqnarray}
We use this operator in our calculations in order to find the general result for the phase sensitivity using the SKS as one of the inputs of the MZI. It is to be noted that the SKS, as shown in Fig. \ref{fig:1s}, can be used to generate different states under different conditions.

\section{Coherent state and SKS as the inputs of MZI under the lossless condition} \label{section 3}
In this section, we will discuss the phase sensitivity of the MZI using the coherent state, $|\alpha\rangle$, and SKS, $|\psi_{SK}\rangle$, as the inputs in $1^{st}$ and $2^{nd}$ input ports, respectively (Fig. \ref{fig:1}). The phase sensitivity, i.e., $\Delta\phi$, for different detection schemes are the functions of different six variables ($\phi$, $\alpha$, $\beta$, $\theta$, $\gamma$ and $r$). In order to optimize the parameters, we divide our results into different sub-sections by considering the special cases.

For the lossless case, the relation between output and input annihilation operators of the MZI is given by Eqs. \eqref{2q} and \eqref{1q}. For the case of squeezed Kerr state and coherent state as the inputs of the MZI, i.e., $|\psi\rangle_{in}=|\alpha\rangle_1\otimes|\psi_{SK}\rangle_2$, the detailed expressions of the corresponding phase sensitivity associated with SID, IDD and HD schemes are derived in Appendix \ref{appendix A} and can be written as
\begin{widetext}
 \begin{equation}
\begin{split}
    \Delta\phi_{sid}=\frac{2}{\left|(g_1-g_2)sin\phi-g_3\cos\phi\right|}\left(g_1\sin^2\left(\frac{\phi}{2}\right)+g_2\cos^2\left(\frac{\phi}{2}\right)+(g_4-g_2^2)\cos^4\left(\frac{\phi}{2}\right)+(g_5-g_1^2)\sin^4\left(\frac{\phi}{2}\right)\right.\\
    +\left.\frac{1}{4}\left(g_6-g_3^2-2g_1g_2+4g_7\right)\sin^2\phi-\left(\frac{1}{2}g_3+(g_8-g_2g_3) \cos^2\left(\frac{\phi}{2}\right)+(g_9-g_1g_3)\sin^2\left(\frac{\phi}{2}\right)\right)\sin{\phi}\right)^\frac{1}{2},
\end{split}
 \label{B35}
 \end{equation}
    \begin{equation}
    \begin{split}
     \Delta\phi_{idd}=\frac{\sqrt{g_1+g_2+(g_4+g_5-2g_7-(g_2-g_1)^2)\cos^2\phi+(g_6+2g_7-g_3^2)\sin^2\phi-(g_8-g_9-g_3(g_2-g_1))\sin2\phi}}{\left|(g_1-g_2)\sin\phi-g_3\cos\phi\right|}\label{B36},
 \end{split}
 \end{equation}
\begin{equation}
    \begin{split}
        \Delta\phi_{hd}=\frac{\sqrt{2\cos^2\left(\frac{\phi}{2}\right)\left(Re(e^{i\phi}(\Delta\hat{a}_2)^2)+(g_2-\langle\hat{a}_{2}^{\dagger}\rangle\langle\hat{a}_{2}\rangle)\right)+1}}{\left|Re(e^{i\phi}(\langle\hat{a}_1\rangle - i\langle\hat{a}_2\rangle))\right|}.
        \end{split}\label{B37}
\end{equation}
    \end{widetext}
Where, $\Delta\phi_{sid}$, $\Delta\phi_{idd}$ and $\Delta\phi_{hd}$ are the phase sensitivity for SID, IDD and HD schemes in lossless case, respectively. Here, the $g_i$ with $i={1,2,...,9}$ are given in Eq.~\eqref{14} and complete expressions for the expectation values and their relations are given in the Appendix \ref{appendix}. Calculation of QCRB is done by using Eq. \eqref{q16} in Eq. \eqref{q13}.

Note that, throughout the calculations, without loss of generalities, we take $\alpha=|\alpha|$, $\beta=|\beta|$ and $\theta=\pi$, since, analytically we found that $\theta=\pi$ gives better results in comparison to other values of $\theta$ in all the cases.

The central task of our work is to find the effect of Kerr nonlinearity (in terms of $\gamma$) and Kerr nonlinearity with squeezing parameter ($r$) on the $\Delta\phi$ of the MZI for three different detection schemes. So, in each case, we will try to see the variation of $\Delta\phi$ with $\gamma$. To be more clear, we divide our discussions into two parts: (i) $|\psi\rangle_{in}=|0\rangle_1\otimes|\psi_{SK}\rangle_2$; (ii)$|\psi\rangle_{in}=|\alpha\rangle_1\otimes|\psi_{SK}\rangle_2$.
 
\subsection{Vacuum state and SKS as inputs of MZI}\label{subsec iii(a)}
For the case of $|\psi\rangle_{in}=|0\rangle_1\otimes|\psi_{SK}\rangle_2$, Eqs. \eqref{B35}, \eqref{B36} and \eqref{B37} become
 \begin{equation}
    \Delta\phi_{sid}=\frac{\sqrt{g_2+(g_4 -g_2^2)\cos^2\left(\frac{\phi}{2}\right)}}{\left|g_2\sin\left(\frac{\phi}{2}\right)\right|},
    \label{e12}
\end{equation}
\begin{equation}
    \Delta\phi_{idd}=\frac{\sqrt{g_2+(g_4 -g_2^2)\cos^2\phi}}{\left|g_2\sin\phi\right|},
    \label{e121}
\end{equation}
and
\begin{equation}
    \begin{split}
        \Delta\phi_{hd}=\frac{1}{\left|Re(-ie^{i\phi}\langle\hat{a}_2\rangle)\right|}\left(2\cos^2\left(\frac{\phi}{2}\right)\times\right.\\
        \left.\left(Re(e^{i\phi}(\Delta\hat{a}_2)^2)+(g_2-\langle\hat{a}_{2}^{\dagger}\rangle\langle\hat{a}_{2}\rangle)\right)+1\right)^\frac{1}{2},
        \end{split}\label{a18}
\end{equation}
respectively. From Eqs. \eqref{e12} and \eqref{e121}, we can see that in the case of optimal phase, i.e., for which phase sensitivity becomes maximum, phase sensitivity for SID and IDD schemes become
\begin{equation}
    \Delta\phi_{sid}=\Delta\phi_{idd}|_{|\alpha|=0}=1/\sqrt{g_2}.
    \label{e122}
\end{equation}
Here $g_2$, the total photon number in the second input port and given in Eq. \eqref{h2}.

\begin{figure}
\includegraphics[width=8.5cm, height=6.5cm]{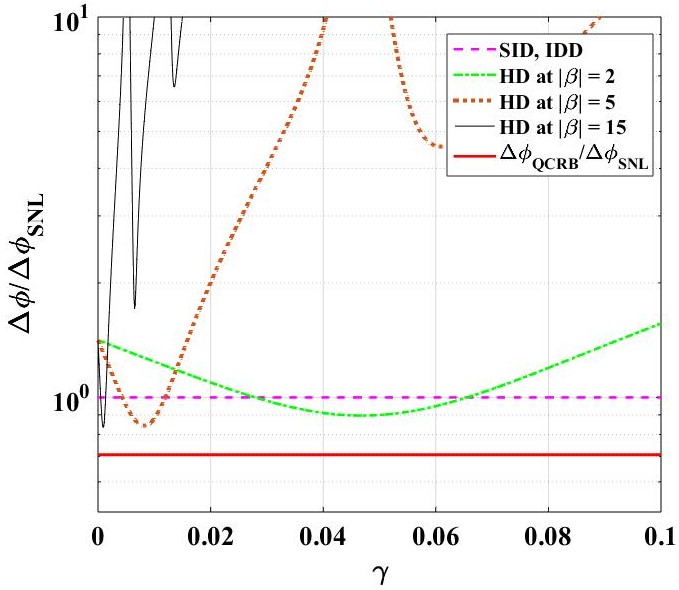}
\caption{\label{fig:2a} Plots of $\Delta\phi/\Delta\phi_{SNL}$ with $\gamma$ for different values of $|\beta|$ (= 2, 5 and 15). One can see that, $\Delta\phi/\Delta\phi_{SNL}$ equal to 1 for both SID and IDD schemes, i.e., SID and IDD saturate the SNL, while for HD scheme $\Delta\phi/\Delta\phi_{SNL}$ $<$ 1, i.e., $\Delta\phi_{hd}$ beats the SNL. Other parameters are $r=0,~\theta=\pi,~|\alpha|=0$ and $\phi=\pi,~\pi/2,~7\pi/4$ for SID, IDD and HD schemes, respectively.}
\end{figure}

\textit{Case (i)}: For $r=0$ case, i.e. $|\psi\rangle_{in}=|0\rangle_1\otimes|\psi_{K}\rangle_2$. In this case $g_2$ becomes $|\beta|^2$. This implies that, for optimal values of $\phi$, $\Delta\phi_{sid}$ and $\Delta\phi_{idd}$ saturates the SNL and is independent from the $\gamma$ (Fig. \ref{fig:2a}). Further, for the HD scheme, we see that in the case of $r=0$, the Eq. \eqref{a18} becomes
\begin{equation}
    \begin{split}
        \Delta\phi_{hd}=\frac{1}{|c\sin(\phi-s)|}\left(\frac{1}{g_2}+2\cos^2\left(\frac{\phi}{2}\right)\left(1-c^2\right.\right.\\
        +\left.c_2\cos(\phi-2\gamma-s_2)-\left.c\cos(\phi-2s)\right)\right)^\frac{1}{2},
        \end{split}\label{h12}
\end{equation}
and for $\gamma=0$,
\begin{equation}
    \begin{split}
        \Delta\phi_{hd}=\frac{1}{\sqrt{g_2}\sin(\phi)}.
        \end{split}\label{}
\end{equation}
At $\phi=\pi/2$, we get the maximum value of $\Delta\phi_{hd}$ and which is nothing but the SNL. Hence, at $r=\gamma=0$, we get $\Delta\phi_{sid}=\Delta\phi_{idd}=\Delta\phi_{hd}=\Delta\phi_{SNL}$ and these are the well know results.

\begin{figure}
\includegraphics[width=8.5cm, height=6.5cm]{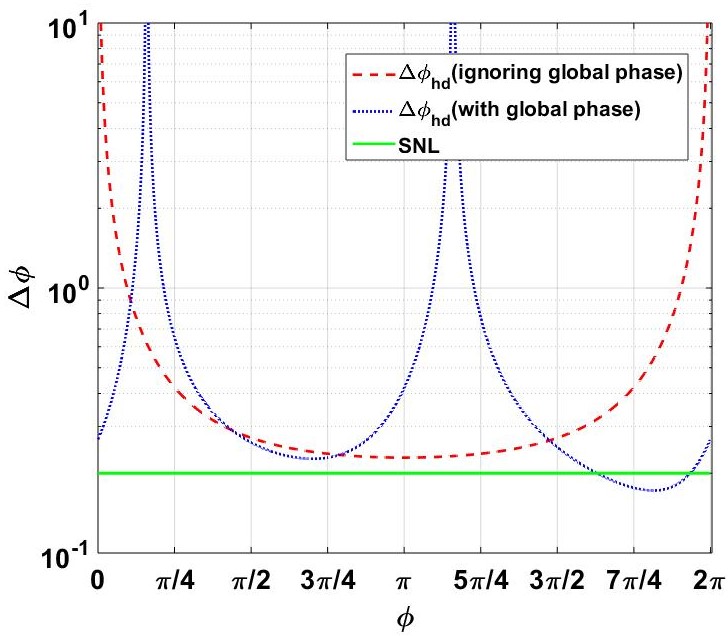}
\caption{\label{fig:2ab} Plots of phase sensitivity, $(\Delta\phi)$, with $\phi$ for $|\beta|=5$ and $\gamma=0.01$. 
$\Delta\phi_{hd}$ beats the SNL only when we consider the global phase, while in the case of without a global phase, we never beat the SNL. Other parameters are $r=0,~\theta=\pi$ and $|\alpha|=0$.}
\end{figure}

From Eq. \eqref{h12}, we find that for $r=0$, $\Delta\phi_{hd}$ is depending on $\gamma$, as one can see in Fig. \ref{fig:2a}. So, for HD scheme with $\gamma\neq0$, optimum value of $\phi$ varies with $|\beta|$. We found analytically that for wide range values of $|\beta|$ $(\sim 1$ to 100) optimum value of $\phi$ is approximately at $7\pi/4$. So, it is interesting to mention that with the optimum value of $\phi=7\pi/4$, $\Delta\phi_{hd}$ beats SNL for some non-zero values of $\gamma$ keeping first input port as vacuum (Fig. \ref{fig:2a}). This is in agreement with the recent study of  Masahiro \textit{et al.} \cite{PhysRevA.96.052118} that a system working with a vacuum state in one port and a nonclassical state on another port can beat the SNL if only one of the arms of the MZI has unknown phase shift (i.e., single parameter estimation case) and the detector uses any external phase reference and power resource during the detection process. In our work, we are taking Kerr state as a nonclassical state \cite{sizmann1999v} and we are not ignoring the global phase factor as is obvious in Eqs. \eqref{2q} and \eqref{1q}. If we ignore the global phase factor, phase sensitivity never beats the SNL, we can see this in Fig. \ref{fig:2ab}. This means that the global phase factor acts as an external phase source for the HD scheme and beating of the SNL, in Fig. \ref{fig:2a}, is not the violation of the ``no-go theorem" \cite{PhysRevA.96.052118, PhysRevD.23.1693}.

\begin{figure}
\includegraphics[width=8.5cm, height=6.5cm]{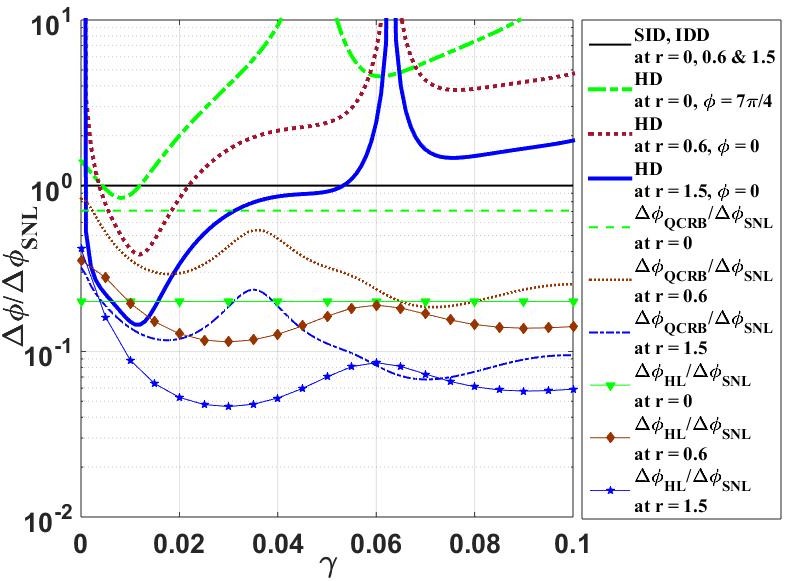}
\caption{\label{fig:3} Plots of $\Delta\phi/\Delta\phi_{SNL}$ with $\gamma$ for different values of $r$. One can see that phase sensitivity saturates the SNL for both SID and IDD schemes for $r=0$ and $r\neq0$, while phase sensitivity for the HD scheme beats the SNL. QCRB, for different values of $r$, shows the lower limit achieved by the system. Other parameters are $|\beta|=5,~\theta=\pi,~|\alpha|=0$ and $\phi=\pi,~\pi/2$ for SID, IDD schemes, respectively.}
\end{figure}

\begin{figure}
\includegraphics[width=8.5cm, height=6.5cm]{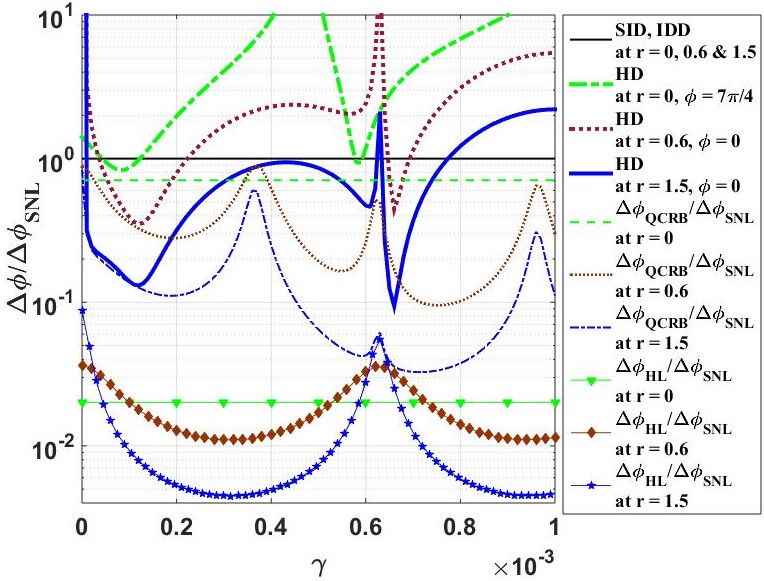}
\caption{\label{fig:3b} Plots of $\Delta\phi/\Delta\phi_{SNL}$ with $\gamma$ for different values of $r$. In the case of the SID and IDD scheme, phase sensitivity saturates the SNL for $r=0$ and $r\neq0$ both, while phase sensitivity for the HD scheme beats the SNL. QCRB, for different values of $r$, shows the lower limit achieved by the system. Other parameters are $|\beta|=50,~\theta=\pi,~|\alpha|=0$ and $\phi=\pi,~\pi/2$ for SID, IDD schemes, respectively.}
\end{figure}

We note that for the case of the HD scheme, normalised phase sensitivity ($(\Delta\phi)/SNL$) can not reach up to 1 for $\gamma=0$ (Fig. \ref{fig:2a}). This is because, here, we plot the graph at $\phi=7\pi/4$ while the optimal value of $\phi$ is $\pi/2$ for the case of $\gamma=0$. Similar cases arise for the Figs. \ref{fig:3} \& \ref{fig:3b}.

\textit{Case (ii)}: Consider $r\neq0$ case, i.e., $|\psi\rangle_{in}=|0\rangle_1\otimes|\psi_{SK}\rangle_2$. From Eqs. \eqref{e122} and \eqref{h2}, we find that $\Delta\phi_{sid}$ and $\Delta\phi_{idd}$ are dependent on $\gamma$. That means, the Kerr medium plays a role in the variation of phase sensitivity in the case of SID and IDD schemes when $r\neq0$, but it should be noted that still phase sensitivity only saturates the SNL (Figs. \ref{fig:3} \& \ref{fig:3b}). In order to visualize the effect of $\gamma$ on $\Delta\phi$ we consider two values (lower and higher energies) of $|\beta|=5$ \& 100 and see the variation of $\Delta\phi/\Delta\phi_{SNL}$ with $\gamma$ for different values of the squeezing parameter $r$ (plots are shown in Figs. \ref{fig:3} \& \ref{fig:3b}). In Figs. \ref{fig:3} \& \ref{fig:3b}, we can see that phase sensitivity for SID and IDD saturates the SNL for all values of $\gamma$. While enhancement in phase sensitivity occurs for the HD scheme significantly. We also plot $\Delta\phi_{QCRB}/\Delta\phi_{SNL}$ and $\Delta\phi_{HL}/\Delta\phi_{SNL}$ and we can see that for some values of $\gamma$ phase sensitivity for HD scheme saturates QCRB and approaches HL. As we can see phase sensitivity is enhanced with an increase in $r$, so, to enhance the precision one can use the higher squeezing for better phase sensitivity. It is important to mention here that the current record for the squeezing factor (15.3 dB or r =
1.7) is reported in \cite{PhysRevLett.117.110801}.

\begin{figure}
\includegraphics[width=8.5cm, height=6cm]{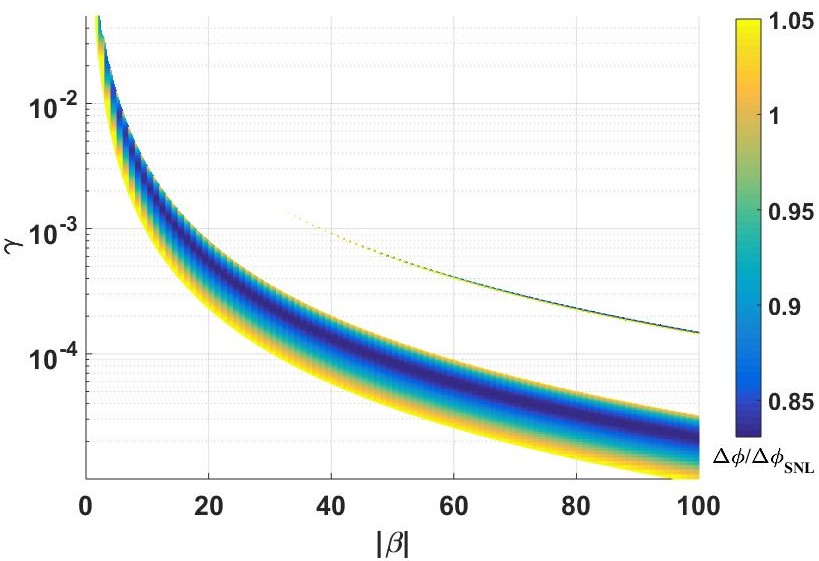}
\caption{\label{fig:3c} Colour graph, for HD scheme, in between $|\beta|$ and $\gamma$ where colour variation shows $\Delta\phi/\Delta\phi_{SNL}$. This graph gives two results, one is on increasing the value of $|\beta|$ enhancement in $\Delta\phi$ occurs, and the second is for higher values of $|\beta|$ the corresponding optimal value of $\gamma$ decreases. Other parameters are $\phi=7\pi/4,~\theta=\pi,~|\alpha|=0,~r=0$}
\end{figure}

Further, in Fig. \ref{fig:2a} and also in Figs. \ref{fig:3} \& \ref{fig:3b}, we can see that maximum phase sensitivity depends on $|\beta|$ as well as on $\gamma$ in case of HD scheme. As $|\beta|$ changes the corresponding optimal value of $\gamma$ for which we get maximum phase sensitivity also changes. So, in order to explore the effect of $|\beta|$ and $\gamma$ on $\Delta\phi_{hd}$, we plot a graph between $|\beta|$ and $\gamma$ and show the variation in phase sensitivity via colour change in the graph (Fig. \ref{fig:3c}). Fig. \ref{fig:3c} gives two important results from an experimental point of view, (i) enhancement in $\Delta\phi_{hd}$ with an increase in the value of $|\beta|$; and (ii) for higher values of $|\beta|$, optimal value of $\gamma$ decreases. Value of $\gamma$ decreases means interaction time of light with Kerr medium decreases (Eq. \eqref{K19}). Thus, we can say that the Kerr medium plays an important role in the enhancement of the phase sensitivity with the HD scheme.


\subsection{Coherent and SKS as inputs of MZI}\label{subsec iii(b)}
For the case of SKS and coherent state as the inputs of the MZI, i.e., $|\psi\rangle_{in}=|\alpha\rangle_1\otimes|\psi_{SK}\rangle_2$, the expressions of the corresponding phase sensitivity associated with SID, IDD and HD schemes are given in Eqs. \eqref{B35}, \eqref{B36} and \eqref{B37}, respectively.

\begin{figure}
\includegraphics[width=8cm, height=7cm]{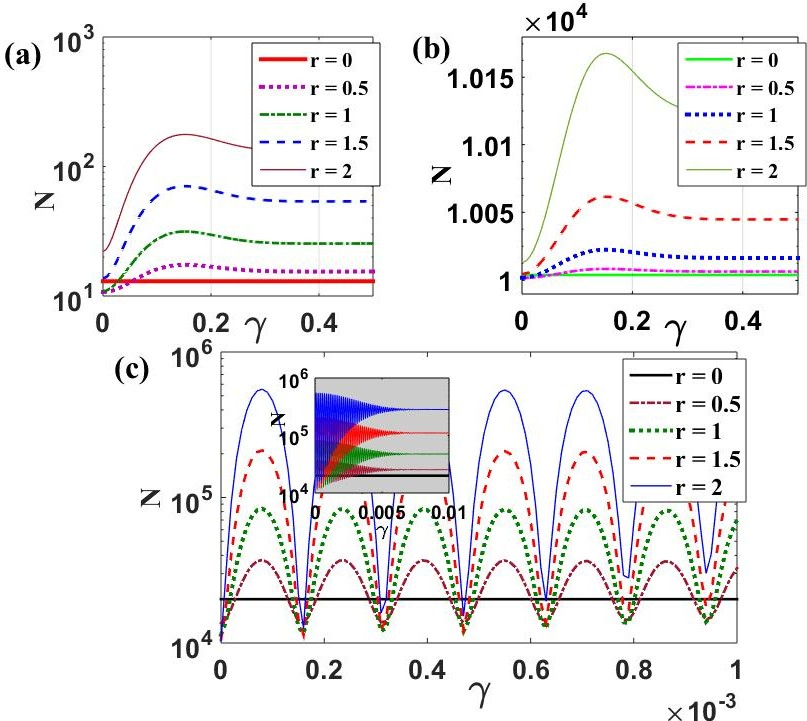}
\caption{\label{fig:4} Variation in $N$ with $\gamma$ by considering the different values of $r$ for (a) $|\alpha|=3$, $|\beta|=2$, (b) $|\alpha|=100$, $|\beta|=2$ and (c) $|\alpha|=100$, $|\beta|=100$. We find that $N$ does not vary with $\gamma$ in the case of $r=0$ but for the case $r\neq0$ rapid growth is seen. Here we take $\theta=\pi$.}
\end{figure}

Since, $\Delta\phi$ is dependent on the input number of photons, $N=g_1+g_2$, and $N$ depends on the parameters $|\alpha|,~|\beta|,~\gamma,~\theta$ and $r$, so, we can figure out the role of these parameters on $\Delta\phi$ by looking the variation of $N$ with these parameters. We have seen in the previous cases that the phase sensitivity is better for $\theta=\pi$ in comparison to the other values of $\theta$ and, so, here we consider $\theta=\pi$. Now, we see the variation in $N$ with $\gamma$ for different values of $r$ in three different cases by considering low and high energy limits at inputs, viz., (i) $|\alpha|=3$ and $|\beta|=2$, (ii) $|\alpha|=100$ and $|\beta|=2$ and (iii) $|\alpha|=100$ and $|\beta|=100$, as shown in Fig. \ref{fig:4}. We find that $N$ is independent of $\gamma$ in the case of $r=0$ but for the case $r\neq0$, rapid growth in $N$ is seen as $\gamma$ is increasing. This is because the nonlinear index of refraction of the medium modifies the phases of the number states in the initial coherent state but photon statistics of the field remain Poissonian \cite{PhysRevA.49.2033}. The effect of the Kerr medium on photon statistics becomes readily apparent by application of the squeeze operator \cite{PhysRevA.49.2033}. As we can see that, in Fig. \ref{fig:4}, in case of squeezed coherent state, $|\psi_S\rangle$ ($r\neq0$ and $\gamma=0$), photon number decreases, while for $\gamma\neq0$ it varies rapidly. That is, variation in the input number of photons, $N$, is dependent on the interaction of photons in the Kerr medium.

As a particular case, let us consider two cases: (i) $r=0$, i.e., $|\psi\rangle_{in}=|\alpha\rangle_1\otimes|\psi_{K}\rangle_2$ (when second input is in Kerr state), and (ii) $r\neq0$, i.e., $|\psi\rangle_{in}=|\alpha\rangle_1\otimes|\psi_{SK}\rangle_2$ (when second input is squeezed Kerr state).
\subsubsection{Kerr state at the second input port}\label{subsection B1}
\begin{figure}
\includegraphics[width=8cm, height=7cm]{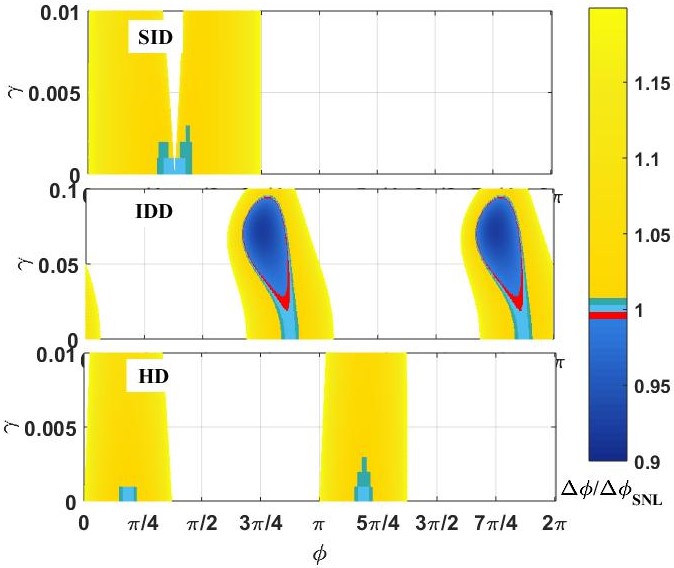}
\caption{\label{fig:3d} Colour graphs show how optimal phase $\phi_{opt}$ varies with $\phi$ and $\gamma$ in different detection schemes. We can see that phase sensitivity is better in the IDD scheme having optimal phase $3\pi/4$ or $7\pi/4$. Other parameters are $\theta=\pi,~|\alpha|=3,~|\beta|=2,~r=0$.}
\end{figure}

\begin{figure}
\includegraphics[width=8cm, height=7cm]{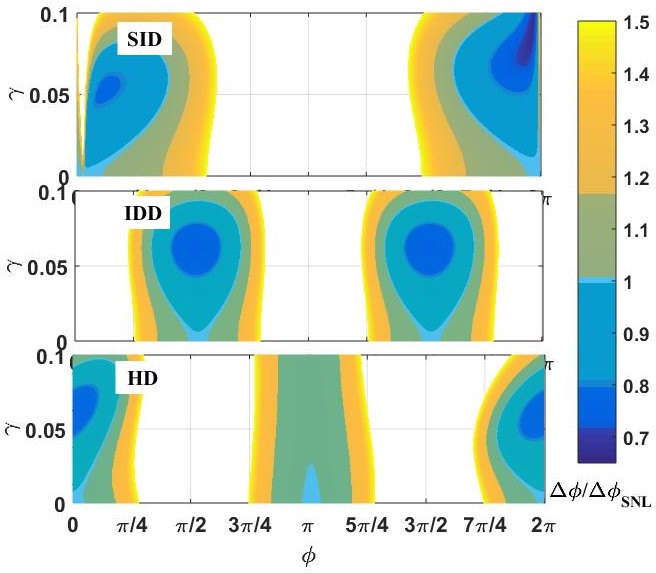}
\caption{\label{fig:3e} Colour graphs show how optimal phase $\phi_{opt}$ varies with $\phi$ and $\gamma$ in different detection schemes. We can see that phase sensitivity is better in all the detection schemes having different optimal phases. Other parameters are $\theta=\pi,~|\alpha|=50,~|\beta|=2,~r=0$}
\end{figure}

\begin{figure}
\includegraphics[width=8cm, height=7.5cm]{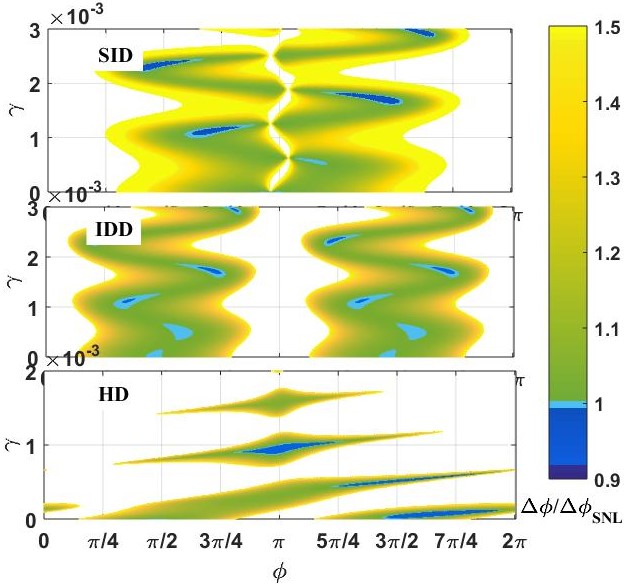}
\caption{\label{fig:3ea} Colour graphs show how optimal phase $\phi_{opt}$ varies with $\phi$ and $\gamma$ in different detection schemes. We can see that, phase sensitivity is better in HD scheme having optimal phases 0 or $2\pi$. Other parameters are $\theta=\pi,~|\alpha|=3,~|\beta|=50,~r=0$}
\end{figure}

\begin{figure}
\includegraphics[width=8.5cm, height=8cm]{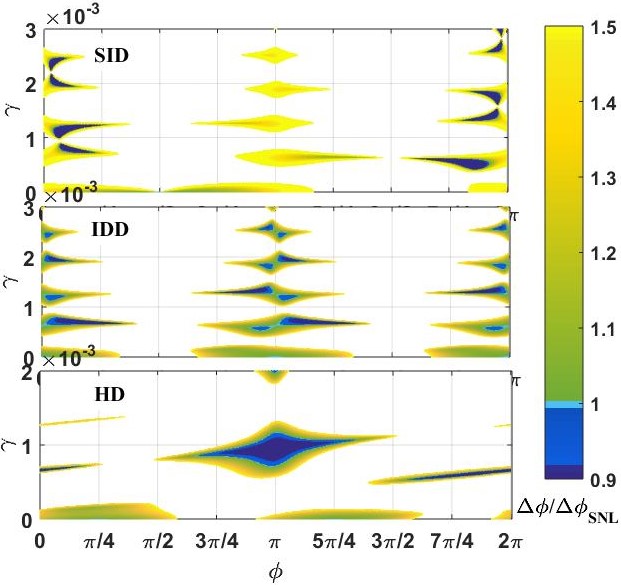}
\caption{\label{fig:3f} Colour graphs show how optimal phase $\phi_{opt}$ varies with $\phi$ and $\gamma$ in different detection schemes. We can see that phase sensitivity is better in all the detection schemes having broad optimal phase range. Other parameters are $\theta=\pi,~|\alpha|=50,~|\beta|=50,~r=0$}
\end{figure}

It is obvious from the Eqs. \eqref{B35}, \eqref{B36} \& \eqref{B37} that for $r=0$, $\Delta\phi$ still depends on $|\alpha|,~|\beta|,~\gamma$ and $\phi$. In the case of $|\alpha|=0$, the calculation of the optimal value of $\phi$ for different detection cases was straightforward but, here, it is relatively hard. So, keeping in mind the lower and high energy inputs, we consider four cases: (i) $|\alpha|=3$ and $|\beta|=2$, (ii) $|\alpha|=50$ and $|\beta|=2$, (iii) $|\alpha|=3$ and $|\beta|=50$ and (iv) $|\alpha|=50$ and $|\beta|=50$; and plot the $\phi$ \textit{vs} $\gamma$ graph as shown by Fig. \ref{fig:3d}, Fig. \ref{fig:3e}, Fig. \ref{fig:3ea} and Fig. \ref{fig:3f}, respectively, where colour variation shows $\Delta\phi/\Delta\phi_{SNL}$. We can see that phase sensitivity is better in the IDD scheme as compared to the SID and HD schemes having optimal phase $3\pi/4$ or $7\pi/4$ in case (i), as shown in Fig. \ref{fig:3d}. In cases (ii), (iii) \& (iv), all three detection schemes perform better phase sensitivity with different optimal phases, as shown in Figs. \ref{fig:3e}, \ref{fig:3ea} \& \ref{fig:3f}, respectively. Multiple colour regions in phase sensitivity pattern in case (iii) \& (iv) is because of fluctuation in $N$, as previously we saw in Fig. \ref{fig:4}(c).


So in conclusion we find that, in enhancing the phase sensitivity of the interferometer, coherent input with Kerr state is more useful than double coherent input, in approximately all the situations.

\subsubsection{Squeezed Kerr state at the second input port}\label{}
\begin{figure}[]
\includegraphics[width=8cm, height=7cm]{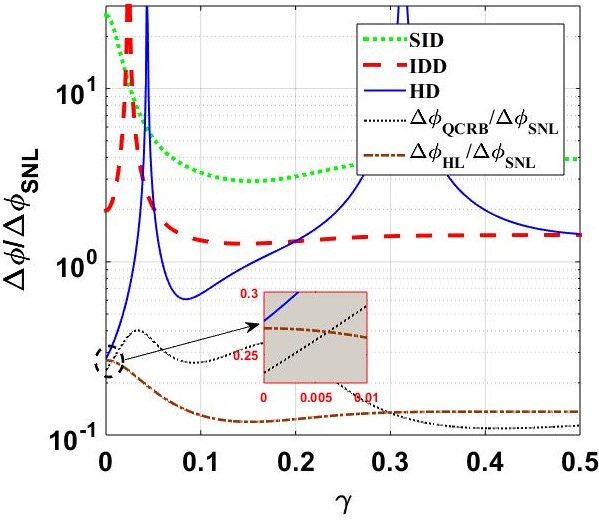}
\caption{\label{fig:5} Variation in phase sensitivity, $\Delta\phi$, with $\gamma$ when the inputs are $|\alpha|=3,~|\beta|=2$. If we see the performance of the three detection schemes, we find that the HD scheme is dominant than the IDD scheme which is doing better than the SID scheme. Other parameters are $\theta=\pi$, $r=1.5$ and $\phi=0,~\frac{\pi}{2},~\frac{9\pi}{8}$ for HD, IDD and SID schemes, respectively.}
\end{figure}

\begin{figure}[]
\includegraphics[width=7.5cm, height=6.5cm]{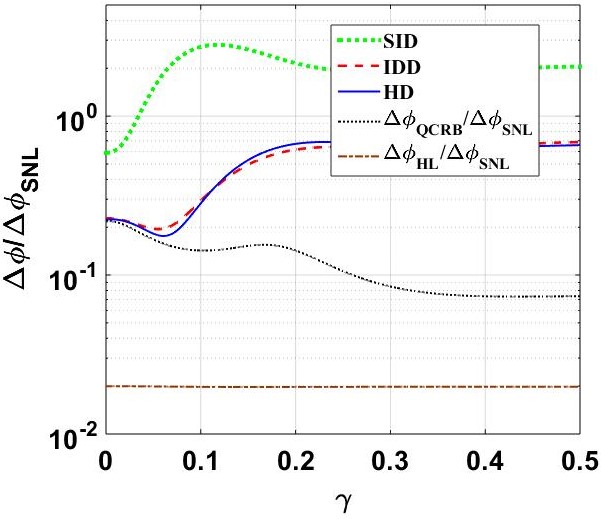}
\caption{\label{fig:5a} Variation in phase sensitivity, $\Delta\phi$, with $\gamma$ when the inputs are $|\alpha|=50,~|\beta|=2$. We can see that the Kerr medium enhances the phase sensitivity. We find that the HD scheme and IDD scheme are doing better than the SID scheme. Other parameters are $\theta=\pi$, $r=1.5$ and $\phi=0,~\frac{\pi}{2},~\frac{\pi}{4}$ for HD, IDD and SID schemes, respectively.}
\end{figure}

\begin{figure}[]
\includegraphics[width=8cm, height=7cm]{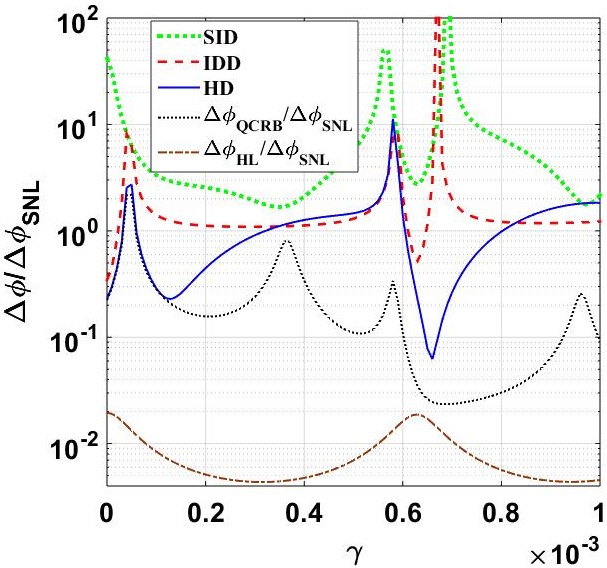}
\caption{\label{fig:5a1} Variation in phase sensitivity, $\Delta\phi$, with $\gamma$ when the inputs are $|\alpha|=50,~|\beta|=50$. We can see that the Kerr medium enhances the phase sensitivity remarkably. HD scheme is more dominant than the IDD scheme which is doing better than the SID scheme. Other parameters are $\theta=\pi$, $r=1.5$ and $\phi=0,~\frac{\pi}{2},~\frac{9\pi}{8}$ for HD, IDD and SID schemes, respectively.}
\end{figure}

\begin{figure}[]
\includegraphics[width=7.5cm, height=7cm]{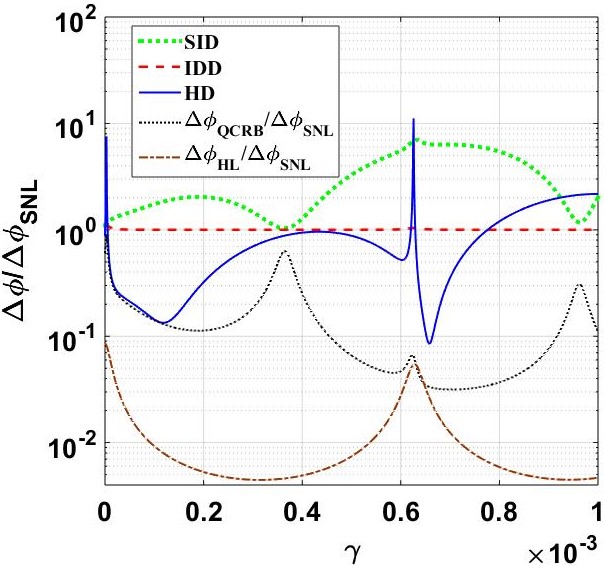}
\caption{\label{fig:5a12} Variation in phase sensitivity, $\Delta\phi$, with $\gamma$ when the inputs are $|\alpha|=3,~|\beta|=50$. We can see that the Kerr medium enhances the phase sensitivity remarkably. We see the performance of the three detection schemes, we find that the HD scheme is better than the IDD scheme which is better than the SID scheme. Other parameters are $\theta=\pi$, $r=1.5$ and $\phi=0,~\frac{\pi}{2},~\frac{9\pi}{8}$ for HD, IDD and SID schemes, respectively.}
\end{figure}

In this section, we explore the Eqs. \eqref{B35}-\eqref{B37} by considering $r\neq0$. For better phase sensitivity, optimal value of $\phi$ depends on the parameters $|\alpha|$, $|\beta|$, $\gamma$ and $r$ similar to the previous section \ref{subsection B1}. In order to explore the effect of $\gamma$ on $\Delta\phi$, for the three detection schemes, we find a nearly optimal value of $\phi$ by using the analytical method. Analytically, we find that for the cases (i) $|\alpha|=3$, $|\beta|=2$, (ii) $|\alpha|=50$, $|\beta|=2$, (iii) $|\alpha|=50$, $|\beta|=50$ and (iv) $|\alpha|=3$, $|\beta|=50$, optimal phases $\phi_{opt}$ for SID scheme will be $9\pi/8$, $\pi/4$, $9\pi/8$ and $9\pi/8$, respectively. But, all these four cases have  $\phi_{opt}=\pi/2$ for the IDD scheme and  $\phi_{opt}=0$ for the HD scheme. Since we find that squeezing triggers the photon enhancement in Kerr medium at several instances, as shown in Fig. \ref{fig:4}, so here we choose the value of $r=1.5$ for our convenience. So, we plot $\Delta\phi/\Delta\phi_{SNL}$ \textit{versus} $\gamma$ by taking $\phi_{opt}$ for $r=1.5$. Figs. \ref{fig:5}, Fig. \ref{fig:5a}, Fig. \ref{fig:5a1} and Fig. \ref{fig:5a12} show the phase sensitivity for the four cases (i), (ii), (iii) and (iv), respectively. We can see that, the Kerr medium enhances the phase sensitivity remarkably. If we look at the performance of the three detection schemes, we find that the HD scheme is dominant in all four cases than the IDD scheme which, in turn, is doing better than the SID scheme.

On comparison of the four cases (i), (ii), (iii) and (iv),  we find that increased values of $|\alpha|$ and $|\beta|$ enhance the phase sensitivity but it is important to note that variation in $|\beta|$ affects more than that for $|\alpha|$. As we can see in Fig. \ref{fig:5a1} and Fig. \ref{fig:5a12}, increase in $|\alpha|$ is less effective for the larger values of $|\beta|$. On the other hand, increase in  $|\beta|$ gets more effective even though $|\alpha|$ is large enough (Fig. \ref{fig:5a} and Fig. \ref{fig:5a1}).

\begin{figure}[]
\includegraphics[width=8cm, height=7cm]{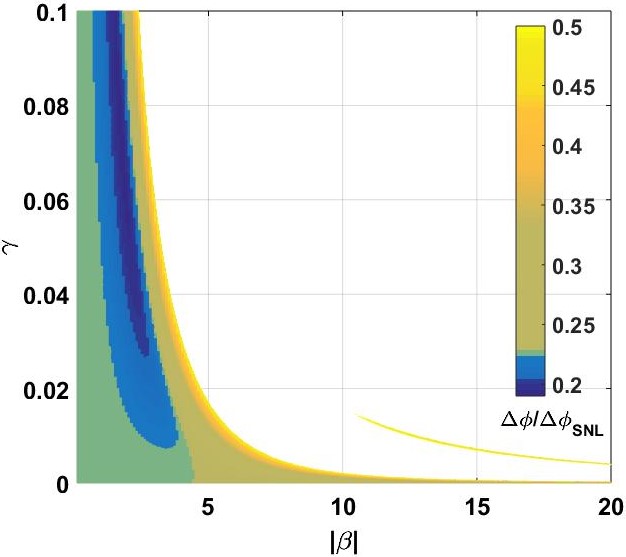}
\caption{\label{fig:8a1} Plot shows the variation in $\Delta\phi/\Delta\phi_{SNL}$ with $\gamma$ and $|\beta|$ for IDD scheme. The dark region shows the maximum phase sensitivity and we can see that the combination of coherent plus squeezed Kerr ($\gamma\neq0, |\beta|\neq0)$ states as inputs give better phase sensitivity than the coherent plus squeezed vacuum ($\gamma=0, |\beta|=0$) states as inputs. Other parameters are $\theta=\pi$, $r=1.5$, $|\alpha|=50$, $\phi=\pi/2$ for IDD detection schemes.}
\end{figure}

\begin{figure}[]
\includegraphics[width=8cm, height=7.5cm]{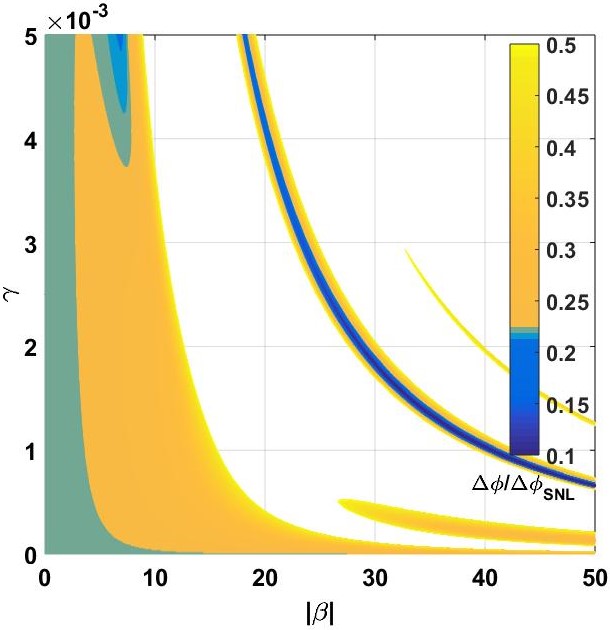}
\caption{\label{fig:8a12} Plot shows the variation in $\Delta\phi/\Delta\phi_{SNL}$ with $\gamma$ and $|\beta|$ for HD scheme. The dark region shows the maximum phase sensitivity and we can see that the combination of coherent plus squeezed Kerr ($\gamma\neq0, |\beta|\neq0)$ states as inputs give better phase sensitivity than the coherent plus squeezed vacuum ($\gamma=0, |\beta|=0$) states as inputs. Other parameters are $\theta=\pi$, $r=1.5$, $|\alpha|=50$, $\phi=0$ for HD detection schemes.}
\end{figure}

So, from these results, we can compare the sensitivity of MZI for the two cases of inputs: (i) coherent plus SKS, i.e., $|\alpha\rangle_1\otimes|\psi_{SK}\rangle_2$ and (ii) coherent plus squeezed vacuum, i.e., $|\alpha\rangle_1\otimes|\psi_{SV}\rangle_2$. To do this, we plot a colour graph in between $\gamma$ and $|\beta|$, where colour variation shows $\Delta\phi/\Delta\phi_{SNL}$, for IDD and HD schemes only. Since, analytically, we know that $\Delta\phi_{opt}$ for IDD and HD schemes are $\pi/2$ and 0, respectively, for all the values of $|\alpha|$ and $|\beta|$ but in case of SID, $\Delta\phi_{opt}$ varies with $|\beta|$. Fig. \ref{fig:8a1} and Fig. \ref{fig:8a12} show the variation for IDD and HD, respectively, in which the dark region shows the maximum phase sensitivity and we can see that, the combination of coherent plus squeezed Kerr ($\gamma\neq0, |\beta|\neq0)$ states as inputs give better phase sensitivity than the coherent plus squeezed vacuum ($\gamma=0, |\beta|=0$) states as inputs. This improvement can also be seen easily from the Eq. \eqref{B36} and \eqref{B37} if we put the optimal values of $\phi$ and $\theta$ for the respective cases. For example, for HD scheme: $\phi=0$, $\theta=\pi$, Eq. \eqref{B37} becomes 
\begin{equation}
    \Delta\phi_{hd}=\frac{\sqrt{e^{-2r}(1+A)}}{||\alpha| +|\beta|ce^r\sin{(s-\pi)}|}\label{37i},
\end{equation}
with 
\begin{equation}
A=2|\beta|^{2}\left(1 + c_2\cos{(2\gamma+s_2)} - 2c^2\cos^2{(s)}\right).\label{37n}    
\end{equation}
Here, $c=e^{|\beta|^2(\cos{2\gamma}-1)},~c_2=e^{|\beta|^2(\cos{4\gamma}-1)},~s={|\beta|^2\sin{2\gamma}}$ and $s_2={|\beta|^2\sin{4\gamma}}$. With $|\beta|=0$, Eq. \eqref{37i} gives $\Delta\phi_{hd}={\sqrt{e^{-2r}}}/{|\alpha|}$ and this is the phase sensitivity for coherent plus squeezed vacuum state as inputs. In the case of non-zero $|\beta|$ (i.e., in SKS case), for some values of $\gamma$ (shown in Fig. \ref{fig:8a12}), $s$ is greater than $\pi$ and $A\rightarrow0$. For example, for $|\beta|=30$, we have corresponding $\gamma\approx0.0018$ which gives $s\approx3.24$, $s_2\approx9.72$, $c\approx0.99$, $c_2\approx0.95$ and $A\rightarrow0$, indicating the improvement in phase sensitivity as compared to the case of squeezed vacuum state as input. Similarly, we can also get a simpler relation for the IDD case.

\section{Coherent state and squeezed Kerr state as the inputs of MZI under the lossy condition}\label{section 4}
In this section, we see the effect of internal and external losses on the phase sensitivity of the interferometer and the behaviour of the factor $\gamma$ under realistic scenarios. So, for this purpose, we consider the cases of best performances under lossless conditions (Section \ref{section 3}).

\begin{figure}
\includegraphics[width=5cm, height=4.5cm]{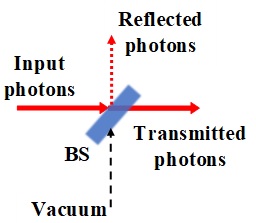}
\caption{\label{fig:1ab} The schematic diagram of a fictitious beam splitter mimicking the photon loss.}
\end{figure}

\begin{figure}
\includegraphics[width=8cm, height=6.5cm]{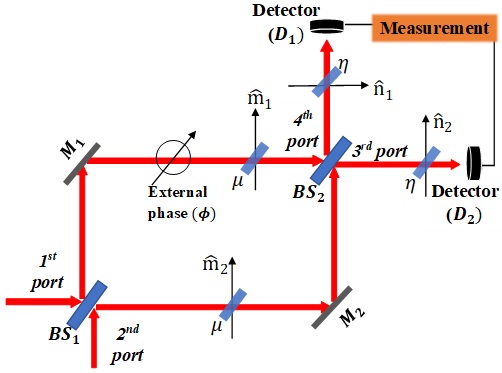}
\caption{\label{fig:1a} The schematic diagram of MZI having two inputs and two outputs associated with two 50:50 beam splitters ($BS_1$ and $BS_2$), two mirrors ($M_1$ and $M_2$) and two detectors ($D_1$ and $D_2$). In order to consider the internal (external) loss, we consider the fictitious beam splitters having a transmission coefficient is $\mu(\eta)$ in the internal (external) arms of the interferometer. $\hat{m}_1,~\hat{m}_2,~\hat{n}_1$ and $\hat{n}_2$ are the vacuum annihilation operators of the corresponding modes.}
\end{figure}

In quantum optics, photon loss can be occurred in two ways. In the first way, reflections or refractions of photons in undesired directions by means of optical elements are used in the setup of the quantum optics experiment (we call it internal photon loss). In the second way, the sensitivity of detectors is used for the detection of photons in the experiment (we call it external photon loss). Typically, photon loss can be modelled by considering a fictitious beam splitter that routes photons out of the interferometer, Fig. \ref{fig:1ab} \cite{loudon2000quantum}. Suppose there is a fictitious beam splitter having transmitivity $\tau$. Let us take the annihilation operator $\hat{i}$ corresponding to the input photons and $\hat{v}$ corresponds to the annihilation operator for the vacuum. Therefore, annihilation operator $\hat{t}$ for the transmitted photons can be written in terms of $\hat{i}$ and $\hat{v}$ as
\begin{equation}
    \hat{t}=\sqrt{\tau} \hat{i}+\sqrt{1-\tau} \hat{v}.
\end{equation}
In order to consider the photon loss, we simply take the transmitted photons from the fictitious beam splitter as our main signal and reflected photons as loss (Fig. \ref{fig:1ab}).
Therefore, for internal photon loss, we use a fictitious beam splitter, with transmitivity $\mu$, in both arms of the interferometer (Fig. \ref{fig:1a}) and for external photon loss, we use a fictitious beam splitter, with transmitivity $\eta$, at the outputs of the interferometer. The relation between the input and output annihilation operators under the photon loss (both \textit{internal} and \textit{external}) conditions, as shown in Fig. \ref{fig:1a} are,
\begin{eqnarray}
\hat{a}'_{3}=-\sqrt{\eta\mu}e^{\frac{i\phi}{2}}(-\sin\left(\frac{\phi}{2}\right)\hat{a}_{1}+\cos\left(\frac{\phi}{2}\right)\hat{a}_{2})\nonumber\\
+\sqrt{\frac{\eta(1-\mu)}{2}}(i\hat{m}_1+\hat{m}_2)+\sqrt{1-\eta}\hat{n}_2,\label{q1l}\\
\hat{a}'_{4}=-\sqrt{\eta\mu}e^{\frac{i\phi}{2}}(\cos\left(\frac{\phi}{2}\right)\hat{a}_{1}+\sin\left(\frac{\phi}{2}\right)\hat{a}_{2})\nonumber\\
+\sqrt{\frac{\eta(1-\mu)}{2}}(\hat{m}_1+i\hat{m}_2)+\sqrt{1-\eta}\hat{n}_1\label{q2l}.
\end{eqnarray}
Where $\hat{a}'_{3}$ and $\hat{a}'_{4}$ are the output annihilation operators corresponding to $3^{rd}$ and $4^{th}$ port, respectively. Here, $\hat{m}_1,~\hat{m}_2,~\hat{n}_1$ and $\hat{n}_2$ are the vacuum annihilation operators associated with fictitious beam splitter of the corresponding modes, as shown in Fig. \ref{fig:1a}.
Therefore, for the lossy case, phase sensitivity associated with SID, IDD and HD schemes, when squeezed Kerr state and coherent state taken as the inputs of the MZI, i.e., $|\psi\rangle_{in}=|\alpha\rangle_1\otimes|\psi_{SK}\rangle_2$, can be written as
\begin{widetext}
 \begin{equation}
\begin{split}
    \Delta\phi'_{sid}=\frac{2}{\left|(g_1-g_2)sin\phi-g_3\cos\phi\right|}\left(\frac{1}{\mu\eta}\left(g_1\sin^2\left(\frac{\phi}{2}\right)+g_2\cos^2\left(\frac{\phi}{2}\right)-\frac{1}{2}g_3\sin{\phi}\right)+(g_4-g_2^2)\cos^4\left(\frac{\phi}{2}\right)\right.\\
    +\left.(g_5-g_1^2)\sin^4\left(\frac{\phi}{2}\right)+\frac{1}{4}\left(g_6-g_3^2-2g_1g_2+4g_7\right)\sin^2\phi-\left((g_8-g_2g_3) \cos^2\left(\frac{\phi}{2}\right)+(g_9-g_1g_3)\sin^2\left(\frac{\phi}{2}\right)\right)\sin{\phi}\right)^\frac{1}{2},
\end{split}
 \label{B351}
 \end{equation}
    \begin{equation}
    \begin{split}
     \Delta\phi'_{idd}=\frac{\sqrt{\frac{1}{\mu\eta}\left(g_1+g_2\right)+(g_4+g_5-2g_7-(g_2-g_1)^2)\cos^2\phi+(g_6+2g_7-g_3^2)\sin^2\phi-(g_8-g_9-g_3(g_2-g_1))\sin2\phi}}{\left|(g_1-g_2)\sin\phi-g_3\cos\phi\right|}\label{B361},
 \end{split}
 \end{equation}
\begin{equation}
    \begin{split}
        \Delta\phi'_{hd}=\frac{\sqrt{2\cos^2\left(\frac{\phi}{2}\right)\left(Re(e^{i\phi}(\Delta\hat{a}_2)^2)+(g_2-\langle\hat{a}_{2}^{\dagger}\rangle\langle\hat{a}_{2}\rangle)\right)+\left(\frac{1}{\mu\eta}\right)}}{\left|Re(e^{i\phi}(\langle\hat{a}_1\rangle - i\langle\hat{a}_2\rangle))\right|}.
        \end{split}\label{B371}
\end{equation}
    \end{widetext}
Here, the $g_i$ with $i={1,2,...,9}$ are given in Eq.~\eqref{14}. Detailed expressions of the corresponding phase sensitivity associated with SID, IDD and HD schemes in lossy cases are derived in Appendix \ref{appendix c}. 

We can see that Eqs. \eqref{B351}, \eqref{B361} and \eqref{B371} depict that internal ($\mu$) and external ($\eta$) losses show an equal effect on the phase sensitivity in the case of SID, IDD and HD schemes respectively. So, we can consider either internal or external loss and can see the variation in $\Delta\phi$ with $\gamma$.

\begin{figure}
\includegraphics[width=8.5cm, height=6.5cm]{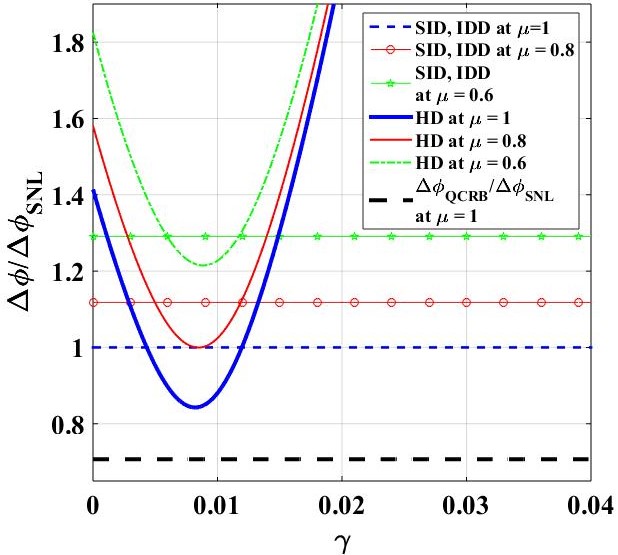}
\caption{\label{fig:2al} 
Plots show the variation of $\Delta\phi/\Delta\phi_{SNL}$ with $\gamma$ for $r=0$. We can see that in the case of the HD scheme, we can surpass the SNL in the lossy case (for $<20\%$ loss) for some values of $\gamma$. While for SID and IDD schemes we are getting worse phase sensitivity compared to SNL under lossy conditions. Other parameters are $\phi=\pi,~\pi/2,~7\pi/4$ for SID, IDD and HD, respectively, and $r=0,~\theta=\pi,~|\alpha|=0,~|\beta|=5$ and $\eta=1$.}
\end{figure}

\begin{figure}
\includegraphics[width=8.5cm, height=7cm]{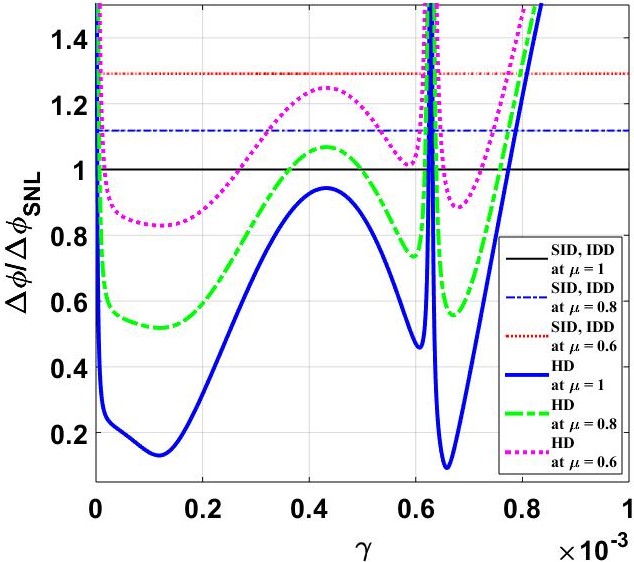}
\caption{\label{fig:2al2} 
Plots show the variation of $\Delta\phi/\Delta\phi_{SNL}$ with $\gamma$ for $r=1.5$. We can surpass the SNL even for more than 40\% photon loss in the HD scheme, while SID and IDD schemes give worse phase sensitivity with loss compared to SNL. Other parameters are $\phi=\pi,~\pi/2,~0$ for SID, IDD and HD, respectively, and $r=1.5,~\theta=\pi,~|\alpha|=0,~|\beta|=50$ and $\eta=1$.}
\end{figure}

Here, we are considering the case of photon loss under different combinations as discussed in Section \ref{section 3}. So, let us start with $|\psi\rangle_{in}=|0\rangle_1\otimes|\psi_{SK}\rangle_2$, Eqs. \eqref{B351}-\eqref{B371} can be written as,
 \begin{equation}
    \Delta\phi'_{sid}=\frac{\sqrt{\left(\frac{g_2}{\mu\eta}\right)+(g_4 -g_2^2)\cos^2\left(\frac{\phi}{2}\right)}}{\left|g_2\sin\left(\frac{\phi}{2}\right)\right|},
    \label{l1}
\end{equation}
\begin{equation}
    \Delta\phi'_{idd}=\frac{\sqrt{\left(\frac{g_2}{\mu\eta}\right)+(g_4 -g_2^2)\cos^2\phi}}{\left|g_2\sin\phi\right|},
    \label{l2}
\end{equation}
\begin{equation}
    \begin{split}
        \Delta\phi'_{hd}=\frac{1}{\left|Re(-ie^{i\phi}\langle\hat{a}_2\rangle)\right|}\left(2\cos^2\left(\frac{\phi}{2}\right)\times\right.\\
        \left(Re(e^{i\phi}(\Delta\hat{a})^2)+\left.(F_2-\langle\hat{a}_{2}^{\dagger}\rangle\langle\hat{a}_{2}\rangle)\right)+\left(\frac{1}{\mu\eta}\right)\right)^\frac{1}{2}.
    \label{l3h}
    \end{split}
\end{equation}
Fig. \ref{fig:2al} and Fig. \ref{fig:2al2} show the variation of $\Delta\phi/\Delta\phi_{SNL}$ with $\gamma$ for $r=0$ and $r=1.5$, respectively. For Kerr state, in Fig. \ref{fig:2al}, we can see that in the case of HD scheme, we can surpass the SNL in lossy case (for $<20\%$ loss) for some values of $\gamma$. While for SID and IDD schemes we are getting worse phase sensitivity compared to SNL under lossy conditions. On the other hand, if we take SKS we can surpass the SNL even for more than 40\% photon loss in the HD scheme (Fig. \ref{fig:2al2}), while SID and IDD schemes are yet giving worse phase sensitivity with loss compared to SNL.

\begin{figure}
\includegraphics[width=8.5cm, height=7cm]{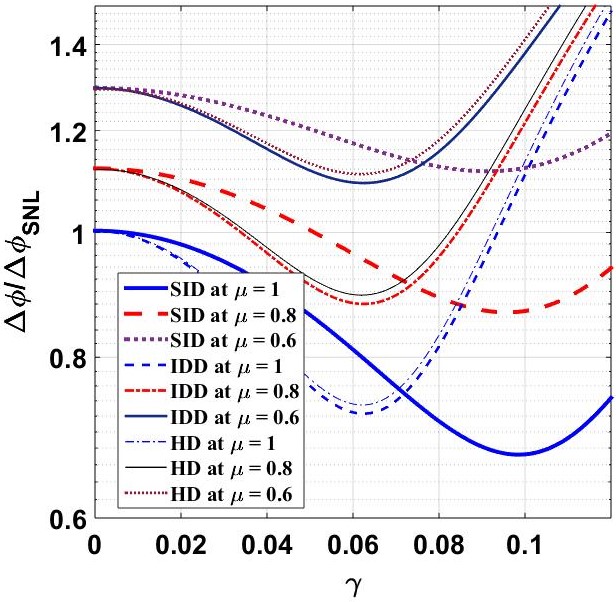}
\caption{\label{fig:2al3} 
 Plots show the variation of $\Delta\phi/\Delta\phi_{SNL}$ with $\gamma$ for $r=0$. We can see that in all three cases, we can surpass the SNL in the lossy case (for $<30\%$ loss) for some values of $\gamma$. Other parameters are $\phi=6.2,~\pi/2,~0$ for SID, IDD and HD, respectively, and $r=0,~\theta=\pi,~|\alpha|=50,~|\beta|=2$ and $\eta=1$.}
\end{figure}

\begin{figure}
\includegraphics[width=8.5cm, height=7cm]{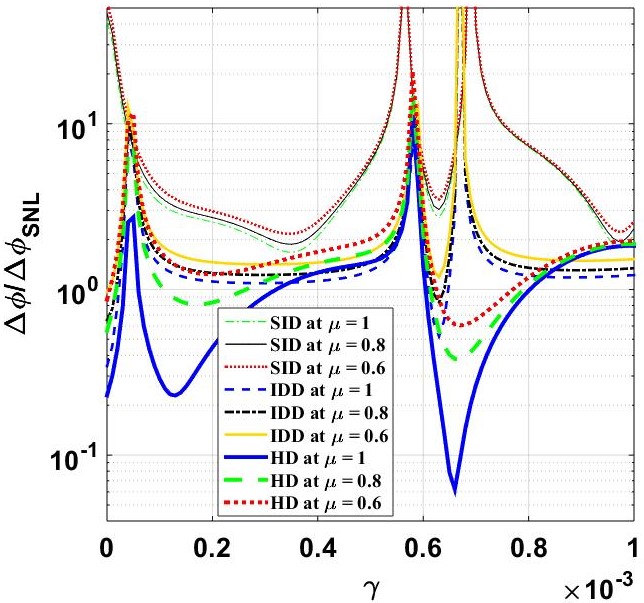}
\caption{\label{fig:2al4} Plots show the variation of $\Delta\phi/\Delta\phi_{SNL}$ with $\gamma$ for $r=1.5$. We can surpass the SNL even for more than 40\% photon loss in the HD scheme, and the IDD scheme gives phase sensitivity below the SNL for more than $20\%$ photon loss. For the SID scheme, even for a lossless case, we cannot beat the SNL. Other parameters are} $\phi=\frac{9\pi}{8},~\pi/2,~0$ for SID, IDD and HD, respectively, and $r=1.5,~\theta=\pi,~|\alpha|=50,~|\beta|=50$ and $\eta=1$.
\end{figure}

Now, we are considering the case in which the input state is $|\psi\rangle_{in}=|\alpha\rangle_1\otimes|\psi_{SK}\rangle_2$. Phase sensitivity for this case is given in Eqs. \eqref{B351}-\eqref{B371} for all three cases. Fig. \ref{fig:2al3} and Fig. \ref{fig:2al4} show the variation of $\Delta\phi/\Delta\phi_{SNL}$ with $\gamma$ for $r=0$ and $r=1.5$, respectively. For Kerr state, in Fig. \ref{fig:2al3}, we can see that in all three cases, we can surpass the SNL in the lossy case (for $<30\%$ loss) for some values of $\gamma$. If we take SKS we can surpass the SNL even for more than 40\% photon loss in the HD scheme (Fig. \ref{fig:2al4}), and the IDD scheme gives phase sensitivity below the SNL for more than $20\%$ photon loss. For the SID scheme, even for lossless cases, we can not beat the SNL.

We can also see this improvement by putting optimal values of $\phi$ and $\theta$ in Eq. \eqref{B361} and Eq. \eqref{B371}. For example in HD case, $\phi=0$ and $\theta=\pi$ are the optimal values, so, Eq. \eqref{B371} becomes



\begin{equation}
    \Delta\phi'_{hd}=\frac{\sqrt{(\frac{1}{\eta\mu} - 1) + e^{-2r}(1+A)}}{||\alpha| + |\beta|ce^r\sin{(s-\pi)}|}.\label{47n}
\end{equation}
Here, $A$ is given in Eq. \eqref{37n} and $c=e^{|\beta|^2(\cos{2\gamma}-1)},~s={|\beta|^2\sin{2\gamma}}$. 
With $|\beta|=0$, Eq. \eqref{47n} gives $\Delta\phi_{hd}={\sqrt{(\frac{1}{\eta\mu} - 1) + e^{-2r}}}/{|\alpha|}$ and this is the phase sensitivity for coherent plus squeezed vacuum state as inputs in the lossy case. In the case of non-zero $|\beta|$ (i.e., in SKS case), there are instances of the values of $\gamma$ corresponding to the $|\beta|$ (shown in Fig. \ref{fig:8a12}) for which $s$ is greater than $\pi$ and $A\rightarrow0$ indicates improvement in the phase sensitivity as compared to the squeezed vacuum state as input case. Similarly, we can also get a simpler equation for the IDD case.

\section{Conclusion and results}\label{section 5}

We studied the phase super-sensitivity of MZI using SKS as one of the inputs. We studied the effect of the Kerr medium on the phase sensitivity of a MZI with squeezing (i.e., $|\psi_{SK}\rangle$) and without squeezing state (i.e., $|\psi_{K}\rangle$). We found several conditions under which SKS gives better sensitivity than other combinations of input states, viz, coherent plus squeezed vacuum state, coherent plus vacuum state, coherent plus coherent state, etc. We discuss our results in Section \ref{section 3} and Section \ref{section 4} for lossless and lossy conditions, respectively. These sections are further divided into sub-sections based on the input combinations of the light.

To conclude the results found in section \ref{section 3} and section \ref{section 4} we made two tables TABLE \ref{tab:table1} and TABLE \ref{tab:table2}, respectively. These tables contain the approximated best values of $\Delta\phi/\Delta\phi_{SNL}$ for different cases from their respective plots for all three detection schemes, namely, SID, IDD and HD in order to conclude the results step by step.

In Section \ref{subsec iii(a)}, we discussed the vacuum and SKS as the inputs of MZI, i.e., $|\psi\rangle_{in}=|0\rangle_1\otimes|\psi_{SK}\rangle$. We found that by using the SKS along with the vacuum state as the inputs of the MZI we can surpass the SNL with significant amount for the HD scheme (Figs. \ref{fig:3} and \ref{fig:3b}). Not only for SKS but also for Kerr state, phase sensitivity surpasses the SNL in the HD scheme (Fig. \ref{fig:2a}). 

In Section \ref{subsec iii(b)}, we discussed the coherent and SKS as the inputs of MZI, i.e., $|\psi\rangle_{in}=|\alpha\rangle_1\otimes|\psi_{SK}\rangle$. Firstly we investigated the phase sensitivity of MZI by choosing the Kerr state with coherent state as the input, i.e., $|\psi\rangle_{in}=|\alpha\rangle_1\otimes|\psi_{K}\rangle$. We found that coherent input with Kerr state is more useful than double coherent input, in approximately all the situations. After that, we investigated $|\psi\rangle_{in}=|\alpha\rangle_1\otimes|\psi_{SK}\rangle$ case and found that in the case of squeezing Kerr medium enhances the phase sensitivity remarkably. The interesting finding was, that coherent plus SKS as inputs give better phase sensitivity than the coherent plus squeezed vacuum state as the inputs. If we look at the performance of the three detection schemes, we find that the HD scheme is dominant in all four cases than the IDD scheme which, in turn, is doing better than the SID scheme.

In Section \ref{section 4} we investigated the tolerances property of Kerr state against the photon loss. For the Kerr state along with the vacuum state, in Fig. \ref{fig:2al}, we can see that in the case of the HD scheme, we can surpass the SNL in the lossy case with $<20\%$ loss. If we take SKS along with vacuum, we can surpass the SNL even for more than 40\% photon loss in the HD scheme (Fig. \ref{fig:2al2}). For the Kerr state along with the coherent state, we found that in all three cases, we can surpass the SNL in the lossy case for $<30\%$ photon loss (Fig. \ref{fig:2al3}). If we take SKS along with coherent state, we can surpass the SNL even for more than 40\% photon loss in the HD scheme, and the IDD scheme gives phase sensitivity below the SNL for more than $20\%$ photon loss (Fig. \ref{fig:2al4}).

Recently, we found that Kalinin \textit{et al.} \cite{KalininDirmeierSorokinAnashkinaSánchezSotoCorneyLeuchsAndrianov+2023} reported the first experimental attempt of improving interferometric sensitivity utilising the Kerr effect. In Ref. \cite{kalinin2023observation}, a squeezing of $-6.5$ dB was observed for 20\% photon loss, while an estimated $-8.4$ dB was reported for lossless conditions by using the optical Kerr effect. They showed that the length of the fibre placed a restriction on the amount of squeezing possible, as longer pieces of fibre can obtain larger values of squeezing \cite{kalinin2023observation}. Since, factor $\gamma$ tells about the interaction time of coherent state with Kerr medium in order to produce the required Kerr state. From relation \eqref{K19}, we can see that $\gamma$ is proportional to Kerr medium length or interaction time. Larger $\gamma$ means larger medium length and vice-versa. Analytically, we found that for larger value of $|\beta|$ phase sensitivity increases as well as factor $\gamma$ is decreased, as we can see in Figs. \ref{fig:3c} and \ref{fig:8a12}.

In summary, SKS can be used to improve the phase sensitivity of a MZI. Importantly, we found some alternate states in place of the squeezed vacuum state for the phase super-sensitivity under both lossless as well as lossy conditions.  

\begin{table*}
\caption{\label{tab:table1}This table contains the best values of $\Delta\phi/\Delta\phi_{SNL}$ of all three detection schemes for different cases in lossless conditions. Values listed here are approximated values taken from the graphs plotted in the section \ref{section 3}.}
\begin{ruledtabular}
\begin{tabular}{cccccc}
$|\psi\rangle_{in}$&$|\alpha|$&$|\beta|$&$\Delta\phi_{sid}/\Delta\phi_{SNL}$&$\Delta\phi_{idd}/\Delta\phi_{SNL}$&$\Delta\phi_{hd}/\Delta\phi_{SNL}$\\ \hline
$|0\rangle\otimes|\psi_{K}\rangle$&$0$&$2$ &$=1$&$=1$&$\approx0.9$ \\
 &$0$&$5$ &$=1$&$=1$&$\approx0.85$ \\
 &$0$&$15$ &$=1$&$=1$&$\approx0.85$ \\ \hline
$|0\rangle\otimes|\psi_{SK}\rangle$\footnote{For squeezing we use $r=1.5$.}&$0$&$5$ &$=1$&$=1$&$\approx0.15$\\
 &$0$&$50$ &$=1$&$=1$&$\approx0.09$ \\ \hline
 $|\alpha\rangle\otimes|\psi_{K}\rangle$&$3$&$2$ &$>1$&$=1$&$>1$\\
 &$50$&$2$ &$\approx0.65$&$\approx0.72$&$\approx0.73$ \\
 &$3$&$50$ &$\leq1$&$\leq1$&$\leq1$ \\
 &$50$&$50$ &$\approx0.8$&$\approx0.8$&$\approx0.8$ \\ \hline
 $|\alpha\rangle\otimes|\beta\rangle$&$3$&$2$ &$=1$&$=1$&$=1$\\
 &$50$&$2$ &$=1$&$=1$&$=1$\\
 &$3$&$50$ &$=1$&$=1$&$=1$\\
 &$50$&$50$ &$=1$&$=1$&$=1$\\ \hline
 $|\alpha\rangle\otimes|\psi_{SK}\rangle$\footnotemark[1]&$3$&$2$ &$\approx3$&$\approx1.2$&$\approx0.27$\\
 &$50$&$2$ &$\approx0.6$&$\approx0.2$&$\approx0.18$ \\
 &$50$&$50$ &$\approx1.8$&$\approx0.35$&$\approx0.23$ \\
 &$3$&$50$ &$\approx1$&$=1$&$\approx0.085$\\ \hline
 $|\alpha\rangle\otimes|\psi_{SV}\rangle$\footnotemark[1]&$50$&$0$ &&$\approx0.23$&$\approx0.23$
\end{tabular}
\end{ruledtabular}
\end{table*}

\begin{table*}
\caption{\label{tab:table2}This table contains the best values of $\Delta\phi/\Delta\phi_{SNL}$ of all three detection schemes for different cases in lossy conditions. Values listed here are approximated values taken from the graphs plotted in the section \ref{section 4}.}
\begin{ruledtabular}
\begin{tabular}{ccccccc}
$|\psi\rangle_{in}$&$|\alpha|$&$|\beta|$&$\mu$&$\Delta\phi_{sid}/\Delta\phi_{SNL}\approx$&$\Delta\phi_{idd}/\Delta\phi_{SNL}\approx$&$\Delta\phi_{hd}/\Delta\phi_{SNL}\approx$\\ \hline
$|0\rangle\otimes|\psi_{K}\rangle$&$0$&$5$ &$1$&$=1$&$=1$&$\approx0.85$ \\
 &&&$0.8$&$\approx1.12$&$\approx1.12$&$\approx1$ \\
 &&&$0.6$&$\approx1.3$&$\approx1.3$&$\approx1.21$ \\ \hline
$|0\rangle\otimes|\psi_{SK}\rangle$\footnote{For squeezing we use $r=1.5$.}&$0$&$50$&$1$&$=1$&$=1$&$\approx0.1$\\
 &&&$0.8$&$\approx1.12$&$\approx1.12$&$\approx0.35$\\
 &&&$0.6$&$\approx1.3$&$\approx1.3$&$\approx0.82$\\ \hline
 $|\alpha\rangle\otimes|\psi_{K}\rangle$&$50$&$2$&$1$&$\approx0.67$&$\approx0.72$&$\approx0.74$\\
 &&&$0.8$&$\approx0.87$&$\approx0.88$&$\approx0.9$\\
 &&&$0.6$&$\approx1.12$&$\approx1.1$&$\approx1.12$\\ \hline
 $|\alpha\rangle\otimes|\psi_{SK}\rangle$\footnotemark[1]&$50$&$50$&$1$&$\approx1.6$&$\approx0.35$&$\approx0.06$\\
 &&&$0.8$&$\approx1.8$&$\approx0.7$&$\approx0.4$\\
 &&&$0.6$&$\approx2.1$&$\approx1$&$\approx0.6$\\
\end{tabular}
\end{ruledtabular}
\end{table*}

\section{Acknowledgements}
DY, GS and PS acknowledge UGC for the UGC Research Fellowship. DKM acknowledges financial supports from the Science \& Engineering Research Board (SERB), New Delhi for CRG Grant (CRG/2021/005917) and Incentive Grant under Institution of Eminence (IoE), Banaras Hindu University, Varanasi, India (R/Dev/D/IoE/Seed and Incentive Grant-II/2022-23).

\appendix

\section{Phase sensitivity under lossless case}\label{appendix A}
In order to find the phase sensitivity from Eq. \eqref{a1}, we derive the expression for $\langle\hat{L}\rangle$, $\langle\hat{L}^2\rangle$ and $|\partial\langle\hat{L}\rangle/\partial\phi|$ for different detection schemes. So, for SID scheme, from Eq. \eqref{2q} and Eq. \eqref{6l}, we get
\begin{equation}
     \begin{split}
         \langle\hat{L}_{sid}\rangle=g_1\sin^2\left(\frac{\phi}{2}\right)+ g_2\cos^2\left(\frac{\phi}{2}\right)-\frac{1}{2}g_3\sin\phi,
     \end{split}
     \label{eq:19}
 \end{equation}
\begin{equation}
\begin{split}
    \langle\hat{L}^2_{sid}\rangle=g_1\sin^2\left(\frac{\phi}{2}\right)+g_2\cos^2\left(\frac{\phi}{2}\right)-\frac{1}{2}g_3\sin\phi\\
    +g_4\cos^4\left(\frac{\phi}{2}\right)+g_5\sin^4\left(\frac{\phi}{2}\right)+\frac{1}{4}g_6\sin^2\phi\\
    +g_7\sin^2\phi-\left(g_8 \cos^2\left(\frac{\phi}{2}\right)+g_9\sin^2\left(\frac{\phi}{2}\right)\right)\sin\phi,
 \end{split}
 \label{eq:21}
 \end{equation}
 and variation of $\langle\hat{L}_{sid}(\phi)\rangle$ with $\phi$
  \begin{equation}
     \begin{split}
          \left|\frac{\partial\langle\hat{L}_{sid}\rangle}{\partial\phi}\right|=\frac{1}{2}\left|(g_1-g_2)sin\phi-g_3\cos\phi\right|.
     \end{split}
     \label{eq:20}
  \end{equation}
For the IDD scheme, from Eqs. \eqref{2q}, \eqref{1q} and \eqref{7l}, we get
 \begin{equation}
     \begin{split}
         \langle\hat{L}_{idd}\rangle=(g_2-g_1)\cos\phi-g_3\sin\phi,\label{b4}
     \end{split}
 \end{equation}
    \begin{equation}
    \begin{split}
     \langle\hat{L}^2_{idd}\rangle=g_1+g_2+(g_4+g_5)\cos^2\phi+g_6\sin^2\phi\\
     -2g_7\cos2\phi-(g_8-g_9)\sin2\phi\label{b6},
 \end{split}
 \end{equation}
 and variation of $\langle\hat{L}_{idd}(\phi)\rangle$ with $\phi$
  \begin{equation}
     \begin{split}
          \left|\frac{\partial\langle\hat{L}_{idd}\rangle}{\partial\phi}\right|=\left|(g_1-g_2)\sin\phi-g_3\cos\phi\right|.\label{b5}
     \end{split}
  \end{equation}
For the HD scheme, from Eqs. \eqref{2q}, \eqref{1q} and \eqref{8l}, we get
\begin{equation}
    \begin{split}
        (\Delta\hat{L}_{hd})^2=\langle\hat{L}^2_{hd}\rangle-\langle\hat{L}_{hd}\rangle=\frac{1}{2}+\cos^2\left(\frac{\phi}{2}\right)\\
        \times\left(Re(e^{i\phi}(\Delta\hat{a}_{2})^2)+(g_2-\langle\hat{a}_{2}^{\dagger}\rangle\langle\hat{a}_{2}\rangle)\right),
        \end{split}\label{a22}
\end{equation}
\begin{equation}
\begin{split}
\left|\frac{\partial\langle\hat{L}_{hd}\rangle}{\partial\phi}\right|=\frac{1}{\sqrt{2}}\left|Re(e^{i\phi}(\langle\hat{a}_1\rangle - i\langle\hat{a}_2\rangle))\right|.
\end{split}\label{a23}
\end{equation}
Where,
\begin{equation}
    \begin{split}
        g_1=\langle\hat{a}^\dagger_1\hat{a}_1\rangle,~g_2=\langle\hat{a}^\dagger_2\hat{a}_2\rangle,~g_3=\langle\hat{a}_1\hat{a}^\dagger_2\rangle+\langle\hat{a}^\dagger_1\hat{a}_2\rangle,\\
        g_4=\langle\hat{a}^{\dagger2}_2\hat{a}^2_2\rangle,~g_5=\langle\hat{a}^{\dagger2}_1\hat{a}^2_1\rangle,~g_6=\langle\hat{a}^2_1\hat{a}^{\dagger2}_2\rangle\\
        +\langle\hat{a}^{\dagger2}_1\hat{a}^2_2\rangle,~g_7=\langle\hat{a}^\dagger_1\hat{a}_1\hat{a}^\dagger_2\hat{a}_2\rangle,~g_8=\langle\hat{a}_1\hat{a}^{\dagger2}_2\hat{a}_2\rangle\\
        +\langle\hat{a}^\dagger_1\hat{a}^\dagger_2\hat{a}_2^2\rangle,~g_9=\langle\hat{a}^\dagger_1\hat{a}^2_1\hat{a}^\dagger_2\rangle+\langle\hat{a}^{\dagger2}_1\hat{a}_1\hat{a}_2\rangle,\\
        g_{10}=\langle\hat{a}^{\dagger}_1\hat{a}_2\rangle-\langle\hat{a}_1\hat{a}^{\dagger}_2\rangle,  ~g_{11}=\langle\hat{a}^{\dagger}_1\hat{a}^2_1\hat{a}^{\dagger}_2\rangle\\
        -\langle\hat{a}^{\dagger2}_1\hat{a}_1\hat{a}_2\rangle, ~g_{12}=\langle\hat{a}_1\hat{a}^{\dagger2}_2\hat{a}_2\rangle-\langle\hat{a}^{\dagger}_1\hat{a}^{\dagger}_2\hat{a}^2_2\rangle.
    \end{split}
    \label{14}
\end{equation}
Appendix \ref{appendix} contains the separate expressions of the expectation value of the operators given in Eq. (\ref{14}).

\section{Expectation values of the operators w.r.t. state $|\psi\rangle_{in}=|\alpha\rangle\otimes|\psi_{SK}\rangle$}\label{appendix}
\begin{equation}
\begin{split}
    \langle\hat{a}_2\rangle = |\beta|c(Ce^{-is} + Se^{i(s+\theta)}),
\end{split}
\label{}
\end{equation}
\begin{equation}
\begin{split}
    \langle\hat{a}^2_2\rangle= C^2|\beta|^2c_2e^{-i(2\gamma + s_2)} + CS(2|\beta|^2 + 1)e^{i\theta} \\
    + S^2|\beta|^2c_2e^{i(2\gamma + 2\theta + s_2)},
\end{split}
\label{h123}
\end{equation}

\begin{equation}
    \langle\hat{a}^\dagger_1\hat{a}_1\rangle=\mid\alpha\mid^{2},
\end{equation}
\begin{equation}
    \langle\hat{a}^\dagger_2\hat{a}_2\rangle=|\beta|^{2}(C^2+S^2+2c_2CS\cos{(2\gamma+\theta+s_2)})+S^2,
    \label{h2}
\end{equation}
\begin{equation}
    \begin{split}
    \langle\hat{a}_1\hat{a}^\dagger_2\rangle+\langle\hat{a}^\dagger_1\hat{a}_2\rangle=2\mid\alpha\mid\mid\beta\mid c\left(C\cos(s)+S\cos(\theta+s)\right),
\end{split}
\end{equation}
\begin{equation}
    \begin{split}
    \langle\hat{a}^{\dagger2}_2\hat{a}^2_2\rangle=|\beta|^{4}C^{4}+\left(|\beta|^{4}S^4+4|\beta|^{2}S^4+2S^4\right)\\
    +2|\beta|^{4}C^2S^2c_4\cos{(2\theta+12\gamma+s_4)}\\
    +4|\beta|^{4}c_2CS(C^2+S^2)\cos{(\theta+6\gamma+s_2)}\\
    +2CS|\beta|^{2}c_2(C^2+5S^2)\cos{(\theta+2\gamma+s_2)}\\
    +\left(4|\beta|^{4}+8|\beta|^{2}+1\right)C^2S^2,
\end{split}
\end{equation}

\begin{equation}
 \langle\hat{a}^{\dagger2}_1\hat{a}^2_1\rangle=|\alpha|^{4},
\end{equation}

\begin{equation}
    \begin{split}
    \langle\hat{a}^2_1\hat{a}^{\dagger2}_2\rangle+\langle\hat{a}^{\dagger2}_1\hat{a}^2_2\rangle=2|\alpha|^{2}|\beta|^{2}\left(c_2C^2\cos{(2\gamma+s_2)}\right.\\
    \left.+2CS\cos{\theta}+c_2S^2\cos{(2\theta+2\gamma+s_2)}\right)\\
    +2|\alpha|^{2}CS\cos{\theta},
\end{split}
\end{equation}

\begin{equation}
    \begin{split}
    \langle\hat{a}^\dagger_1\hat{a}_1\hat{a}^\dagger_2\hat{a}_2\rangle=|\alpha|^{2}\left(|\beta|^{2}C^2+(|\beta|^{2}+1)S^2\right.\\
    +\left.2\mid\beta\mid^{2}c_2CS\cos{(\theta+2\gamma+s_2)}\right),
\end{split}
\end{equation}

\begin{equation}
    \begin{split}
    \langle\hat{a}_1\hat{a}^{\dagger2}_2\hat{a}_2\rangle+\langle\hat{a}^\dagger_1\hat{a}^\dagger_2\hat{a}_2^2\rangle=\\
    2|\alpha|\left(|\beta|^3c((C^3+2CS^2)\cos{(2\gamma+s)}\right.\\
    +(S^3+2C^2S)\cos{(\theta+2\gamma+s))}\\
    +|\beta|c((2S^3+C^2S)\cos{(\theta+s)}+3CS^2\cos{s})\\
    +|\beta|^3c_3CS(C\cos{(\theta+6\gamma+s_3)}\\
    +\left.S\cos{(2\theta+6\gamma+s_3)})\right),
\end{split}
\end{equation}

\begin{equation}
    \begin{split}
    \langle\hat{a}^\dagger_1\hat{a}^2_1\hat{a}^\dagger_2\rangle+\langle\hat{a}^{\dagger2}_1\hat{a}_1\hat{a}_2\rangle=2|\alpha|^{3}|\beta|c\left(C\cos(s)\right.\\
    \left.+S\cos(\theta+s)\right),
\end{split}
\end{equation}

\begin{equation}
        \langle\hat{a}^{\dagger}_1\hat{a}_2\rangle-\langle\hat{a}_1\hat{a}^{\dagger}_2\rangle=2ic|\alpha||\beta|\left(-C\sin{s}+S\sin{(\theta+s)}\right),
\end{equation}

\begin{equation}
        \langle\hat{a}^{\dagger}_1\hat{a}^2_1\hat{a}^{\dagger}_2\rangle-\langle\hat{a}^{\dagger2}_1\hat{a}_1\hat{a}_2\rangle=2ic|\alpha|^3|\beta|\left(C\sin{s}-S\sin{(\theta+s)}\right),
\end{equation}

\begin{equation}
    \begin{split}
    \langle\hat{a}_1\hat{a}^{\dagger2}_2\hat{a}_2\rangle-\langle\hat{a}^{\dagger}_1\hat{a}^{\dagger}_2\hat{a}^2_2\rangle=\\
    2i|\alpha|\left(|\beta|^3c((C^3+2CS^2)\sin{(2\gamma+s)}\right.\\
    -(S^3+2C^2S)\sin{(\theta+2\gamma+s))}\\
    -|\beta|c((2S^3+C^2S)\sin{(\theta+s)}-3CS^2\sin{s})\\
    +|\beta|^3c_3CS(C\sin{(\theta+6\gamma+s_3)}\\
    -\left.S\sin{(2\theta+6\gamma+s_3)})\right).
\end{split}
\end{equation}
Where, $C=\cosh{r},~S=\sinh{r},~c=e^{|\beta|^2(\cos{2\gamma}-1)},~c_2=e^{|\beta|^2(\cos{4\gamma}-1)},~c_3=e^{|\beta|^2(\cos{6\gamma}-1)},~c_4=e^{|\beta|^2(\cos{8\gamma}-1)},~s={|\beta|^2\sin{2\gamma}},~s_2={|\beta|^2\sin{4\gamma}},~s_3={|\beta|^2\sin{6\gamma}},~s_4={|\beta|^2\sin{8\gamma}}$.

\section{Phase sensitivity in lossy condition}\label{appendix c}
In order to find the phase sensitivity from Eq. \eqref{a1}, we derive the expression for $\langle\hat{L}\rangle$, $\langle\hat{L}^2\rangle$ and $|\partial\langle\hat{L}\rangle/\partial\phi|$ for different detection schemes. So, for SID scheme, from Eq. \eqref{q1l} and Eq. \eqref{6l}, we get 
\begin{equation}
     \begin{split}
         \langle\hat{L}_{sid}\rangle=\mu\eta\left(g_1\sin^2\left(\frac{\phi}{2}\right)+ g_2\cos^2\left(\frac{\phi}{2}\right)-\frac{1}{2}g_3\sin\phi\right),
     \end{split}
     \label{eq:19m}
 \end{equation}
\begin{equation}
\begin{split}
    \langle\hat{L}^2_{sid}\rangle=\langle\hat{L}_{sid}\rangle+\mu^2\eta^2\left(g_4\cos^4\left(\frac{\phi}{2}\right)\right.\\
    +g_5\sin^4\left(\frac{\phi}{2}\right)+g_7\sin^2\phi+\frac{1}{4}g_6\sin^2\phi\\
    -\left.\left(g_8 \cos^2\left(\frac{\phi}{2}\right)+g_9\sin^2\left(\frac{\phi}{2}\right)\right)\sin{\phi}\right),
 \end{split}
 \label{eq:21m}
 \end{equation}
and variation of $\langle\hat{L}_{sid}(\phi)\rangle$ with $\phi$
  \begin{equation}
     \begin{split}
          \left|\frac{\partial\langle\hat{L}_{sid}\rangle}{\partial\phi}\right|=\frac{\mu\eta}{2}\left|(g_1-g_2)sin\phi-g_3\cos\phi\right|.
     \end{split}
     \label{eq:20m}
  \end{equation}
  For the IDD scheme, from Eqs. \eqref{q1l}, \eqref{q2l} and \eqref{7l}, we get
 \begin{equation}
     \begin{split}
         \langle\hat{L}_{idd}\rangle=\mu\eta\left((g_2-g_1)\cos\phi-g_3\sin\phi\right),
     \end{split}
 \end{equation}
 \begin{equation}
    \begin{split}
     \langle\hat{L}^2_{idd}\rangle=\mu\eta(g_1+g_2)+\mu^2\eta^2((g_4+g_5)\cos^2\phi\\
     +g_6\sin^2\phi-2g_7\cos2\phi-(g_8-g_9)\sin2\phi).
 \end{split}
 \end{equation}
 and variation of $\langle\hat{L}_{idd}(\phi)\rangle$ with $\phi$
  \begin{equation}
     \begin{split}
          \left|\frac{\partial\langle\hat{L}_{idd}\rangle}{\partial\phi}\right|=\mu\eta\left|(g_1-g_2)\sin\phi-g_3\cos\phi\right|.
     \end{split}
  \end{equation}
For the HD scheme, from Eqs. \eqref{q1l}, \eqref{q2l} and \eqref{8l}, we get
\begin{equation}
    \begin{split}
        (\Delta\hat{L}_{hd})^2=\langle\hat{L}^2_{hd}\rangle-\langle\hat{L}_{hd}\rangle=\frac{1}{2}+\frac{\mu\eta}{2}\cos^2\left(\frac{\phi}{2}\right)\\
        \times\left(Re(e^{i\phi}(\Delta\hat{a}_{2})^2)+(g_2-\langle\hat{a}_{2}^{\dagger}\rangle\langle\hat{a}_{2}\rangle)\right)
        \end{split}\label{a22a}
\end{equation}
\begin{equation}
\begin{split}
\left|\frac{\partial\langle\hat{L}_{hd}\rangle}{\partial\phi}\right|=\sqrt{\frac{\mu\eta}{2}}\left|Re(e^{i\phi}(\langle\hat{a}_1\rangle - i\langle\hat{a}_2\rangle))\right|.
\end{split}\label{a23a}
\end{equation}

\nocite{*}
\bibliography{aipsamp}

\providecommand{\noopsort}[1]{}\providecommand{\singleletter}[1]{#1}%
\begin{thebibliography}{81}%
\makeatletter
\providecommand \@ifxundefined [1]{%
 \@ifx{#1\undefined}
}%
\providecommand \@ifnum [1]{%
 \ifnum #1\expandafter \@firstoftwo
 \else \expandafter \@secondoftwo
 \fi
}%
\providecommand \@ifx [1]{%
 \ifx #1\expandafter \@firstoftwo
 \else \expandafter \@secondoftwo
 \fi
}%
\providecommand \natexlab [1]{#1}%
\providecommand \enquote  [1]{``#1''}%
\providecommand \bibnamefont  [1]{#1}%
\providecommand \bibfnamefont [1]{#1}%
\providecommand \citenamefont [1]{#1}%
\providecommand \href@noop [0]{\@secondoftwo}%
\providecommand \href [0]{\begingroup \@sanitize@url \@href}%
\providecommand \@href[1]{\@@startlink{#1}\@@href}%
\providecommand \@@href[1]{\endgroup#1\@@endlink}%
\providecommand \@sanitize@url [0]{\catcode `\\12\catcode `\$12\catcode `\&12\catcode `\#12\catcode `\^12\catcode `\_12\catcode `\%12\relax}%
\providecommand \@@startlink[1]{}%
\providecommand \@@endlink[0]{}%
\providecommand \url  [0]{\begingroup\@sanitize@url \@url }%
\providecommand \@url [1]{\endgroup\@href {#1}{\urlprefix }}%
\providecommand \urlprefix  [0]{URL }%
\providecommand \Eprint [0]{\href }%
\providecommand \doibase [0]{http://dx.doi.org/}%
\providecommand \selectlanguage [0]{\@gobble}%
\providecommand \bibinfo  [0]{\@secondoftwo}%
\providecommand \bibfield  [0]{\@secondoftwo}%
\providecommand \translation [1]{[#1]}%
\providecommand \BibitemOpen [0]{}%
\providecommand \bibitemStop [0]{}%
\providecommand \bibitemNoStop [0]{.\EOS\space}%
\providecommand \EOS [0]{\spacefactor3000\relax}%
\providecommand \BibitemShut  [1]{\csname bibitem#1\endcsname}%
\let\auto@bib@innerbib\@empty
\bibitem [{\citenamefont {Helstrom}(1976)}]{1976Helstrom}%
  \BibitemOpen
  \bibfield  {author} {\bibinfo {author} {\bibfnamefont {C.~W.}\ \bibnamefont {Helstrom}},\ }\href {https://books.google.co.in/books?id=Ne3iT\_QLcsMC} {\emph {\bibinfo {title} {Quantum {D}etection and {E}stimation {T}heory}}}\ (\bibinfo  {publisher} {Academic Press, San Diego, CA},\ \bibinfo {year} {1976})\BibitemShut {NoStop}%
\bibitem [{\citenamefont {Pezz\`e}\ \emph {et~al.}(2018)\citenamefont {Pezz\`e}, \citenamefont {Smerzi}, \citenamefont {Oberthaler}, \citenamefont {Schmied},\ and\ \citenamefont {Treutlein}}]{RevModPhys.90.035005}%
  \BibitemOpen
  \bibfield  {author} {\bibinfo {author} {\bibfnamefont {L.}~\bibnamefont {Pezz\`e}}, \bibinfo {author} {\bibfnamefont {A.}~\bibnamefont {Smerzi}}, \bibinfo {author} {\bibfnamefont {M.~K.}\ \bibnamefont {Oberthaler}}, \bibinfo {author} {\bibfnamefont {R.}~\bibnamefont {Schmied}}, \ and\ \bibinfo {author} {\bibfnamefont {P.}~\bibnamefont {Treutlein}},\ }\bibfield  {title} {\enquote {\bibinfo {title} {Quantum metrology with nonclassical states of atomic ensembles},}\ }\href {\doibase 10.1103/RevModPhys.90.035005} {\bibfield  {journal} {\bibinfo  {journal} {Rev. Mod. Phys.}\ }\textbf {\bibinfo {volume} {90}},\ \bibinfo {pages} {035005} (\bibinfo {year} {2018})}\BibitemShut {NoStop}%
\bibitem [{\citenamefont {{Demkowicz-Dobrzanski}}, \citenamefont {{Jarzyna}},\ and\ \citenamefont {{Kolodynski}}(2015)}]{2015PrOpt..60..345D}%
  \BibitemOpen
  \bibfield  {author} {\bibinfo {author} {\bibfnamefont {R.}~\bibnamefont {{Demkowicz-Dobrzanski}}}, \bibinfo {author} {\bibfnamefont {M.}~\bibnamefont {{Jarzyna}}}, \ and\ \bibinfo {author} {\bibfnamefont {J.}~\bibnamefont {{Kolodynski}}},\ }\bibfield  {title} {\enquote {\bibinfo {title} {{Quantum limits in optical interferometry}},}\ }\href {\doibase 10.1016/bs.po.2015.02.003} {\bibfield  {journal} {\bibinfo  {journal} {Progess in Optics}\ }\textbf {\bibinfo {volume} {60}},\ \bibinfo {pages} {345} (\bibinfo {year} {2015})},\ \Eprint {http://arxiv.org/abs/1405.7703} {arXiv:1405.7703 [quant-ph]} \BibitemShut {NoStop}%
\bibitem [{\citenamefont {Lawrie}\ \emph {et~al.}(2016)\citenamefont {Lawrie}, \citenamefont {Lett}, \citenamefont {Marino},\ and\ \citenamefont {Pooser}}]{Lawrie2019}%
  \BibitemOpen
  \bibfield  {author} {\bibinfo {author} {\bibfnamefont {B.~J.}\ \bibnamefont {Lawrie}}, \bibinfo {author} {\bibfnamefont {P.~D.}\ \bibnamefont {Lett}}, \bibinfo {author} {\bibfnamefont {A.~M.}\ \bibnamefont {Marino}}, \ and\ \bibinfo {author} {\bibfnamefont {R.~C.}\ \bibnamefont {Pooser}},\ }\bibfield  {title} {\enquote {\bibinfo {title} {Quantum {S}ensing with {S}queezed {L}ight},}\ }\href {\doibase 10.1021/acsphotonics.9b00250} {\bibfield  {journal} {\bibinfo  {journal} {ACS Photonics}\ }\textbf {\bibinfo {volume} {6}},\ \bibinfo {pages} {1307--1318} (\bibinfo {year} {2016})}\BibitemShut {NoStop}%
\bibitem [{\citenamefont {Dowling}(2008)}]{doi:10.1080/00107510802091298}%
  \BibitemOpen
  \bibfield  {author} {\bibinfo {author} {\bibfnamefont {J.~P.}\ \bibnamefont {Dowling}},\ }\bibfield  {title} {\enquote {\bibinfo {title} {Quantum optical metrology – the lowdown on high-{N00N} states},}\ }\href {\doibase 10.1080/00107510802091298} {\bibfield  {journal} {\bibinfo  {journal} {Contemporary Physics}\ }\textbf {\bibinfo {volume} {49}},\ \bibinfo {pages} {125--143} (\bibinfo {year} {2008})}\BibitemShut {NoStop}%
\bibitem [{\citenamefont {Yurke}, \citenamefont {McCall},\ and\ \citenamefont {Klauder}(1986)}]{PhysRevA.33.4033}%
  \BibitemOpen
  \bibfield  {author} {\bibinfo {author} {\bibfnamefont {B.}~\bibnamefont {Yurke}}, \bibinfo {author} {\bibfnamefont {S.~L.}\ \bibnamefont {McCall}}, \ and\ \bibinfo {author} {\bibfnamefont {J.~R.}\ \bibnamefont {Klauder}},\ }\bibfield  {title} {\enquote {\bibinfo {title} {{S}{U}(2) and {S}{U}(1,1) interferometers},}\ }\href {\doibase 10.1103/PhysRevA.33.4033} {\bibfield  {journal} {\bibinfo  {journal} {Phys. Rev. A}\ }\textbf {\bibinfo {volume} {33}},\ \bibinfo {pages} {4033--4054} (\bibinfo {year} {1986})}\BibitemShut {NoStop}%
\bibitem [{\citenamefont {Ou}\ and\ \citenamefont {Li}(2020)}]{doi:10.1063/5.0004873}%
  \BibitemOpen
  \bibfield  {author} {\bibinfo {author} {\bibfnamefont {Z.~Y.}\ \bibnamefont {Ou}}\ and\ \bibinfo {author} {\bibfnamefont {X.}~\bibnamefont {Li}},\ }\bibfield  {title} {\enquote {\bibinfo {title} {Quantum {SU}(1,1) interferometers: {B}asic principles and applications},}\ }\href {\doibase 10.1063/5.0004873} {\bibfield  {journal} {\bibinfo  {journal} {APL Photonics}\ }\textbf {\bibinfo {volume} {5}},\ \bibinfo {pages} {080902} (\bibinfo {year} {2020})}\BibitemShut {NoStop}%
\bibitem [{\citenamefont {Hudelist}\ \emph {et~al.}(2014)\citenamefont {Hudelist}, \citenamefont {Kong}, \citenamefont {Liu}, \citenamefont {Jing}, \citenamefont {Ou},\ and\ \citenamefont {Zhang}}]{Hudelist2014}%
  \BibitemOpen
  \bibfield  {author} {\bibinfo {author} {\bibfnamefont {F.}~\bibnamefont {Hudelist}}, \bibinfo {author} {\bibfnamefont {J.}~\bibnamefont {Kong}}, \bibinfo {author} {\bibfnamefont {C.}~\bibnamefont {Liu}}, \bibinfo {author} {\bibfnamefont {J.}~\bibnamefont {Jing}}, \bibinfo {author} {\bibfnamefont {Z.}~\bibnamefont {Ou}}, \ and\ \bibinfo {author} {\bibfnamefont {W.}~\bibnamefont {Zhang}},\ }\bibfield  {title} {\enquote {\bibinfo {title} {Quantum metrology with parametric amplifier-based photon correlation interferometers},}\ }\href {\doibase 10.1038/ncomms4049} {\bibfield  {journal} {\bibinfo  {journal} {Nature Communications}\ }\textbf {\bibinfo {volume} {5}},\ \bibinfo {pages} {3049} (\bibinfo {year} {2014})}\BibitemShut {NoStop}%
\bibitem [{\citenamefont {Linnemann}\ \emph {et~al.}(2016)\citenamefont {Linnemann}, \citenamefont {Strobel}, \citenamefont {Muessel}, \citenamefont {Schulz}, \citenamefont {Lewis-Swan}, \citenamefont {Kheruntsyan},\ and\ \citenamefont {Oberthaler}}]{PhysRevLett.117.013001}%
  \BibitemOpen
  \bibfield  {author} {\bibinfo {author} {\bibfnamefont {D.}~\bibnamefont {Linnemann}}, \bibinfo {author} {\bibfnamefont {H.}~\bibnamefont {Strobel}}, \bibinfo {author} {\bibfnamefont {W.}~\bibnamefont {Muessel}}, \bibinfo {author} {\bibfnamefont {J.}~\bibnamefont {Schulz}}, \bibinfo {author} {\bibfnamefont {R.~J.}\ \bibnamefont {Lewis-Swan}}, \bibinfo {author} {\bibfnamefont {K.~V.}\ \bibnamefont {Kheruntsyan}}, \ and\ \bibinfo {author} {\bibfnamefont {M.~K.}\ \bibnamefont {Oberthaler}},\ }\bibfield  {title} {\enquote {\bibinfo {title} {Quantum-{E}nhanced {S}ensing {B}ased on {T}ime {R}eversal of {N}onlinear {D}ynamics},}\ }\href {\doibase 10.1103/PhysRevLett.117.013001} {\bibfield  {journal} {\bibinfo  {journal} {Phys. Rev. Lett.}\ }\textbf {\bibinfo {volume} {117}},\ \bibinfo {pages} {013001} (\bibinfo {year} {2016})}\BibitemShut {NoStop}%
\bibitem [{\citenamefont {Shukla}\ \emph {et~al.}(2022)\citenamefont {Shukla}, \citenamefont {Mishra}, \citenamefont {Yadav}, \citenamefont {Pandey},\ and\ \citenamefont {Mishra}}]{Shukla:22}%
  \BibitemOpen
  \bibfield  {author} {\bibinfo {author} {\bibfnamefont {G.}~\bibnamefont {Shukla}}, \bibinfo {author} {\bibfnamefont {K.~K.}\ \bibnamefont {Mishra}}, \bibinfo {author} {\bibfnamefont {D.}~\bibnamefont {Yadav}}, \bibinfo {author} {\bibfnamefont {R.~K.}\ \bibnamefont {Pandey}}, \ and\ \bibinfo {author} {\bibfnamefont {D.~K.}\ \bibnamefont {Mishra}},\ }\bibfield  {title} {\enquote {\bibinfo {title} {Quantum-enhanced super-sensitivity of a {M}ach--{Z}ehnder interferometer with superposition of {S}chr\"{o}dinger's cat-like state and {F}ock state as inputs using a two-channel detection},}\ }\href {\doibase 10.1364/JOSAB.434967} {\bibfield  {journal} {\bibinfo  {journal} {J. Opt. Soc. Am. B}\ }\textbf {\bibinfo {volume} {39}},\ \bibinfo {pages} {59--68} (\bibinfo {year} {2022})}\BibitemShut {NoStop}%
\bibitem [{\citenamefont {Mishra}\ \emph {et~al.}(2021)\citenamefont {Mishra}, \citenamefont {Yadav}, \citenamefont {Shukla},\ and\ \citenamefont {Mishra}}]{Mishra_2021}%
  \BibitemOpen
  \bibfield  {author} {\bibinfo {author} {\bibfnamefont {K.~K.}\ \bibnamefont {Mishra}}, \bibinfo {author} {\bibfnamefont {D.}~\bibnamefont {Yadav}}, \bibinfo {author} {\bibfnamefont {G.}~\bibnamefont {Shukla}}, \ and\ \bibinfo {author} {\bibfnamefont {D.~K.}\ \bibnamefont {Mishra}},\ }\bibfield  {title} {\enquote {\bibinfo {title} {Non-classicalities exhibited by the superposition of {S}chrödinger's cat state with the vacuum of the optical field},}\ }\href {\doibase 10.1088/1402-4896/abe00f} {\bibfield  {journal} {\bibinfo  {journal} {Physica Scripta}\ }\textbf {\bibinfo {volume} {96}},\ \bibinfo {pages} {045102} (\bibinfo {year} {2021})}\BibitemShut {NoStop}%
\bibitem [{\citenamefont {Clauser}(1974)}]{PhysRevD.9.853}%
  \BibitemOpen
  \bibfield  {author} {\bibinfo {author} {\bibfnamefont {J.~F.}\ \bibnamefont {Clauser}},\ }\bibfield  {title} {\enquote {\bibinfo {title} {Experimental distinction between the quantum and classical field-theoretic predictions for the photoelectric effect},}\ }\href {\doibase 10.1103/PhysRevD.9.853} {\bibfield  {journal} {\bibinfo  {journal} {Phys. Rev. D}\ }\textbf {\bibinfo {volume} {9}},\ \bibinfo {pages} {853--860} (\bibinfo {year} {1974})}\BibitemShut {NoStop}%
\bibitem [{\citenamefont {Walls}(1983)}]{Walls1983}%
  \BibitemOpen
  \bibfield  {author} {\bibinfo {author} {\bibfnamefont {D.~F.}\ \bibnamefont {Walls}},\ }\bibfield  {title} {\enquote {\bibinfo {title} {Squeezed states of light},}\ }\href {\doibase 10.1038/306141a0} {\bibfield  {journal} {\bibinfo  {journal} {Nature}\ }\textbf {\bibinfo {volume} {306}},\ \bibinfo {pages} {141--146} (\bibinfo {year} {1983})}\BibitemShut {NoStop}%
\bibitem [{\citenamefont {Friedrichs}\ and\ \citenamefont {Oppenheim}(1954)}]{Friedrichs1954MathematicalAO}%
  \BibitemOpen
  \bibfield  {author} {\bibinfo {author} {\bibfnamefont {K.~O.}\ \bibnamefont {Friedrichs}}\ and\ \bibinfo {author} {\bibfnamefont {I.}~\bibnamefont {Oppenheim}},\ }\bibfield  {title} {\enquote {\bibinfo {title} {Mathematical aspects of the quantum theory of fields},}\ }\href@noop {} {\bibfield  {journal} {\bibinfo  {journal} {Physics Today}\ }\textbf {\bibinfo {volume} {7}},\ \bibinfo {pages} {23--23} (\bibinfo {year} {1954})}\BibitemShut {NoStop}%
\bibitem [{\citenamefont {Yurke}\ and\ \citenamefont {Stoler}(1986)}]{PhysRevLett.57.13}%
  \BibitemOpen
  \bibfield  {author} {\bibinfo {author} {\bibfnamefont {B.}~\bibnamefont {Yurke}}\ and\ \bibinfo {author} {\bibfnamefont {D.}~\bibnamefont {Stoler}},\ }\bibfield  {title} {\enquote {\bibinfo {title} {Generating quantum mechanical superpositions of macroscopically distinguishable states via amplitude dispersion},}\ }\href {\doibase 10.1103/PhysRevLett.57.13} {\bibfield  {journal} {\bibinfo  {journal} {Phys. Rev. Lett.}\ }\textbf {\bibinfo {volume} {57}},\ \bibinfo {pages} {13--16} (\bibinfo {year} {1986})}\BibitemShut {NoStop}%
\bibitem [{\citenamefont {Sanders}(1992)}]{PhysRevA.45.6811}%
  \BibitemOpen
  \bibfield  {author} {\bibinfo {author} {\bibfnamefont {B.~C.}\ \bibnamefont {Sanders}},\ }\bibfield  {title} {\enquote {\bibinfo {title} {Entangled coherent states},}\ }\href {\doibase 10.1103/PhysRevA.45.6811} {\bibfield  {journal} {\bibinfo  {journal} {Phys. Rev. A}\ }\textbf {\bibinfo {volume} {45}},\ \bibinfo {pages} {6811--6815} (\bibinfo {year} {1992})}\BibitemShut {NoStop}%
\bibitem [{\citenamefont {Wineland}(2013)}]{RevModPhys.85.1103}%
  \BibitemOpen
  \bibfield  {author} {\bibinfo {author} {\bibfnamefont {D.~J.}\ \bibnamefont {Wineland}},\ }\bibfield  {title} {\enquote {\bibinfo {title} {{N}obel {L}ecture: {S}uperposition, entanglement, and raising schr\"{o}dinger's cat},}\ }\href {\doibase 10.1103/RevModPhys.85.1103} {\bibfield  {journal} {\bibinfo  {journal} {Rev. Mod. Phys.}\ }\textbf {\bibinfo {volume} {85}},\ \bibinfo {pages} {1103--1114} (\bibinfo {year} {2013})}\BibitemShut {NoStop}%
\bibitem [{\citenamefont {Boto}\ \emph {et~al.}(2000)\citenamefont {Boto}, \citenamefont {Kok}, \citenamefont {Abrams}, \citenamefont {Braunstein}, \citenamefont {Williams},\ and\ \citenamefont {Dowling}}]{PhysRevLett.85.2733}%
  \BibitemOpen
  \bibfield  {author} {\bibinfo {author} {\bibfnamefont {A.~N.}\ \bibnamefont {Boto}}, \bibinfo {author} {\bibfnamefont {P.}~\bibnamefont {Kok}}, \bibinfo {author} {\bibfnamefont {D.~S.}\ \bibnamefont {Abrams}}, \bibinfo {author} {\bibfnamefont {S.~L.}\ \bibnamefont {Braunstein}}, \bibinfo {author} {\bibfnamefont {C.~P.}\ \bibnamefont {Williams}}, \ and\ \bibinfo {author} {\bibfnamefont {J.~P.}\ \bibnamefont {Dowling}},\ }\bibfield  {title} {\enquote {\bibinfo {title} {Quantum {I}nterferometric {O}ptical {L}ithography: {E}xploiting {E}ntanglement to {B}eat the {D}iffraction {L}imit},}\ }\href {\doibase 10.1103/PhysRevLett.85.2733} {\bibfield  {journal} {\bibinfo  {journal} {Phys. Rev. Lett.}\ }\textbf {\bibinfo {volume} {85}},\ \bibinfo {pages} {2733--2736} (\bibinfo {year} {2000})}\BibitemShut {NoStop}%
\bibitem [{\citenamefont {Horodecki}\ \emph {et~al.}(2009)\citenamefont {Horodecki}, \citenamefont {Horodecki}, \citenamefont {Horodecki},\ and\ \citenamefont {Horodecki}}]{RevModPhys.81.865}%
  \BibitemOpen
  \bibfield  {author} {\bibinfo {author} {\bibfnamefont {R.}~\bibnamefont {Horodecki}}, \bibinfo {author} {\bibfnamefont {P.}~\bibnamefont {Horodecki}}, \bibinfo {author} {\bibfnamefont {M.}~\bibnamefont {Horodecki}}, \ and\ \bibinfo {author} {\bibfnamefont {K.}~\bibnamefont {Horodecki}},\ }\bibfield  {title} {\enquote {\bibinfo {title} {Quantum entanglement},}\ }\href {\doibase 10.1103/RevModPhys.81.865} {\bibfield  {journal} {\bibinfo  {journal} {Rev. Mod. Phys.}\ }\textbf {\bibinfo {volume} {81}},\ \bibinfo {pages} {865--942} (\bibinfo {year} {2009})}\BibitemShut {NoStop}%
\bibitem [{\citenamefont {Huver}, \citenamefont {Wildfeuer},\ and\ \citenamefont {Dowling}(2008)}]{PhysRevA.78.063828}%
  \BibitemOpen
  \bibfield  {author} {\bibinfo {author} {\bibfnamefont {S.~D.}\ \bibnamefont {Huver}}, \bibinfo {author} {\bibfnamefont {C.~F.}\ \bibnamefont {Wildfeuer}}, \ and\ \bibinfo {author} {\bibfnamefont {J.~P.}\ \bibnamefont {Dowling}},\ }\bibfield  {title} {\enquote {\bibinfo {title} {Entangled {F}ock states for robust quantum optical metrology, imaging, and sensing},}\ }\href {\doibase 10.1103/PhysRevA.78.063828} {\bibfield  {journal} {\bibinfo  {journal} {Phys. Rev. A}\ }\textbf {\bibinfo {volume} {78}},\ \bibinfo {pages} {063828} (\bibinfo {year} {2008})}\BibitemShut {NoStop}%
\bibitem [{\citenamefont {Wildfeuer}\ and\ \citenamefont {Schiller}(2003)}]{PhysRevA.67.053801}%
  \BibitemOpen
  \bibfield  {author} {\bibinfo {author} {\bibfnamefont {C.}~\bibnamefont {Wildfeuer}}\ and\ \bibinfo {author} {\bibfnamefont {D.~H.}\ \bibnamefont {Schiller}},\ }\bibfield  {title} {\enquote {\bibinfo {title} {Generation of entangled {N}-photon states in a two-mode {J}aynes-{C}ummings model},}\ }\href {\doibase 10.1103/PhysRevA.67.053801} {\bibfield  {journal} {\bibinfo  {journal} {Phys. Rev. A}\ }\textbf {\bibinfo {volume} {67}},\ \bibinfo {pages} {053801} (\bibinfo {year} {2003})}\BibitemShut {NoStop}%
\bibitem [{\citenamefont {Sanders}\ and\ \citenamefont {Gerry}(2014)}]{PhysRevA.90.045804}%
  \BibitemOpen
  \bibfield  {author} {\bibinfo {author} {\bibfnamefont {B.~C.}\ \bibnamefont {Sanders}}\ and\ \bibinfo {author} {\bibfnamefont {C.~C.}\ \bibnamefont {Gerry}},\ }\bibfield  {title} {\enquote {\bibinfo {title} {Connection between the {NOON} state and a superposition of {S}{U}(2) coherent states},}\ }\href {\doibase 10.1103/PhysRevA.90.045804} {\bibfield  {journal} {\bibinfo  {journal} {Phys. Rev. A}\ }\textbf {\bibinfo {volume} {90}},\ \bibinfo {pages} {045804} (\bibinfo {year} {2014})}\BibitemShut {NoStop}%
\bibitem [{\citenamefont {Caves}(1981)}]{PhysRevD.23.1693}%
  \BibitemOpen
  \bibfield  {author} {\bibinfo {author} {\bibfnamefont {C.~M.}\ \bibnamefont {Caves}},\ }\bibfield  {title} {\enquote {\bibinfo {title} {Quantum-mechanical noise in an interferometer},}\ }\href {\doibase 10.1103/PhysRevD.23.1693} {\bibfield  {journal} {\bibinfo  {journal} {Phys. Rev. D}\ }\textbf {\bibinfo {volume} {23}},\ \bibinfo {pages} {1693--1708} (\bibinfo {year} {1981})}\BibitemShut {NoStop}%
\bibitem [{\citenamefont {Pezz\'e}\ \emph {et~al.}(2007)\citenamefont {Pezz\'e}, \citenamefont {Smerzi}, \citenamefont {Khoury}, \citenamefont {Hodelin},\ and\ \citenamefont {Bouwmeester}}]{PhysRevLett.99.223602}%
  \BibitemOpen
  \bibfield  {author} {\bibinfo {author} {\bibfnamefont {L.}~\bibnamefont {Pezz\'e}}, \bibinfo {author} {\bibfnamefont {A.}~\bibnamefont {Smerzi}}, \bibinfo {author} {\bibfnamefont {G.}~\bibnamefont {Khoury}}, \bibinfo {author} {\bibfnamefont {J.~F.}\ \bibnamefont {Hodelin}}, \ and\ \bibinfo {author} {\bibfnamefont {D.}~\bibnamefont {Bouwmeester}},\ }\bibfield  {title} {\enquote {\bibinfo {title} {Phase {D}etection at the {Q}uantum {L}imit with {M}ultiphoton {M}ach-{Z}ehnder {I}nterferometry},}\ }\href {\doibase 10.1103/PhysRevLett.99.223602} {\bibfield  {journal} {\bibinfo  {journal} {Phys. Rev. Lett.}\ }\textbf {\bibinfo {volume} {99}},\ \bibinfo {pages} {223602} (\bibinfo {year} {2007})}\BibitemShut {NoStop}%
\bibitem [{\citenamefont {Barnett}, \citenamefont {Fabre},\ and\ \citenamefont {Maıtre}(2003)}]{Barnett2003}%
  \BibitemOpen
  \bibfield  {author} {\bibinfo {author} {\bibfnamefont {S.}~\bibnamefont {Barnett}}, \bibinfo {author} {\bibfnamefont {C.}~\bibnamefont {Fabre}}, \ and\ \bibinfo {author} {\bibfnamefont {A.}~\bibnamefont {Maıtre}},\ }\bibfield  {title} {\enquote {\bibinfo {title} {Ultimate quantum limits for resolution of beam displacements},}\ }\href {\doibase 10.1140/epjd/e2003-00003-3} {\bibfield  {journal} {\bibinfo  {journal} {The European Physical Journal D - Atomic, Molecular, Optical and Plasma Physics}\ }\textbf {\bibinfo {volume} {22}},\ \bibinfo {pages} {513--519} (\bibinfo {year} {2003})}\BibitemShut {NoStop}%
\bibitem [{\citenamefont {Gard}\ \emph {et~al.}(2017)\citenamefont {Gard}, \citenamefont {You}, \citenamefont {Mishra}, \citenamefont {Singh}, \citenamefont {Lee}, \citenamefont {Corbitt},\ and\ \citenamefont {Dowling}}]{Gard2017}%
  \BibitemOpen
  \bibfield  {author} {\bibinfo {author} {\bibfnamefont {B.~T.}\ \bibnamefont {Gard}}, \bibinfo {author} {\bibfnamefont {C.}~\bibnamefont {You}}, \bibinfo {author} {\bibfnamefont {D.~K.}\ \bibnamefont {Mishra}}, \bibinfo {author} {\bibfnamefont {R.}~\bibnamefont {Singh}}, \bibinfo {author} {\bibfnamefont {H.}~\bibnamefont {Lee}}, \bibinfo {author} {\bibfnamefont {T.~R.}\ \bibnamefont {Corbitt}}, \ and\ \bibinfo {author} {\bibfnamefont {J.~P.}\ \bibnamefont {Dowling}},\ }\bibfield  {title} {\enquote {\bibinfo {title} {Nearly optimal measurement schemes in a noisy {M}ach-{Z}ehnder interferometer with coherent and squeezed vacuum},}\ }\href {\doibase 10.1140/epjqt/s40507-017-0058-8} {\bibfield  {journal} {\bibinfo  {journal} {EPJ Quantum Technology}\ }\textbf {\bibinfo {volume} {4}},\ \bibinfo {pages} {4--4} (\bibinfo {year} {2017})}\BibitemShut {NoStop}%
\bibitem [{\citenamefont {Ataman}(2019)}]{PhysRevA.100.063821}%
  \BibitemOpen
  \bibfield  {author} {\bibinfo {author} {\bibfnamefont {S.}~\bibnamefont {Ataman}},\ }\bibfield  {title} {\enquote {\bibinfo {title} {Optimal {M}ach-{Z}ehnder phase sensitivity with {G}aussian states},}\ }\href {\doibase 10.1103/PhysRevA.100.063821} {\bibfield  {journal} {\bibinfo  {journal} {Phys. Rev. A}\ }\textbf {\bibinfo {volume} {100}},\ \bibinfo {pages} {063821} (\bibinfo {year} {2019})}\BibitemShut {NoStop}%
\bibitem [{\citenamefont {Lang}\ and\ \citenamefont {Caves}(2013)}]{PhysRevLett.111.173601}%
  \BibitemOpen
  \bibfield  {author} {\bibinfo {author} {\bibfnamefont {M.~D.}\ \bibnamefont {Lang}}\ and\ \bibinfo {author} {\bibfnamefont {C.~M.}\ \bibnamefont {Caves}},\ }\bibfield  {title} {\enquote {\bibinfo {title} {Optimal {Q}uantum-{E}nhanced {I}nterferometry {U}sing a {L}aser {P}ower {S}ource},}\ }\href {\doibase 10.1103/PhysRevLett.111.173601} {\bibfield  {journal} {\bibinfo  {journal} {Phys. Rev. Lett.}\ }\textbf {\bibinfo {volume} {111}},\ \bibinfo {pages} {173601} (\bibinfo {year} {2013})}\BibitemShut {NoStop}%
\bibitem [{\citenamefont {Acernese}\ and\ \citenamefont {\textit{et al}}(2014)}]{Acernese_2014}%
  \BibitemOpen
  \bibfield  {author} {\bibinfo {author} {\bibfnamefont {F.}~\bibnamefont {Acernese}}\ and\ \bibinfo {author} {\bibnamefont {\textit{et al}}},\ }\bibfield  {title} {\enquote {\bibinfo {title} {Advanced virgo: a second-generation interferometric gravitational wave detector},}\ }\href {\doibase 10.1088/0264-9381/32/2/024001} {\bibfield  {journal} {\bibinfo  {journal} {Classical and Quantum Gravity}\ }\textbf {\bibinfo {volume} {32}},\ \bibinfo {pages} {024001} (\bibinfo {year} {2014})}\BibitemShut {NoStop}%
\bibitem [{\citenamefont {Oelker}\ \emph {et~al.}(2014)\citenamefont {Oelker}, \citenamefont {Barsotti}, \citenamefont {Dwyer}, \citenamefont {Sigg},\ and\ \citenamefont {Mavalvala}}]{Oelker:14}%
  \BibitemOpen
  \bibfield  {author} {\bibinfo {author} {\bibfnamefont {E.}~\bibnamefont {Oelker}}, \bibinfo {author} {\bibfnamefont {L.}~\bibnamefont {Barsotti}}, \bibinfo {author} {\bibfnamefont {S.}~\bibnamefont {Dwyer}}, \bibinfo {author} {\bibfnamefont {D.}~\bibnamefont {Sigg}}, \ and\ \bibinfo {author} {\bibfnamefont {N.}~\bibnamefont {Mavalvala}},\ }\bibfield  {title} {\enquote {\bibinfo {title} {Squeezed light for advanced gravitational wave detectors and beyond},}\ }\href {\doibase 10.1364/OE.22.021106} {\bibfield  {journal} {\bibinfo  {journal} {Opt. Express}\ }\textbf {\bibinfo {volume} {22}},\ \bibinfo {pages} {21106--21121} (\bibinfo {year} {2014})}\BibitemShut {NoStop}%
\bibitem [{\citenamefont {Mehmet}\ and\ \citenamefont {Vahlbruch}(2018)}]{Mehmet_2018}%
  \BibitemOpen
  \bibfield  {author} {\bibinfo {author} {\bibfnamefont {M.}~\bibnamefont {Mehmet}}\ and\ \bibinfo {author} {\bibfnamefont {H.}~\bibnamefont {Vahlbruch}},\ }\bibfield  {title} {\enquote {\bibinfo {title} {High-efficiency squeezed light generation for gravitational wave detectors},}\ }\href {\doibase 10.1088/1361-6382/aaf448} {\bibfield  {journal} {\bibinfo  {journal} {Classical and Quantum Gravity}\ }\textbf {\bibinfo {volume} {36}},\ \bibinfo {pages} {015014} (\bibinfo {year} {2018})}\BibitemShut {NoStop}%
\bibitem [{\citenamefont {Vahlbruch}\ \emph {et~al.}(2018)\citenamefont {Vahlbruch}, \citenamefont {Wilken}, \citenamefont {Mehmet},\ and\ \citenamefont {Willke}}]{PhysRevLett.121.173601}%
  \BibitemOpen
  \bibfield  {author} {\bibinfo {author} {\bibfnamefont {H.}~\bibnamefont {Vahlbruch}}, \bibinfo {author} {\bibfnamefont {D.}~\bibnamefont {Wilken}}, \bibinfo {author} {\bibfnamefont {M.}~\bibnamefont {Mehmet}}, \ and\ \bibinfo {author} {\bibfnamefont {B.}~\bibnamefont {Willke}},\ }\bibfield  {title} {\enquote {\bibinfo {title} {Laser power stabilization beyond the shot noise limit using squeezed light},}\ }\href {\doibase 10.1103/PhysRevLett.121.173601} {\bibfield  {journal} {\bibinfo  {journal} {Phys. Rev. Lett.}\ }\textbf {\bibinfo {volume} {121}},\ \bibinfo {pages} {173601} (\bibinfo {year} {2018})}\BibitemShut {NoStop}%
\bibitem [{\citenamefont {Tse}\ and\ \citenamefont {et~al.}(2019)}]{PhysRevLett.123.231107}%
  \BibitemOpen
  \bibfield  {author} {\bibinfo {author} {\bibfnamefont {M.}~\bibnamefont {Tse}}\ and\ \bibinfo {author} {\bibnamefont {et~al.}},\ }\bibfield  {title} {\enquote {\bibinfo {title} {Quantum-enhanced advanced ligo detectors in the era of gravitational-wave astronomy},}\ }\href {\doibase 10.1103/PhysRevLett.123.231107} {\bibfield  {journal} {\bibinfo  {journal} {Phys. Rev. Lett.}\ }\textbf {\bibinfo {volume} {123}},\ \bibinfo {pages} {231107} (\bibinfo {year} {2019})}\BibitemShut {NoStop}%
\bibitem [{\citenamefont {Breitenbach}, \citenamefont {Schiller},\ and\ \citenamefont {Mlynek}(1997)}]{Breitenbach1997}%
  \BibitemOpen
  \bibfield  {author} {\bibinfo {author} {\bibfnamefont {G.}~\bibnamefont {Breitenbach}}, \bibinfo {author} {\bibfnamefont {S.}~\bibnamefont {Schiller}}, \ and\ \bibinfo {author} {\bibfnamefont {J.}~\bibnamefont {Mlynek}},\ }\bibfield  {title} {\enquote {\bibinfo {title} {Measurement of the quantum states of squeezed light},}\ }\href {\doibase 10.1038/387471a0} {\bibfield  {journal} {\bibinfo  {journal} {Nature}\ }\textbf {\bibinfo {volume} {387}},\ \bibinfo {pages} {471--475} (\bibinfo {year} {1997})}\BibitemShut {NoStop}%
\bibitem [{\citenamefont {Andersen}\ \emph {et~al.}(2016)\citenamefont {Andersen}, \citenamefont {Gehring}, \citenamefont {Marquardt},\ and\ \citenamefont {Leuchs}}]{Andersen_2016}%
  \BibitemOpen
  \bibfield  {author} {\bibinfo {author} {\bibfnamefont {U.~L.}\ \bibnamefont {Andersen}}, \bibinfo {author} {\bibfnamefont {T.}~\bibnamefont {Gehring}}, \bibinfo {author} {\bibfnamefont {C.}~\bibnamefont {Marquardt}}, \ and\ \bibinfo {author} {\bibfnamefont {G.}~\bibnamefont {Leuchs}},\ }\bibfield  {title} {\enquote {\bibinfo {title} {30 years of squeezed light generation},}\ }\href {\doibase 10.1088/0031-8949/91/5/053001} {\bibfield  {journal} {\bibinfo  {journal} {Physica Scripta}\ }\textbf {\bibinfo {volume} {91}},\ \bibinfo {pages} {053001} (\bibinfo {year} {2016})}\BibitemShut {NoStop}%
\bibitem [{\citenamefont {Shukla}\ \emph {et~al.}(2021)\citenamefont {Shukla}, \citenamefont {Salykina}, \citenamefont {Frascella}, \citenamefont {Mishra}, \citenamefont {Chekhova},\ and\ \citenamefont {Khalili}}]{Shukla:21}%
  \BibitemOpen
  \bibfield  {author} {\bibinfo {author} {\bibfnamefont {G.}~\bibnamefont {Shukla}}, \bibinfo {author} {\bibfnamefont {D.}~\bibnamefont {Salykina}}, \bibinfo {author} {\bibfnamefont {G.}~\bibnamefont {Frascella}}, \bibinfo {author} {\bibfnamefont {D.~K.}\ \bibnamefont {Mishra}}, \bibinfo {author} {\bibfnamefont {M.~V.}\ \bibnamefont {Chekhova}}, \ and\ \bibinfo {author} {\bibfnamefont {F.~Y.}\ \bibnamefont {Khalili}},\ }\bibfield  {title} {\enquote {\bibinfo {title} {Broadening the high sensitivity range of squeezing-assisted interferometers by means of two-channel detection},}\ }\href {\doibase 10.1364/OE.413391} {\bibfield  {journal} {\bibinfo  {journal} {Opt. Express}\ }\textbf {\bibinfo {volume} {29}},\ \bibinfo {pages} {95--104} (\bibinfo {year} {2021})}\BibitemShut {NoStop}%
\bibitem [{\citenamefont {Ataman}, \citenamefont {Preda},\ and\ \citenamefont {Ionicioiu}(2018)}]{PhysRevA.98.043856}%
  \BibitemOpen
  \bibfield  {author} {\bibinfo {author} {\bibfnamefont {S.}~\bibnamefont {Ataman}}, \bibinfo {author} {\bibfnamefont {A.}~\bibnamefont {Preda}}, \ and\ \bibinfo {author} {\bibfnamefont {R.}~\bibnamefont {Ionicioiu}},\ }\bibfield  {title} {\enquote {\bibinfo {title} {Phase sensitivity of a {M}ach-{Z}ehnder interferometer with single-intensity and difference-intensity detection},}\ }\href {\doibase 10.1103/PhysRevA.98.043856} {\bibfield  {journal} {\bibinfo  {journal} {Phys. Rev. A}\ }\textbf {\bibinfo {volume} {98}},\ \bibinfo {pages} {043856} (\bibinfo {year} {2018})}\BibitemShut {NoStop}%
\bibitem [{\citenamefont {Takeoka}\ \emph {et~al.}(2017)\citenamefont {Takeoka}, \citenamefont {Seshadreesan}, \citenamefont {You}, \citenamefont {Izumi},\ and\ \citenamefont {Dowling}}]{PhysRevA.96.052118}%
  \BibitemOpen
  \bibfield  {author} {\bibinfo {author} {\bibfnamefont {M.}~\bibnamefont {Takeoka}}, \bibinfo {author} {\bibfnamefont {K.~P.}\ \bibnamefont {Seshadreesan}}, \bibinfo {author} {\bibfnamefont {C.}~\bibnamefont {You}}, \bibinfo {author} {\bibfnamefont {S.}~\bibnamefont {Izumi}}, \ and\ \bibinfo {author} {\bibfnamefont {J.~P.}\ \bibnamefont {Dowling}},\ }\bibfield  {title} {\enquote {\bibinfo {title} {Fundamental precision limit of a {M}ach-{Z}ehnder interferometric sensor when one of the inputs is the vacuum},}\ }\href {\doibase 10.1103/PhysRevA.96.052118} {\bibfield  {journal} {\bibinfo  {journal} {Phys. Rev. A}\ }\textbf {\bibinfo {volume} {96}},\ \bibinfo {pages} {052118} (\bibinfo {year} {2017})}\BibitemShut {NoStop}%
\bibitem [{Note1()}]{Note1}%
  \BibitemOpen
  \bibinfo {note} {In experiment, decibel [dB] is a common unit of squeezing. The degree of squeezing in dB is calculated according to $10\ \protect \text {log}_{10}e^{-2r}$, where $r$ is the squeezing parameter.}\BibitemShut {Stop}%
\bibitem [{\citenamefont {Schnabel}(2017)}]{SCHNABEL20171}%
  \BibitemOpen
  \bibfield  {author} {\bibinfo {author} {\bibfnamefont {R.}~\bibnamefont {Schnabel}},\ }\bibfield  {title} {\enquote {\bibinfo {title} {Squeezed states of light and their applications in laser interferometers},}\ }\href {\doibase https://doi.org/10.1016/j.physrep.2017.04.001} {\bibfield  {journal} {\bibinfo  {journal} {Physics Reports}\ }\textbf {\bibinfo {volume} {684}},\ \bibinfo {pages} {1--51} (\bibinfo {year} {2017})}\BibitemShut {NoStop}%
\bibitem [{\citenamefont {Vahlbruch}\ \emph {et~al.}(2016)\citenamefont {Vahlbruch}, \citenamefont {Mehmet}, \citenamefont {Danzmann},\ and\ \citenamefont {Schnabel}}]{PhysRevLett.117.110801}%
  \BibitemOpen
  \bibfield  {author} {\bibinfo {author} {\bibfnamefont {H.}~\bibnamefont {Vahlbruch}}, \bibinfo {author} {\bibfnamefont {M.}~\bibnamefont {Mehmet}}, \bibinfo {author} {\bibfnamefont {K.}~\bibnamefont {Danzmann}}, \ and\ \bibinfo {author} {\bibfnamefont {R.}~\bibnamefont {Schnabel}},\ }\bibfield  {title} {\enquote {\bibinfo {title} {Detection of 15 db squeezed states of light and their application for the absolute calibration of photoelectric quantum efficiency},}\ }\href {\doibase 10.1103/PhysRevLett.117.110801} {\bibfield  {journal} {\bibinfo  {journal} {Phys. Rev. Lett.}\ }\textbf {\bibinfo {volume} {117}},\ \bibinfo {pages} {110801} (\bibinfo {year} {2016})}\BibitemShut {NoStop}%
\bibitem [{\citenamefont {Sch\"{o}nbeck}, \citenamefont {Thies},\ and\ \citenamefont {Schnabel}(2018)}]{Schonbeck:18}%
  \BibitemOpen
  \bibfield  {author} {\bibinfo {author} {\bibfnamefont {A.}~\bibnamefont {Sch\"{o}nbeck}}, \bibinfo {author} {\bibfnamefont {F.}~\bibnamefont {Thies}}, \ and\ \bibinfo {author} {\bibfnamefont {R.}~\bibnamefont {Schnabel}},\ }\bibfield  {title} {\enquote {\bibinfo {title} {13 d{B} squeezed vacuum states at 1550 nm from 12 m{W} external pump power at 775 nm},}\ }\href {\doibase 10.1364/OL.43.000110} {\bibfield  {journal} {\bibinfo  {journal} {Opt. Lett.}\ }\textbf {\bibinfo {volume} {43}},\ \bibinfo {pages} {110--113} (\bibinfo {year} {2018})}\BibitemShut {NoStop}%
\bibitem [{\citenamefont {Pezz\'e}\ and\ \citenamefont {Smerzi}(2008)}]{PhysRevLett.100.073601}%
  \BibitemOpen
  \bibfield  {author} {\bibinfo {author} {\bibfnamefont {L.}~\bibnamefont {Pezz\'e}}\ and\ \bibinfo {author} {\bibfnamefont {A.}~\bibnamefont {Smerzi}},\ }\bibfield  {title} {\enquote {\bibinfo {title} {Mach-{Z}ehnder {I}nterferometry at the {H}eisenberg {L}imit with {C}oherent and {S}queezed-{V}acuum {L}ight},}\ }\href {\doibase 10.1103/PhysRevLett.100.073601} {\bibfield  {journal} {\bibinfo  {journal} {Phys. Rev. Lett.}\ }\textbf {\bibinfo {volume} {100}},\ \bibinfo {pages} {073601} (\bibinfo {year} {2008})}\BibitemShut {NoStop}%
\bibitem [{\citenamefont {Zhang}\ \emph {et~al.}(2018)\citenamefont {Zhang}, \citenamefont {Um}, \citenamefont {Lv}, \citenamefont {Zhang}, \citenamefont {Duan},\ and\ \citenamefont {Kim}}]{PhysRevLett.121.160502}%
  \BibitemOpen
  \bibfield  {author} {\bibinfo {author} {\bibfnamefont {J.}~\bibnamefont {Zhang}}, \bibinfo {author} {\bibfnamefont {M.}~\bibnamefont {Um}}, \bibinfo {author} {\bibfnamefont {D.}~\bibnamefont {Lv}}, \bibinfo {author} {\bibfnamefont {J.-N.}\ \bibnamefont {Zhang}}, \bibinfo {author} {\bibfnamefont {L.-M.}\ \bibnamefont {Duan}}, \ and\ \bibinfo {author} {\bibfnamefont {K.}~\bibnamefont {Kim}},\ }\bibfield  {title} {\enquote {\bibinfo {title} {Noon states of nine quantized vibrations in two radial modes of a trapped ion},}\ }\href {\doibase 10.1103/PhysRevLett.121.160502} {\bibfield  {journal} {\bibinfo  {journal} {Phys. Rev. Lett.}\ }\textbf {\bibinfo {volume} {121}},\ \bibinfo {pages} {160502} (\bibinfo {year} {2018})}\BibitemShut {NoStop}%
\bibitem [{\citenamefont {Vanhaele}\ and\ \citenamefont {Schlagheck}(2021)}]{PhysRevA.103.013315}%
  \BibitemOpen
  \bibfield  {author} {\bibinfo {author} {\bibfnamefont {G.}~\bibnamefont {Vanhaele}}\ and\ \bibinfo {author} {\bibfnamefont {P.}~\bibnamefont {Schlagheck}},\ }\bibfield  {title} {\enquote {\bibinfo {title} {Noon states with ultracold bosonic atoms via resonance- and chaos-assisted tunneling},}\ }\href {\doibase 10.1103/PhysRevA.103.013315} {\bibfield  {journal} {\bibinfo  {journal} {Phys. Rev. A}\ }\textbf {\bibinfo {volume} {103}},\ \bibinfo {pages} {013315} (\bibinfo {year} {2021})}\BibitemShut {NoStop}%
\bibitem [{\citenamefont {Mishra}\ \emph {et~al.}(2020)\citenamefont {Mishra}, \citenamefont {Shukla}, \citenamefont {Yadav},\ and\ \citenamefont {Mishra}}]{Mishra2020}%
  \BibitemOpen
  \bibfield  {author} {\bibinfo {author} {\bibfnamefont {K.~K.}\ \bibnamefont {Mishra}}, \bibinfo {author} {\bibfnamefont {G.}~\bibnamefont {Shukla}}, \bibinfo {author} {\bibfnamefont {D.}~\bibnamefont {Yadav}}, \ and\ \bibinfo {author} {\bibfnamefont {D.~K.}\ \bibnamefont {Mishra}},\ }\bibfield  {title} {\enquote {\bibinfo {title} {Generation of sum- and difference-squeezing by the beam splitter having third-order nonlinear material},}\ }\href {\doibase 10.1007/s11082-020-02303-x} {\bibfield  {journal} {\bibinfo  {journal} {Optical and Quantum Electronics}\ }\textbf {\bibinfo {volume} {52}},\ \bibinfo {pages} {186} (\bibinfo {year} {2020})}\BibitemShut {NoStop}%
\bibitem [{\citenamefont {Yadav}\ \emph {et~al.}(2021)\citenamefont {Yadav}, \citenamefont {Mishra}, \citenamefont {Shukla},\ and\ \citenamefont {Mishra}}]{Yadav2021}%
  \BibitemOpen
  \bibfield  {author} {\bibinfo {author} {\bibfnamefont {D.}~\bibnamefont {Yadav}}, \bibinfo {author} {\bibfnamefont {K.~K.}\ \bibnamefont {Mishra}}, \bibinfo {author} {\bibfnamefont {G.}~\bibnamefont {Shukla}}, \ and\ \bibinfo {author} {\bibfnamefont {D.~K.}\ \bibnamefont {Mishra}},\ }\bibfield  {title} {\enquote {\bibinfo {title} {Enhancement of amplitude-squared squeezing of light with the {SU}(3) multiport beam splitters},}\ }\href {\doibase 10.1007/s11082-021-02773-7} {\bibfield  {journal} {\bibinfo  {journal} {Optical and Quantum Electronics}\ }\textbf {\bibinfo {volume} {53}},\ \bibinfo {pages} {133} (\bibinfo {year} {2021})}\BibitemShut {NoStop}%
\bibitem [{\citenamefont {Dirmeier}\ \emph {et~al.}(2020)\citenamefont {Dirmeier}, \citenamefont {Tiedau}, \citenamefont {Khan}, \citenamefont {Ansari}, \citenamefont {M\"{u}ller}, \citenamefont {Silberhorn}, \citenamefont {Marquardt},\ and\ \citenamefont {Leuchs}}]{Dirmeier:20}%
  \BibitemOpen
  \bibfield  {author} {\bibinfo {author} {\bibfnamefont {T.}~\bibnamefont {Dirmeier}}, \bibinfo {author} {\bibfnamefont {J.}~\bibnamefont {Tiedau}}, \bibinfo {author} {\bibfnamefont {I.}~\bibnamefont {Khan}}, \bibinfo {author} {\bibfnamefont {V.}~\bibnamefont {Ansari}}, \bibinfo {author} {\bibfnamefont {C.~R.}\ \bibnamefont {M\"{u}ller}}, \bibinfo {author} {\bibfnamefont {C.}~\bibnamefont {Silberhorn}}, \bibinfo {author} {\bibfnamefont {C.}~\bibnamefont {Marquardt}}, \ and\ \bibinfo {author} {\bibfnamefont {G.}~\bibnamefont {Leuchs}},\ }\bibfield  {title} {\enquote {\bibinfo {title} {Distillation of squeezing using an engineered pulsed parametric down-conversion source},}\ }\href {\doibase 10.1364/OE.402178} {\bibfield  {journal} {\bibinfo  {journal} {Opt. Express}\ }\textbf {\bibinfo {volume} {28}},\ \bibinfo {pages} {30784--30796} (\bibinfo {year} {2020})}\BibitemShut {NoStop}%
\bibitem [{\citenamefont {Chaba}, \citenamefont {Collett},\ and\ \citenamefont {Walls}(1992)}]{PhysRevA.46.1499}%
  \BibitemOpen
  \bibfield  {author} {\bibinfo {author} {\bibfnamefont {A.~N.}\ \bibnamefont {Chaba}}, \bibinfo {author} {\bibfnamefont {M.~J.}\ \bibnamefont {Collett}}, \ and\ \bibinfo {author} {\bibfnamefont {D.~F.}\ \bibnamefont {Walls}},\ }\bibfield  {title} {\enquote {\bibinfo {title} {Quantum-nondemolition-measurement scheme using a kerr medium},}\ }\href {\doibase 10.1103/PhysRevA.46.1499} {\bibfield  {journal} {\bibinfo  {journal} {Phys. Rev. A}\ }\textbf {\bibinfo {volume} {46}},\ \bibinfo {pages} {1499--1506} (\bibinfo {year} {1992})}\BibitemShut {NoStop}%
\bibitem [{\citenamefont {Bocko}\ and\ \citenamefont {Onofrio}(1996)}]{RevModPhys.68.755}%
  \BibitemOpen
  \bibfield  {author} {\bibinfo {author} {\bibfnamefont {M.~F.}\ \bibnamefont {Bocko}}\ and\ \bibinfo {author} {\bibfnamefont {R.}~\bibnamefont {Onofrio}},\ }\bibfield  {title} {\enquote {\bibinfo {title} {On the measurement of a weak classical force coupled to a harmonic oscillator: experimental progress},}\ }\href {\doibase 10.1103/RevModPhys.68.755} {\bibfield  {journal} {\bibinfo  {journal} {Rev. Mod. Phys.}\ }\textbf {\bibinfo {volume} {68}},\ \bibinfo {pages} {755--799} (\bibinfo {year} {1996})}\BibitemShut {NoStop}%
\bibitem [{\citenamefont {Loudon}\ and\ \citenamefont {Knight}(1987)}]{doi:10.1080/09500348714550721}%
  \BibitemOpen
  \bibfield  {author} {\bibinfo {author} {\bibfnamefont {R.}~\bibnamefont {Loudon}}\ and\ \bibinfo {author} {\bibfnamefont {P.}~\bibnamefont {Knight}},\ }\bibfield  {title} {\enquote {\bibinfo {title} {Squeezed light},}\ }\href {\doibase 10.1080/09500348714550721} {\bibfield  {journal} {\bibinfo  {journal} {Journal of Modern Optics}\ }\textbf {\bibinfo {volume} {34}},\ \bibinfo {pages} {709--759} (\bibinfo {year} {1987})}\BibitemShut {NoStop}%
\bibitem [{\citenamefont {Sundar}(1996)}]{PhysRevA.53.1096}%
  \BibitemOpen
  \bibfield  {author} {\bibinfo {author} {\bibfnamefont {K.}~\bibnamefont {Sundar}},\ }\bibfield  {title} {\enquote {\bibinfo {title} {Amplitude-squeezed quantum states produced by the evolution of a quadrature-squeezed coherent state in a kerr medium},}\ }\href {\doibase 10.1103/PhysRevA.53.1096} {\bibfield  {journal} {\bibinfo  {journal} {Phys. Rev. A}\ }\textbf {\bibinfo {volume} {53}},\ \bibinfo {pages} {1096--1111} (\bibinfo {year} {1996})}\BibitemShut {NoStop}%
\bibitem [{\citenamefont {Gl\"ockl}, \citenamefont {Andersen},\ and\ \citenamefont {Leuchs}(2006)}]{PhysRevA.73.012306}%
  \BibitemOpen
  \bibfield  {author} {\bibinfo {author} {\bibfnamefont {O.}~\bibnamefont {Gl\"ockl}}, \bibinfo {author} {\bibfnamefont {U.~L.}\ \bibnamefont {Andersen}}, \ and\ \bibinfo {author} {\bibfnamefont {G.}~\bibnamefont {Leuchs}},\ }\bibfield  {title} {\enquote {\bibinfo {title} {Verifying continuous-variable entanglement of intense light pulses},}\ }\href {\doibase 10.1103/PhysRevA.73.012306} {\bibfield  {journal} {\bibinfo  {journal} {Phys. Rev. A}\ }\textbf {\bibinfo {volume} {73}},\ \bibinfo {pages} {012306} (\bibinfo {year} {2006})}\BibitemShut {NoStop}%
\bibitem [{\citenamefont {Silberhorn}\ \emph {et~al.}(2001)\citenamefont {Silberhorn}, \citenamefont {Lam}, \citenamefont {Wei\ss{}}, \citenamefont {K\"onig}, \citenamefont {Korolkova},\ and\ \citenamefont {Leuchs}}]{PhysRevLett.86.4267}%
  \BibitemOpen
  \bibfield  {author} {\bibinfo {author} {\bibfnamefont {C.}~\bibnamefont {Silberhorn}}, \bibinfo {author} {\bibfnamefont {P.~K.}\ \bibnamefont {Lam}}, \bibinfo {author} {\bibfnamefont {O.}~\bibnamefont {Wei\ss{}}}, \bibinfo {author} {\bibfnamefont {F.}~\bibnamefont {K\"onig}}, \bibinfo {author} {\bibfnamefont {N.}~\bibnamefont {Korolkova}}, \ and\ \bibinfo {author} {\bibfnamefont {G.}~\bibnamefont {Leuchs}},\ }\bibfield  {title} {\enquote {\bibinfo {title} {Generation of continuous variable einstein-podolsky-rosen entanglement via the kerr nonlinearity in an optical fiber},}\ }\href {\doibase 10.1103/PhysRevLett.86.4267} {\bibfield  {journal} {\bibinfo  {journal} {Phys. Rev. Lett.}\ }\textbf {\bibinfo {volume} {86}},\ \bibinfo {pages} {4267--4270} (\bibinfo {year} {2001})}\BibitemShut {NoStop}%
\bibitem [{\citenamefont {McCormick}\ \emph {et~al.}(2008)\citenamefont {McCormick}, \citenamefont {Marino}, \citenamefont {Boyer},\ and\ \citenamefont {Lett}}]{PhysRevA.78.043816}%
  \BibitemOpen
  \bibfield  {author} {\bibinfo {author} {\bibfnamefont {C.~F.}\ \bibnamefont {McCormick}}, \bibinfo {author} {\bibfnamefont {A.~M.}\ \bibnamefont {Marino}}, \bibinfo {author} {\bibfnamefont {V.}~\bibnamefont {Boyer}}, \ and\ \bibinfo {author} {\bibfnamefont {P.~D.}\ \bibnamefont {Lett}},\ }\bibfield  {title} {\enquote {\bibinfo {title} {Strong low-frequency quantum correlations from a four-wave-mixing amplifier},}\ }\href {\doibase 10.1103/PhysRevA.78.043816} {\bibfield  {journal} {\bibinfo  {journal} {Phys. Rev. A}\ }\textbf {\bibinfo {volume} {78}},\ \bibinfo {pages} {043816} (\bibinfo {year} {2008})}\BibitemShut {NoStop}%
\bibitem [{\citenamefont {Guerrero}\ \emph {et~al.}(2022)\citenamefont {Guerrero}, \citenamefont {Celis}, \citenamefont {Martinelli},\ and\ \citenamefont {Florez}}]{https://doi.org/10.48550/arxiv.2201.10935}%
  \BibitemOpen
  \bibfield  {author} {\bibinfo {author} {\bibfnamefont {A.~M.}\ \bibnamefont {Guerrero}}, \bibinfo {author} {\bibfnamefont {R.~L.~R.}\ \bibnamefont {Celis}}, \bibinfo {author} {\bibfnamefont {M.}~\bibnamefont {Martinelli}}, \ and\ \bibinfo {author} {\bibfnamefont {H.~M.}\ \bibnamefont {Florez}},\ }\href {\doibase 10.48550/ARXIV.2201.10935} {\enquote {\bibinfo {title} {Spectral control of quantum correlations in four wave mixing using dressing fields},}\ } (\bibinfo {year} {2022})\BibitemShut {NoStop}%
\bibitem [{\citenamefont {Hosaka}\ \emph {et~al.}(2015)\citenamefont {Hosaka}, \citenamefont {Hirosawa}, \citenamefont {Sawada},\ and\ \citenamefont {Kannari}}]{Hosaka:15}%
  \BibitemOpen
  \bibfield  {author} {\bibinfo {author} {\bibfnamefont {A.}~\bibnamefont {Hosaka}}, \bibinfo {author} {\bibfnamefont {K.}~\bibnamefont {Hirosawa}}, \bibinfo {author} {\bibfnamefont {R.}~\bibnamefont {Sawada}}, \ and\ \bibinfo {author} {\bibfnamefont {F.}~\bibnamefont {Kannari}},\ }\bibfield  {title} {\enquote {\bibinfo {title} {Generation of photon-number squeezed states with a fiber-optic symmetric interferometer},}\ }\href {\doibase 10.1364/OE.23.018850} {\bibfield  {journal} {\bibinfo  {journal} {Opt. Express}\ }\textbf {\bibinfo {volume} {23}},\ \bibinfo {pages} {18850--18863} (\bibinfo {year} {2015})}\BibitemShut {NoStop}%
\bibitem [{\citenamefont {Rosenbluh}\ and\ \citenamefont {Shelby}(1991)}]{PhysRevLett.66.153}%
  \BibitemOpen
  \bibfield  {author} {\bibinfo {author} {\bibfnamefont {M.}~\bibnamefont {Rosenbluh}}\ and\ \bibinfo {author} {\bibfnamefont {R.~M.}\ \bibnamefont {Shelby}},\ }\bibfield  {title} {\enquote {\bibinfo {title} {Squeezed optical solitons},}\ }\href {\doibase 10.1103/PhysRevLett.66.153} {\bibfield  {journal} {\bibinfo  {journal} {Phys. Rev. Lett.}\ }\textbf {\bibinfo {volume} {66}},\ \bibinfo {pages} {153--156} (\bibinfo {year} {1991})}\BibitemShut {NoStop}%
\bibitem [{\citenamefont {Schmitt}\ \emph {et~al.}(1998)\citenamefont {Schmitt}, \citenamefont {Ficker}, \citenamefont {Wolff}, \citenamefont {K\"onig}, \citenamefont {Sizmann},\ and\ \citenamefont {Leuchs}}]{PhysRevLett.81.2446}%
  \BibitemOpen
  \bibfield  {author} {\bibinfo {author} {\bibfnamefont {S.}~\bibnamefont {Schmitt}}, \bibinfo {author} {\bibfnamefont {J.}~\bibnamefont {Ficker}}, \bibinfo {author} {\bibfnamefont {M.}~\bibnamefont {Wolff}}, \bibinfo {author} {\bibfnamefont {F.}~\bibnamefont {K\"onig}}, \bibinfo {author} {\bibfnamefont {A.}~\bibnamefont {Sizmann}}, \ and\ \bibinfo {author} {\bibfnamefont {G.}~\bibnamefont {Leuchs}},\ }\bibfield  {title} {\enquote {\bibinfo {title} {Photon-{N}umber {S}queezed {S}olitons from an {A}symmetric {F}iber-{O}ptic {S}agnac {I}nterferometer},}\ }\href {\doibase 10.1103/PhysRevLett.81.2446} {\bibfield  {journal} {\bibinfo  {journal} {Phys. Rev. Lett.}\ }\textbf {\bibinfo {volume} {81}},\ \bibinfo {pages} {2446--2449} (\bibinfo {year} {1998})}\BibitemShut {NoStop}%
\bibitem [{\citenamefont {Bergman}\ and\ \citenamefont {Haus}(1991)}]{Bergman:91}%
  \BibitemOpen
  \bibfield  {author} {\bibinfo {author} {\bibfnamefont {K.}~\bibnamefont {Bergman}}\ and\ \bibinfo {author} {\bibfnamefont {H.~A.}\ \bibnamefont {Haus}},\ }\bibfield  {title} {\enquote {\bibinfo {title} {Squeezing in fibers with optical pulses},}\ }\href {\doibase 10.1364/OL.16.000663} {\bibfield  {journal} {\bibinfo  {journal} {Opt. Lett.}\ }\textbf {\bibinfo {volume} {16}},\ \bibinfo {pages} {663--665} (\bibinfo {year} {1991})}\BibitemShut {NoStop}%
\bibitem [{\citenamefont {Bergman}\ \emph {et~al.}(1994)\citenamefont {Bergman}, \citenamefont {Haus}, \citenamefont {Ippen},\ and\ \citenamefont {Shirasaki}}]{Bergman:94}%
  \BibitemOpen
  \bibfield  {author} {\bibinfo {author} {\bibfnamefont {K.}~\bibnamefont {Bergman}}, \bibinfo {author} {\bibfnamefont {H.~A.}\ \bibnamefont {Haus}}, \bibinfo {author} {\bibfnamefont {E.~P.}\ \bibnamefont {Ippen}}, \ and\ \bibinfo {author} {\bibfnamefont {M.}~\bibnamefont {Shirasaki}},\ }\bibfield  {title} {\enquote {\bibinfo {title} {Squeezing in a fiber interferometer with a gigahertz pump},}\ }\href {\doibase 10.1364/OL.19.000290} {\bibfield  {journal} {\bibinfo  {journal} {Opt. Lett.}\ }\textbf {\bibinfo {volume} {19}},\ \bibinfo {pages} {290--292} (\bibinfo {year} {1994})}\BibitemShut {NoStop}%
\bibitem [{\citenamefont {Yu}, \citenamefont {Haus},\ and\ \citenamefont {Ippen}(2001)}]{Yu2001SolitonSA}%
  \BibitemOpen
  \bibfield  {author} {\bibinfo {author} {\bibfnamefont {C.~X.}\ \bibnamefont {Yu}}, \bibinfo {author} {\bibfnamefont {H.~A.}\ \bibnamefont {Haus}}, \ and\ \bibinfo {author} {\bibfnamefont {E.~P.}\ \bibnamefont {Ippen}},\ }\bibfield  {title} {\enquote {\bibinfo {title} {Soliton squeezing at the gigahertz rate in a sagnac loop.}}\ }\href@noop {} {\bibfield  {journal} {\bibinfo  {journal} {Optics letters}\ }\textbf {\bibinfo {volume} {26 10}},\ \bibinfo {pages} {669--71} (\bibinfo {year} {2001})}\BibitemShut {NoStop}%
\bibitem [{\citenamefont {Anashkina}\ \emph {et~al.}(2020)\citenamefont {Anashkina}, \citenamefont {Andrianov}, \citenamefont {Corney},\ and\ \citenamefont {Leuchs}}]{Anashkina:20}%
  \BibitemOpen
  \bibfield  {author} {\bibinfo {author} {\bibfnamefont {E.~A.}\ \bibnamefont {Anashkina}}, \bibinfo {author} {\bibfnamefont {A.~V.}\ \bibnamefont {Andrianov}}, \bibinfo {author} {\bibfnamefont {J.~F.}\ \bibnamefont {Corney}}, \ and\ \bibinfo {author} {\bibfnamefont {G.}~\bibnamefont {Leuchs}},\ }\bibfield  {title} {\enquote {\bibinfo {title} {Chalcogenide fibers for {K}err squeezing},}\ }\href {\doibase 10.1364/OL.400326} {\bibfield  {journal} {\bibinfo  {journal} {Opt. Lett.}\ }\textbf {\bibinfo {volume} {45}},\ \bibinfo {pages} {5299--5302} (\bibinfo {year} {2020})}\BibitemShut {NoStop}%
\bibitem [{\citenamefont {Anashkina}\ \emph {et~al.}(2021)\citenamefont {Anashkina}, \citenamefont {Sorokin}, \citenamefont {Leuchs},\ and\ \citenamefont {Andrianov}}]{ANASHKINA2021104843}%
  \BibitemOpen
  \bibfield  {author} {\bibinfo {author} {\bibfnamefont {E.}~\bibnamefont {Anashkina}}, \bibinfo {author} {\bibfnamefont {A.}~\bibnamefont {Sorokin}}, \bibinfo {author} {\bibfnamefont {G.}~\bibnamefont {Leuchs}}, \ and\ \bibinfo {author} {\bibfnamefont {A.}~\bibnamefont {Andrianov}},\ }\bibfield  {title} {\enquote {\bibinfo {title} {Quantum noise squeezing of {CW} light in tellurite glass fibres},}\ }\href {\doibase https://doi.org/10.1016/j.rinp.2021.104843} {\bibfield  {journal} {\bibinfo  {journal} {Results in Physics}\ }\textbf {\bibinfo {volume} {30}},\ \bibinfo {pages} {104843} (\bibinfo {year} {2021})}\BibitemShut {NoStop}%
\bibitem [{\citenamefont {Sorokin}\ \emph {et~al.}(2022)\citenamefont {Sorokin}, \citenamefont {Leuchs}, \citenamefont {Corney}, \citenamefont {Kalinin}, \citenamefont {Anashkina},\ and\ \citenamefont {Andrianov}}]{math10193477}%
  \BibitemOpen
  \bibfield  {author} {\bibinfo {author} {\bibfnamefont {A.~A.}\ \bibnamefont {Sorokin}}, \bibinfo {author} {\bibfnamefont {G.}~\bibnamefont {Leuchs}}, \bibinfo {author} {\bibfnamefont {J.~F.}\ \bibnamefont {Corney}}, \bibinfo {author} {\bibfnamefont {N.~A.}\ \bibnamefont {Kalinin}}, \bibinfo {author} {\bibfnamefont {E.~A.}\ \bibnamefont {Anashkina}}, \ and\ \bibinfo {author} {\bibfnamefont {A.~V.}\ \bibnamefont {Andrianov}},\ }\bibfield  {title} {\enquote {\bibinfo {title} {Towards {Q}uantum {N}oise {S}queezing for 2-{M}icron {L}ight with {T}ellurite and {C}halcogenide {F}ibers with {L}arge {K}err {N}onlinearity},}\ }\href {\doibase 10.3390/math10193477} {\bibfield  {journal} {\bibinfo  {journal} {Mathematics}\ }\textbf {\bibinfo {volume} {10}} (\bibinfo {year} {2022}),\ 10.3390/math10193477}\BibitemShut {NoStop}%
\bibitem [{\citenamefont {Gerry}\ and\ \citenamefont {Grobe}(1994)}]{PhysRevA.49.2033}%
  \BibitemOpen
  \bibfield  {author} {\bibinfo {author} {\bibfnamefont {C.~C.}\ \bibnamefont {Gerry}}\ and\ \bibinfo {author} {\bibfnamefont {R.}~\bibnamefont {Grobe}},\ }\bibfield  {title} {\enquote {\bibinfo {title} {Statistical properties of squeezed kerr states},}\ }\href {\doibase 10.1103/PhysRevA.49.2033} {\bibfield  {journal} {\bibinfo  {journal} {Phys. Rev. A}\ }\textbf {\bibinfo {volume} {49}},\ \bibinfo {pages} {2033--2039} (\bibinfo {year} {1994})}\BibitemShut {NoStop}%
\bibitem [{\citenamefont {Sizmann}\ and\ \citenamefont {Leuchs}(1999)}]{sizmann1999v}%
  \BibitemOpen
  \bibfield  {author} {\bibinfo {author} {\bibfnamefont {A.}~\bibnamefont {Sizmann}}\ and\ \bibinfo {author} {\bibfnamefont {G.}~\bibnamefont {Leuchs}},\ }\bibfield  {title} {\enquote {\bibinfo {title} {V the optical kerr effect and quantum optics in fibers},}\ }\href@noop {} {\bibfield  {journal} {\bibinfo  {journal} {Progress in optics}\ }\textbf {\bibinfo {volume} {39}},\ \bibinfo {pages} {373--469} (\bibinfo {year} {1999})}\BibitemShut {NoStop}%
\bibitem [{\citenamefont {Kitagawa}\ and\ \citenamefont {Yamamoto}(1986)}]{PhysRevA.34.3974}%
  \BibitemOpen
  \bibfield  {author} {\bibinfo {author} {\bibfnamefont {M.}~\bibnamefont {Kitagawa}}\ and\ \bibinfo {author} {\bibfnamefont {Y.}~\bibnamefont {Yamamoto}},\ }\bibfield  {title} {\enquote {\bibinfo {title} {Number-phase minimum-uncertainty state with reduced number uncertainty in a kerr nonlinear interferometer},}\ }\href {\doibase 10.1103/PhysRevA.34.3974} {\bibfield  {journal} {\bibinfo  {journal} {Phys. Rev. A}\ }\textbf {\bibinfo {volume} {34}},\ \bibinfo {pages} {3974--3988} (\bibinfo {year} {1986})}\BibitemShut {NoStop}%
\bibitem [{\citenamefont {Krylov}\ and\ \citenamefont {Bergman}(1998)}]{Krylov:98}%
  \BibitemOpen
  \bibfield  {author} {\bibinfo {author} {\bibfnamefont {D.}~\bibnamefont {Krylov}}\ and\ \bibinfo {author} {\bibfnamefont {K.}~\bibnamefont {Bergman}},\ }\bibfield  {title} {\enquote {\bibinfo {title} {Amplitude-squeezed solitons from an asymmetric fiber interferometer},}\ }\href {\doibase 10.1364/OL.23.001390} {\bibfield  {journal} {\bibinfo  {journal} {Opt. Lett.}\ }\textbf {\bibinfo {volume} {23}},\ \bibinfo {pages} {1390--1392} (\bibinfo {year} {1998})}\BibitemShut {NoStop}%
\bibitem [{\citenamefont {Mishra}(2010)}]{MISHRA20103284}%
  \BibitemOpen
  \bibfield  {author} {\bibinfo {author} {\bibfnamefont {D.~K.}\ \bibnamefont {Mishra}},\ }\bibfield  {title} {\enquote {\bibinfo {title} {Study of higher order non-classical properties of squeezed kerr state},}\ }\href {\doibase https://doi.org/10.1016/j.optcom.2010.04.007} {\bibfield  {journal} {\bibinfo  {journal} {Optics Communications}\ }\textbf {\bibinfo {volume} {283}},\ \bibinfo {pages} {3284--3290} (\bibinfo {year} {2010})}\BibitemShut {NoStop}%
\bibitem [{\citenamefont {Braunstein}\ and\ \citenamefont {Caves}(1994)}]{PhysRevLett.72.3439}%
  \BibitemOpen
  \bibfield  {author} {\bibinfo {author} {\bibfnamefont {S.~L.}\ \bibnamefont {Braunstein}}\ and\ \bibinfo {author} {\bibfnamefont {C.~M.}\ \bibnamefont {Caves}},\ }\bibfield  {title} {\enquote {\bibinfo {title} {Statistical distance and the geometry of quantum states},}\ }\href {\doibase 10.1103/PhysRevLett.72.3439} {\bibfield  {journal} {\bibinfo  {journal} {Phys. Rev. Lett.}\ }\textbf {\bibinfo {volume} {72}},\ \bibinfo {pages} {3439--3443} (\bibinfo {year} {1994})}\BibitemShut {NoStop}%
\bibitem [{\citenamefont {Braunstein}, \citenamefont {Caves},\ and\ \citenamefont {Milburn}(1996)}]{BRAUNSTEIN1996135}%
  \BibitemOpen
  \bibfield  {author} {\bibinfo {author} {\bibfnamefont {S.~L.}\ \bibnamefont {Braunstein}}, \bibinfo {author} {\bibfnamefont {C.~M.}\ \bibnamefont {Caves}}, \ and\ \bibinfo {author} {\bibfnamefont {G.}~\bibnamefont {Milburn}},\ }\bibfield  {title} {\enquote {\bibinfo {title} {Generalized {U}ncertainty {R}elations: {T}heory, {E}xamples, and {L}orentz {I}nvariance},}\ }\href {\doibase https://doi.org/10.1006/aphy.1996.0040} {\bibfield  {journal} {\bibinfo  {journal} {Annals of Physics}\ }\textbf {\bibinfo {volume} {247}},\ \bibinfo {pages} {135 -- 173} (\bibinfo {year} {1996})}\BibitemShut {NoStop}%
\bibitem [{\citenamefont {Gerry}\ and\ \citenamefont {Knight}(2004)}]{gerry_knight_2004}%
  \BibitemOpen
  \bibfield  {author} {\bibinfo {author} {\bibfnamefont {C.~C.}\ \bibnamefont {Gerry}}\ and\ \bibinfo {author} {\bibfnamefont {P.~L.}\ \bibnamefont {Knight}},\ }\href {\doibase 10.1017/CBO9780511791239} {\emph {\bibinfo {title} {Introductory Quantum Optics}}}\ (\bibinfo  {publisher} {Cambridge University Press},\ \bibinfo {year} {2004})\BibitemShut {NoStop}%
\bibitem [{\citenamefont {Loudon}(2000)}]{loudon2000quantum}%
  \BibitemOpen
  \bibfield  {author} {\bibinfo {author} {\bibfnamefont {R.}~\bibnamefont {Loudon}},\ }\href {https://books.google.co.in/books?id=AEkfajgqldoC} {\emph {\bibinfo {title} {The Quantum Theory of Light}}}\ (\bibinfo  {publisher} {OUP Oxford},\ \bibinfo {year} {2000})\BibitemShut {NoStop}%
\bibitem [{\citenamefont {Cram{\'e}r}(1999)}]{cramer1999mathematical}%
  \BibitemOpen
  \bibfield  {author} {\bibinfo {author} {\bibfnamefont {H.}~\bibnamefont {Cram{\'e}r}},\ }\href@noop {} {\emph {\bibinfo {title} {Mathematical methods of statistics}}},\ Vol.~\bibinfo {volume} {43}\ (\bibinfo  {publisher} {Princeton university press},\ \bibinfo {year} {1999})\BibitemShut {NoStop}%
\bibitem [{\citenamefont {Paris}(2009)}]{doi:10.1142/S0219749909004839}%
  \BibitemOpen
  \bibfield  {author} {\bibinfo {author} {\bibfnamefont {M.~G.~A.}\ \bibnamefont {Paris}},\ }\bibfield  {title} {\enquote {\bibinfo {title} {Quantum estimation for quantum technology},}\ }\href {\doibase 10.1142/S0219749909004839} {\bibfield  {journal} {\bibinfo  {journal} {International Journal of Quantum Information}\ }\textbf {\bibinfo {volume} {07}},\ \bibinfo {pages} {125--137} (\bibinfo {year} {2009})}\BibitemShut {NoStop}%
\bibitem [{\citenamefont {Ataman}(2020)}]{PhysRevA.102.013704}%
  \BibitemOpen
  \bibfield  {author} {\bibinfo {author} {\bibfnamefont {S.}~\bibnamefont {Ataman}},\ }\bibfield  {title} {\enquote {\bibinfo {title} {Single- versus two-parameter fisher information in quantum interferometry},}\ }\href {\doibase 10.1103/PhysRevA.102.013704} {\bibfield  {journal} {\bibinfo  {journal} {Phys. Rev. A}\ }\textbf {\bibinfo {volume} {102}},\ \bibinfo {pages} {013704} (\bibinfo {year} {2020})}\BibitemShut {NoStop}%
\bibitem [{\citenamefont {Agarwal}(2012)}]{Agarwal_2012}%
  \BibitemOpen
  \bibfield  {author} {\bibinfo {author} {\bibfnamefont {G.~S.}\ \bibnamefont {Agarwal}},\ }\href {\doibase 10.1017/cbo9781139035170} {\emph {\bibinfo {title} {Quantum Optics}}}\ (\bibinfo  {publisher} {Cambridge University Press},\ \bibinfo {address} {Cambridge, England},\ \bibinfo {year} {2012})\BibitemShut {NoStop}%
\bibitem [{\citenamefont {Messiah}(1999)}]{messiah1999quantum}%
  \BibitemOpen
  \bibfield  {author} {\bibinfo {author} {\bibfnamefont {A.}~\bibnamefont {Messiah}},\ }\href@noop {} {\emph {\bibinfo {title} {Quantum Mechanics}}}\ (\bibinfo  {publisher} {Dover},\ \bibinfo {year} {1999})\BibitemShut {NoStop}%
\bibitem [{\citenamefont {Kalinin}\ \emph {et~al.}(2023{\natexlab{a}})\citenamefont {Kalinin}, \citenamefont {Dirmeier}, \citenamefont {Sorokin}, \citenamefont {Anashkina}, \citenamefont {Sánchez-Soto}, \citenamefont {Corney}, \citenamefont {Leuchs},\ and\ \citenamefont {Andrianov}}]{KalininDirmeierSorokinAnashkinaSánchezSotoCorneyLeuchsAndrianov+2023}%
  \BibitemOpen
  \bibfield  {author} {\bibinfo {author} {\bibfnamefont {N.}~\bibnamefont {Kalinin}}, \bibinfo {author} {\bibfnamefont {T.}~\bibnamefont {Dirmeier}}, \bibinfo {author} {\bibfnamefont {A.~A.}\ \bibnamefont {Sorokin}}, \bibinfo {author} {\bibfnamefont {E.~A.}\ \bibnamefont {Anashkina}}, \bibinfo {author} {\bibfnamefont {L.~L.}\ \bibnamefont {Sánchez-Soto}}, \bibinfo {author} {\bibfnamefont {J.~F.}\ \bibnamefont {Corney}}, \bibinfo {author} {\bibfnamefont {G.}~\bibnamefont {Leuchs}}, \ and\ \bibinfo {author} {\bibfnamefont {A.~V.}\ \bibnamefont {Andrianov}},\ }\bibfield  {title} {\enquote {\bibinfo {title} {Quantum-enhanced interferometer using {K}err squeezing},}\ }\href {\doibase doi:10.1515/nanoph-2023-0032} {\bibfield  {journal} {\bibinfo  {journal} {Nanophotonics}\ } (\bibinfo {year} {2023}{\natexlab{a}}),\ doi:10.1515/nanoph-2023-0032}\BibitemShut {NoStop}%
\bibitem [{\citenamefont {Kalinin}\ \emph {et~al.}(2023{\natexlab{b}})\citenamefont {Kalinin}, \citenamefont {Dirmeier}, \citenamefont {Sorokin}, \citenamefont {Anashkina}, \citenamefont {S{\'a}nchez-Soto}, \citenamefont {Corney}, \citenamefont {Leuchs},\ and\ \citenamefont {Andrianov}}]{kalinin2023observation}%
  \BibitemOpen
  \bibfield  {author} {\bibinfo {author} {\bibfnamefont {N.}~\bibnamefont {Kalinin}}, \bibinfo {author} {\bibfnamefont {T.}~\bibnamefont {Dirmeier}}, \bibinfo {author} {\bibfnamefont {A.~A.}\ \bibnamefont {Sorokin}}, \bibinfo {author} {\bibfnamefont {E.~A.}\ \bibnamefont {Anashkina}}, \bibinfo {author} {\bibfnamefont {L.~L.}\ \bibnamefont {S{\'a}nchez-Soto}}, \bibinfo {author} {\bibfnamefont {J.~F.}\ \bibnamefont {Corney}}, \bibinfo {author} {\bibfnamefont {G.}~\bibnamefont {Leuchs}}, \ and\ \bibinfo {author} {\bibfnamefont {A.~V.}\ \bibnamefont {Andrianov}},\ }\bibfield  {title} {\enquote {\bibinfo {title} {Observation of robust polarization squeezing via the {K}err nonlinearity in an optical fiber},}\ }\href@noop {} {\bibfield  {journal} {\bibinfo  {journal} {Advanced Quantum Technologies}\ }\textbf {\bibinfo {volume} {6}},\ \bibinfo {pages} {2200143} (\bibinfo {year} {2023}{\natexlab{b}})}\BibitemShut {NoStop}%
\end{thebibliography}%

\end{document}